\documentclass[fleqn,usenatbib]{mnras}

\usepackage{newtxtext,newtxmath}
\usepackage{soul}
\usepackage{hyperref}
\usepackage[export]{adjustbox}

\usepackage[T1]{fontenc}

\DeclareRobustCommand{\VAN}[3]{#2}
\let\VANthebibliography\thebibliography
\def\thebibliography{\DeclareRobustCommand{\VAN}[3]{##3}\VANthebibliography}

\usepackage{graphicx}	
\usepackage{amsmath}	
\usepackage{orcidlink}
\usepackage{siunitx}
\usepackage{multirow}

\newcommand{\LCDM}{$\Lambda$CDM}	

\defcitealias{Labbe2023}{L23}

\title[Double-break galaxies in CANUCS]{$\Lambda$CDM not dead yet: massive high-z Balmer break galaxies are less common than previously reported}

\author[Desprez et al.]{Guillaume Desprez$^{1}$\thanks{e-mail: guillaume.desprez@smu.ca}\orcidlink{0000-0001-8325-1742},
Nicholas S. Martis$^{1,2}$\orcidlink{0000-0003-3243-9969},
Yoshihisa Asada$^{1,3}$\orcidlink{0000-0003-3983-5438},
Marcin Sawicki$^{1}$\thanks{Canada Research Chair}\orcidlink{0000-0002-7712-7857},
\newauthor
Chris J. Willott$^{2}$\orcidlink{0000-0002-4201-7367},
Adam Muzzin$^{4}$,
Roberto G. Abraham$^{5,6}$\orcidlink{0000-0002-4542-921X},
Maru\v{s}a Brada\v{c}$^{7,8}$\orcidlink{0000-0001-5984-0395},
Gabe Brammer$^{9,10}$\orcidlink{0000-0003-2680-005X}, \newauthor
Vicente Estrada-Carpenter$^{1}$\orcidlink{0000-0001-8489-2349},
Kartheik G. Iyer$^{11}$\orcidlink{0000-0001-9298-3523},
Jasleen Matharu$^{9,10}$\orcidlink{0000-0002-7547-3385},
Lamiya Mowla$^{12}$\orcidlink{0000-0002-8530-9765}, \newauthor
Ga\"el Noirot$^{1}$,
Ghassan T. E. Sarrouh$^{4}$\orcidlink{0000-0001-8830-2166},
Victoria Strait$^{9,10}$\orcidlink{0000-0002-6338-7295},
Rachel Gledhill$^{9,10}$,
and Gregor Rihtar\v{s}i\v{c}$^{7}$\orcidlink{0009-0009-4388-898X}\\
$^{1}$ Department of Astronomy \& Physics and Institute for Computational Astrophysics, Saint Mary's University, 923 Robie Street, Halifax, NS B3H 3C3, Canada\\
$^{2}$ NRC Herzberg, 5071 West Saanich Rd, Victoria, BC V9E 2E7, Canada\\
$^{3}$ Department of Astronomy, Kyoto University, Sakyo-ku, Kyoto 606-8502, Japan\\
$^{4}$ Department of Physics and Astronomy, York University, 4700 Keele St. Toronto, Ontario, M3J 1P3, Canada\\
$^{5}$ David A. Dunlap Department of Astronomy and Astrophysics, University of Toronto, 50 St. George Street, Toronto, Ontario, M5S 3H4, Canada \\
$^{6}$ Dunlap Institute for Astronomy and Astrophysics, 50 St. George Street, Toronto, Ontario, M5S 3H4, Canada \\
$^{7}$ Department of Mathematics and Physics, Jadranska ulica 19, SI-1000 Ljubljana, Slovenia \\
$^{8}$ Department of Physics and Astronomy, University of California Davis, 1 Shields Avenue, Davis, CA 95616, USA \\
$^{9}$ Cosmic Dawn Center (DAWN), Denmark\\
$^{10}$ Niels Bohr Institute, University of Copenhagen, Jagtvej 128, DK-2200 Copenhagen N, Denmark \\
$^{11}$ Columbia Astrophysics Laboratory, Columbia University, 550 West 120th Street, New York, NY 10027, USA \\
$^{12}$ Whitin Observatory, Department of Physics and Astronomy, Wellesley College, 106 Central Street, Wellesley, MA 02481, USA
}

\date{Accepted 2024 April 18. Received 2024 April 10; in original form 2023 September 29}

\pubyear{2023}

\begin{document}
\label{firstpage}
\pagerange{\pageref{firstpage}--\pageref{lastpage}}
\maketitle

\begin{abstract}
Early {\it JWST} observations that targeted so-called double-break sources (attributed to Lyman and Balmer breaks at $z>7$), reported a previously unknown population of very massive, evolved high-redshift galaxies. This surprising discovery led to a flurry of attempts to explain these objects' unexpected existence including invoking alternatives to the standard $\Lambda$CDM cosmological paradigm. To test these early results, we adopted the same double-break candidate galaxy selection criteria to search for such objects in the {\it JWST} images of the CAnadian NIRISS Unbiased Cluster Survey (CANUCS), and found a sample of 19 sources over five independent CANUCS fields that cover a total effective area of $\sim$60~{\rm arcmin}$^2$ at $z\sim8$. However, (1) our SED fits do not yield exceptionally high stellar masses for our candidates, while (2) spectroscopy of five of the candidates shows that while all five are  at high redshifts, their red colours are due to high-EW emission lines in star-forming galaxies rather than Balmer breaks in massive, evolved systems.  Additionally, (3) field-to-field variance leads to differences of $\sim 1.5$ dex in the maximum stellar masses measured in the different fields, suggesting that the early single-field  {\it JWST} observations may have suffered from cosmic variance and/or sample bias. Finally, (4) we show that the presence of even a single massive outlier can dominate conclusions from small samples such as those in early {\it JWST} observations.  In conclusion, we find that the double-break sources in CANUCS are not sufficiently massive or numerous to warrant questioning the standard $\Lambda$CDM paradigm.
\end{abstract}




\section{Introduction}

Observations of the cosmic dawn era  are crucial to our comprehension of how the first galaxies form and evolve. In its first year of operation, \emph{JWST} \citep{Gardner2023} opened a new window on this particular epoch, allowing for the observation of galaxies in a universe just a few 100~Myr old \citep{Curtis-Lake2023,Robertson2023,ArrabalHaro2023,Bunker2023,WangB2023}. Such observations have the potential to expand our knowledge on the Epoch of Reionisation (EoR) and its mechanism \citep[e.g.,][]{Robertson2022,Atek2023arXiv} as well as to challenge the current paradigm of $\Lambda$CDM cosmology. 

This is the case with the detection of several massive galaxies at high redshift reported by \citet[][hereafter \citetalias{Labbe2023}]{Labbe2023} in the Cosmic Evolution Early Release Science Survey (CEERS) data \citep{Bagley2023}.  These were selected on the basis of a double-break feature in their photometric spectral energy distributions (SED), attributed to the Lyman and Balmer breaks redshifted to $z>7$.  The very high stellar masses reported by \citetalias{Labbe2023} for these sources -- high in part due to the presence of evolved stellar populations marked by the putative Balmer breaks -- disagree with both model predictions \citep{Menci2022,Boylan-Kolchin2023,Lovell2023} and pre-{\it JWST} observations \citep{Stefanon2021,Laporte2023}.  Their discovery thus spurred a flurry of activity that attempted to explain these objects' surprising existence by invoking alternatives to the standard $\Lambda$CDM cosmological model \citep[][and many others]{
Biagetti2023,
Forconi2023,
Hutsi2023,
Lovell2023,
Malekjani2023arXiv,
Menci2022,
Su2023arXiv,
WangD2022arXiv,
WangSarXiv2023,
Dayal2024}. 

Since these first results, a number of studies have debated whether  {\it JWST} observations of these and other high redshift galaxies do indicate a tension between $\Lambda$CDM and the measured galaxy stellar masses in the early universe. For instance, \citet{Xiao2023arXiv} report, in the First Reionization Epoch Spectroscopic Complete Survey \citep[FRESCO,][]{Oesch2023}, the detection of $z>5$ galaxies whose stellar masses are hard to explain without assuming extreme star-formation efficiencies, at the limit of tension with standard cosmology. Additionally, galaxies with possible photometric redshifts (photo-$z$) larger than $z=10$ have been found in the COSMOS-Web survey \citep{Casey2023} with estimated stellar masses that require larger than expected star-formation efficiencies \citep{Casey2023arXiv}. On the other hand, \citet{Franco2023arXiv} presents a population of $z\geq 9$ sources in the same COSMOS-Web survey , for which the derived masses are not in tension with galaxy formation in $\Lambda$CDM. \citet{Vikaeus2024MNRAS.tmp.} looked at the strength of Balmer breaks in $z>6$ galaxy spectra and did not find severe deviation from the predictions of simulations \citep{Mauerhofer2023,Wilkins2024}.

Additionally, some studies have invoked astrophysical reasons why the reported high masses could be overestimated thereby alleviating or eliminating the tension with \LCDM.
For example, while searching for and characterising high-$z$ Balmer-break candidate galaxies in the Prime Extragalactic Areas for Reionization and Lensing Science \citep[PEARLS,][]{Windhorst2023}, \citet{Trussler2023arXiv} find that the large stellar masses derived in \citetalias{Labbe2023} could be due to the assumptions made in the mass inference, and suggest that the red break in the photometry could be caused by strong emission lines, dusty continuum, and active galactic nuclei (AGN) contributions. Thus an appropriate accounting of these contributions can alleviate the need for an evolved stellar population to fit the Balmer break, leading to stellar masses in better agreement with galaxy formation theory expectations \citep{Lovell2023}. A similar conclusion is reached by \citet{Barro2024} after looking at extremely red objects in CEERS. They find that their sources are well fit whether assuming massive and dusty SEDs or accounting for the contribution of AGN, although they conclude that both scenarios cause problems as either massive galaxies or bright AGN are found to be too numerous at high redshift.

Some sources from the \citetalias{Labbe2023} CEERS sample have already been studied further, mitigating the results presented by \citetalias{Labbe2023}. Two of these sources had spectroscopic follow-up, one being confirmed at redshift $z=7.9932$ \citep{Fujimoto2023} with the presence of H$_{\beta}$ and {\sc [Oiii]} emission lines, and the other being identified as an AGN at $z=5.624$ \citep{Kocevski2023}, thus being inconsistent with the initial assumption of lying at $z>7$. Also, \citet{Endsley2023} refit the most massive object from the CEERS double-break sample, but assuming different star formation histories, emission line and AGN contributions to the photometry, leading to lower stellar masses. 

However, while the debate in the literature continues, it is important to keep in mind that since high-$z$ galaxy selection methods vary and each is incomplete in some way, a conclusive disagreement of \emph{just one} of the different selection methods with \LCDM\ is sufficient to cause problems for the accepted cosmological paradigm. Put another way, if the type of sources reported in \citetalias{Labbe2023} turns out to be both very massive and common, then lack of disagreement between \LCDM\ and sources selected by other techniques is moot. 
To ensure that the tensions with \LCDM\ are real, it is thus important to consider and further test each challenging population. This test is what we aim to do for the double-break sources in order to assess whether their nature as high-mass high-$z$ galaxies holds and thus continues to  challenge \LCDM.

The scope of this work is thus to look for double-break sources similar to those in \citetalias{Labbe2023} and to understand if this population is really as numerous and massive as originally reported.  We do so using the deep imaging data of the CAnadian NIRISS Unbiased Cluster Survey \citep[CANUCS,][]{Willott2022}. CANUCS contains deep {\it JWST}/NIRCam \citep{Rieke2023} imaging over 10 fields along 5 widely separated lines-of-sight with total dual filter integration time of 123 hours, { covering a total area of $\sim100\,{\rm arcmin}^{2}$\footnote{CANUCS contains a further 5 fields of NIRISS spectroscopy and limited imaging although these are not used in the present work.}
The combination of imaging depth, multi-wavelength coverage, and independent sightlines makes CANUCS one of the most comprehensive extragalactic NIRCam imaging projects executed in {\it JWST} Cycle 1. Five of the ten CANUCS fields target strong-lensing cluster cores and are covered as well with {\it JWST}/NIRISS \citep{Doyon2023} slitless spectroscopy \citep{Willott2022} and {\it JWST}/NIRSpec \citep{Jakobsen2022} multi-object prism spectroscopy. The other five fields are parallel fields covered with only NIRCam imaging. }
{ With its design, the CANUCS survey allows for a control of cosmic and sample variances (field-to-field variance) and gives a representative view of the general galaxy population distributions.
It also benefits from deep ancillary {\it HST}/ACS and {\it HST}/WFC3 data providing up to 20 band photometry in some fields, thus allowing for tight fits of source SEDs. For all these reasons, CANUCS offers a prime data set to investigate the nature and properties of the bright double-break sources.}

Throughout this work, we assume flat $\Lambda$CDM cosmology, with $\Omega_\Lambda=0.7$, $\Omega_{\rm m}=0.3$, and $H_0=70~{\rm km\,s^{-1}\,Mpc^{-1}}$. Magnitudes are given in the AB system \citep{Oke1983}. We adopt the \citet{Chabrier2003} stellar initial mass function (IMF) in this paper and all results obtained with other IMFs are rescaled to it \citep{Speagle2014}.


\begin{figure*}
    \centering
    \includegraphics[width=\linewidth]{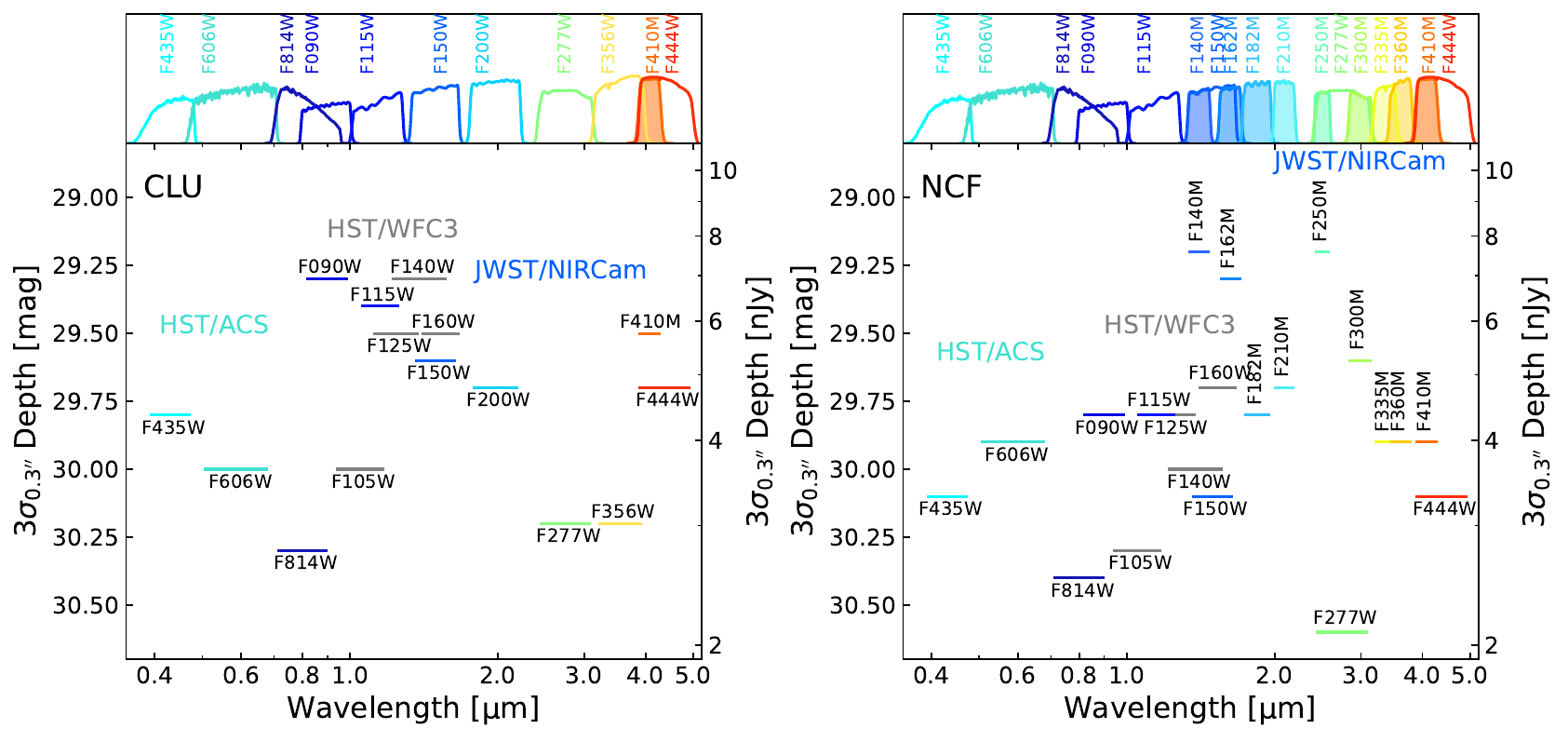}
    \caption{ Summary of the median depths (3$\sigma$ in $0\farcs 3$ apertures) measured across the five different fields in all {\it HST}/ACS and {\it JWST}/NIRCam filters observed, split by \emph{CLU} (left panels) and \emph{NCF} pointings (right panels). {\it HST}/WFC3 measurements have been included, however the coverage with these bands is uneven across the different fields, as shown in Table~\ref{tab:depth}.}
    \label{fig:Depth}
\end{figure*}

\section{Data}

We use the data acquired by the CANUCS NIRISS GTO Program \#1208 \citep[][\href{https://doi.org/10.17909/ph4n-6n76}{DOI}]{Willott2022} which  covers five different strong lensing cluster fields: Abell~370 ($z=0.375$), MACS~J0416.1$-$2403 (hereafter MACS~0416, $z=0.395$), MACS~J0417.5$-$1154 (hereafter MACS~0417, $z=0.443$), MACS~J1149.5$+$2223 (hereafter MACS~1149, $z=0.543$), and MACS~J1423.8$+$2404 (hereafter MACS~1423, $z=0.545$) \citep{Soucail1987, Ebeling2001}. As NIRCam and NIRISS are operated in parallel, each of the five cluster fields is accompanied by a NIRCam flanking field and a NIRISS flanking field. 

The cluster fields (denoted \emph{CLU} in Table~\ref{tab:fields}) are covered with the NIRCam F090W, F115W, F150W, F200W, F277W, F356W, F410M, and F444W filters with exposure times of $6.4~\mbox{ks}$ each. These fields are also covered by NIRISS with both GR150R and GR150C grisms through the F115W, F150W and F200W bands with total exposure $19.2~\mbox{ks}$ in all three filters.

In the NIRCam flanking fields (denoted \emph{NCF} in Table~\ref{tab:fields}), the medium-band F140M, F162M, F182M, F210M, F250M, F300M, F335M, F360M and F410M filters are used in addition to the broad-band F090W, F115W, F150W, F277W, and F444W, with exposure times of $10.3~\mbox{ks}$ for 10 filters and $5.7~\mbox{ks}$ for 4 filters. 

In all fields, we also use the available {\it HST} optical and near-IR imaging data. Depending on the field, there is coverage in some of ACS F435W, F606W, F814W, WFC3/UV F438W, F606W, and WFC3/IR F105W, F110W, F125W, F140W, F160W. These data are predominantly from programs HST-GO-11103 PI Ebeling, HST-GO-12009 PI von der Linden, HST-GO-12065 PI Postman, HST-GO-13504 PI Lotz, HST-GO-14096 PI Coe, HST-GO-15117 PI Steinhardt and HST-GO-16667 PI Brada\v{c}.

NIRSpec spectroscopy has been acquired for four of the five CANUCS clusters at the time of writing, and MACS~1149 spectroscopy is scheduled for December 2023. The spectra are observed using the PRISM/CLEAR disperser and filter, through three Micro-Shutter Assembly (MSA) masks per cluster with total exposure time of $2.9~\mbox{ks}$ per MSA configuration. 

\begin{table}
    \centering
    \caption{ Summary of the CANUCS fields used in this work. Each field has a unique two-digit Field ID code and all source IDs start with the Field ID code of the field they are in. The number of selected double-break galaxies is given in the Sources [\#] column. The sky areas and volumes between $z=7.0$--$8.5$ quoted in the table account for the lensing magnification assuming a redshift $z=8$ and area masked by other sources.}
    \begin{tabular}{l c c c c c c}
        \hline
        \hline
        \rule{0pt}{1.2em} Cluster & Field &  ID &  Sources & Area $z\approx 8$ & Volume \\
        & &  & [\#] & [arcmin$^2$] & [$10^3\,{\rm Mpc}^3$] \\
        \hline
         \rule{0pt}{1.2em} \multirow{2}{*}{MACS~0417}& \emph{CLU} & 11 & 2 & $4.57^{+0.47}_{-0.13}$ & $13.79^{+ 1.42}_{- 0.37}$\\
         \rule{0pt}{1.2em}&\emph{NCF} & 12 &  1 & $6.30^{+0.18}_{-0.09}$ & $19.01^{+0.48}_{-0.13}$ \\
         \hline
         \rule{0pt}{1.2em}\multirow{2}{*}{Abell~370}& \emph{CLU} & 21 & 4 & $3.54^{+0.11}_{-0.12}$ & $10.77^{+0.25}_{-0.3}$\\
         \rule{0pt}{1.2em}&\emph{NCF} & 22 &  1 & $6.30^{+0.03}_{-0.05}$ & $19.01^{+0.06}_{-0.1}$\\
         \hline
         \rule{0pt}{1.2em}\multirow{2}{*}{MACS~0416}& \emph{CLU}& 31 & 3 & $6.05^{+0.05}_{-0.06}$ & $18.23^{+0.09}_{-0.13}$\\
         \rule{0pt}{1.2em}&\emph{NCF}& 32 & 1 & $6.52^{+0.08}_{-0.08}$ & $19.65^{+0.02}_{-0.03}$\\
         \hline
         \rule{0pt}{1.2em}\multirow{2}{*}{MACS~1423}& \emph{CLU}& 41 & 2 & $5.73^{+0.21}_{-0.15}$ & $16.74^{+0.58}_{-0.39}$\\
         \rule{0pt}{1.2em} &\emph{NCF} & 42 & 3 & $7.38^{+0.06}_{-0.05}$ & $22.22^{+0.05}_{-0.04}$\\
         \hline
         \rule{0pt}{1.2em}\multirow{2}{*}{MACS~1149}& \emph{CLU}& 51 & 1 & $6.48^{+0.11}_{-0.11}$ & $19.56^{+0.19}_{-0.17}$\\
         \rule{0pt}{1.2em} &\emph{NCF} & 52 & 1 & $7.59^{+0.09}_{-0.09}$ & $22.88^{+0.01}_{-0.01}$\\
         \hline     
    \end{tabular}
    \label{tab:fields}
\end{table}

\subsection{Imaging}

The CANUCS image reduction, calibration and photometry extraction are done in a manner similar to that in \cite{Noirot2023}, which is further described in { \citet{Asada2024}}. We provide here a brief description of the main steps of this procedure.

The NIRCam images were processed with the standard STScI {\it JWST} pipeline 
and with the Grism redshift \& line analysis software ({\tt Grizli}, \citealt{grizli23b}).The {\it JWST} and {\it HST} images were drizzled onto a common pixel grid with a scale of 40~milli-arcsec\,pixel$^{-1}$, using the {\it Gaia} DR3 astrometry \citep{Gaia2016a,Gaia2023}.

In the cluster fields, the brightest cluster galaxies (BCGs), as well as other bright foreground galaxies, had their light profiles modeled and subtracted from the images { \citep{Martis2024arXiv}}. These BCG-subtracted images were then convolved with empirically-determined kernels to homogenise their point spread functions (PSFs) to that of the F444W filter.

In all fields, $\chi_{\rm mean}$-detection images \citep{Drlica-Wagner2018} were constructed by combining all BCG-subtracted non-PSF homogenised images across all available bands.
Source detection and photometry extraction were done using the python \texttt{photutils} package \citep{Bradley2023}. The detection was run on the $\chi_{\rm mean}$-detection image with a "hot+cold" two-mode strategy, and photometry was then measured in all bands using the BCG-subtracted PSF homogenised images, through matched  Kron \citep{Kron1980} and circular apertures of several diameters. In this project we used the $0\farcs 3$ circular diameter photometry in all bands, corrected to total photometry using Kron fluxes in F444W. { Flux uncertainties are estimated using empty apertures in noise-normalised images in each band. Figure~\ref{fig:Depth} presents the median $3\sigma$ magnitude depths reached in all bands across the five sight-lines. The detailed information of filter availability and depths for the different fields is quoted in Table~\ref{tab:depth}.}

\begin{figure*}
    \centering
    \includegraphics[width=\linewidth]{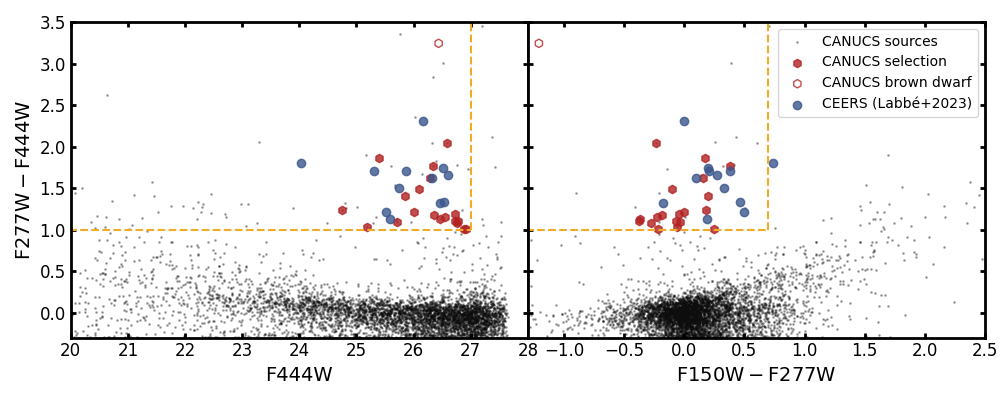} 
    \caption{ Colour-magnitude and colour-colour diagrams used to select the high-$z$ double-break galaxies. The red hexagons represent the selected sources in the CANUCS data, which are compared to the CEERS (\citetalias{Labbe2023}) sources shown as blue circles. The dashed lines indicate the cuts that have been applied to the photometric data. The open red hexagon is a CANUCS brown dwarf, easily identified by its extreme colours in the colour-colour diagram.}
    \label{fig:colorcuts}
\end{figure*}

\subsection{Spectroscopy}

The NIRISS slitless spectroscopy was first processed with \texttt{Grizli} \citep{grizli23b} and subsequent reduction of the grism data was done as in   \citet{Noirot2023} { (see also \citealt{Matharu2023}).}
The NIRSpec data was processed using the STScI {\it JWST} pipeline (software version 1.8.4 and {\tt jwst\_1030.pmap}) and the {\tt msaexp} package \citep{Brammer2022msaexp}. The processing of the raw data is done in a way similar to  \citet{Withers2023}. We used the standard {\it JWST} pipeline for the level 1 processing where we obtained the rate fits files from the raw data. We enabled the jump step option \texttt{expand\_large\_events} to mitigate  contamination by snowball residuals, and also used a custom persistence correction that masked out pixels which approach saturation within the following 1200 s for any readout groups. We then used \texttt{msaexp} to do level 2 processing where we performed the standard wavelength calibration, flat-fielding, path-loss correction, and photometric calibration, and obtained the 2D spectrum for each source. The 1D spectrum was extracted by collapsing the 2D spectrum using an inverse-variance weighted kernel following the prescription by { \citet{Horne_1986}.}

\subsection{Strong lensing models}

The CANUCS survey targets massive cluster fields. Therefore, to measure properly the physical properties of the background sources close to the cluster centres, one must know the magnification due to lensing effects.  Strong lensing models are thus needed to assess the magnification produced by the different clusters. We use strong lensing models we built using the code \texttt{Lenstool} \citep{Kneib1993,Jullo2007}. These models leverage previous models constrained with \textit{HST} data as well as the new CANUCS \textit{JWST} data, adding new multiple image systems that we identified as well as new spectroscopic constraints. 

For MACS~0417, we update the constraints of the \cite{Mahler2019} and \cite{Jauzac2019} models with new multiple image systems and with new spectroscopic redshifts for previously known ones \citep[e.g.,][]{Asada2023,Strait2023}. The details of the \textit{JWST}-updated model will be presented in Desprez et al. (in prep.). In the case of Abell~370, a model is constructed leveraging the constraints of previous models from the literature \citep[e.g.,][]{Strait2018,Lagattuta2019,Lagattuta2022} which are updated and will be presented in Gledhill et al. (in prep.). Rihtar\v{s}i\v{c} et al. (in prep.) will present the update of the \cite{Bergamini2023} model for MACS~0416. The MACS~1423 model  uses the constraints presented in \cite{Hoag2017} and will be detailed in Desprez et al. (in prep.). Finally, in the case of MACS~1149, our lensing model uses constraints of \cite{Desprez2018} and all the details will be reported in Rihtar\v{s}i\v{c} et al. (in prep.). All these models include cluster-size and galaxy-size massive halos, described as double Pseudo-Isothermal Elliptical (dPIE) profiles \citep{Eliasdottir2007arXiv} whose properties are optimised through $\chi^2$ minimisation of the distance between observed multiple images and model predicted ones.

From these models, convergence and shear maps are derived for 100 randomly drawn samples from the optimisation of their parameters, allowing us to account for errors in magnification. The maps created have a size of $20\arcmin\times20$ \arcmin\ centred on the cluster BCGs and a pixel resolution of 600~milli-arcsec\,pixel$^{-1}$.

\subsection{Area and completeness}

{ To compute the effective area that the CANUCS survey covers at high-$z$, we need to account for the magnification due to the clusters and any masked area occupied by other sources. This is done by simulating images of galaxies at different random positions in the fields and recovering them with the CANUCS photometry pipeline. Our selection criteria (Section~\ref{sec:selection}) require galaxies to have F444W magnitudes $<27$, which corresponds to ${\rm S/N}>30$ in this filter for typical high-$z$ galaxies in all the CANUCS fields. Hence there is no incompleteness of the sample due to faint objects not appearing in the detection images. Consequently, we simulate galaxies with flux-density equal to magnitude=28 in all filters. Such galaxies are detected at high S/N in the detection images and are only absent when masked by other sources. }

{ The images of the galaxies presented in \citetalias{Labbe2023} are very compact, so we use simulated Sersic profiles with ellipticities distributed normally with a mean of 0.3 and standard deviation $\sigma=0.2$, Sersic indices distributed normally with a mean of 1.5 and $\sigma=0.3$, and effective radii with a flat distribution ranging from 0.5 to 0.8 pixels (20 to 32 milli-arcsec). As described previously, the completeness calculations are not sensitive to these details, because both the sample and simulated galaxies are bright enough that they always appear in the detection catalog unless in a region of sky occupied by other galaxies. Simulated galaxies are inserted in the survey images at random positions in $6\arcsec \times 6\arcsec$ boxes covering the full images. The data is then processed through the photometry pipeline, allowing us to account for incompleteness in the source detection due to overlap with other sources. The median magnifications and associated 16th and 84th percentile limits are computed for the simulated source positions assuming $z=8$. The area in which the sources have been placed are then corrected for magnification. The full area of each field is then computed by summing the corrected areas of all simulated sources that are recovered after the pipeline detection. The full process is repeated three times for each field with a different random number seed for the placement of simulated galaxies. The difference between each of the three runs is $\approx 1$\%, showing that further iterations to increase statistics are not required. The effective areas from the three runs are averaged, with uncertainties computed by adding the standard deviation of the three results to the results obtained with the 16th and 84th percentile of magnification in quadrature. The results are quoted in Table~\ref{tab:fields} for all fields.  Volumes are computed between $z=7.0$--$8.5$ for all fields, with details also given in Table~\ref{tab:fields}, and add up to a total volume of $185^{+3}_{-2}\times10^3\,{\rm Mpc}^3$ over the full survey. }


\section{Analysis}

In all CANUCS \emph{CLU} and \emph{NCF} fields double-break candidates are selected in a manner similar to \citetalias{Labbe2023}. The selected sources are then fit to obtain redshifts and { stellar} mass measurements. Our population of double-break sources are then compared to that reported by \citetalias{Labbe2023}.  The details of this procedure are given in the following sections.

\subsection{Selection}
\label{sec:selection}

We adopt similar selection criteria as those in \citetalias{Labbe2023} for the double-break sources. These consist of two colour cuts,
\begin{align*}
    {\rm F150W}-{\rm F277W} <0.7, \\
    {\rm F277W}-{\rm F444W} >1.0,
\end{align*}
in addition to two magnitude cuts,
\begin{align*}
    {\rm F444W} < 27, \\
    {\rm F150W} < 29.    
\end{align*}
To ensure proper detection of sources, we also imposed a signal-to-noise ratio (S/N) cut in F444W > 8. Finally, in the F090W band as well as in all the available {\it HST} optical bands non-detections were defined as sources with S/N~$<2$ in order to
ensure a selection of high-$z$ sources. We note that the S/N criterion on the F090W band makes our sample less likely to include $z\sim7$ sources compared to the CEERS sample that does not posses this band. The colour and magnitude cuts are illustrated in Fig.~\ref{fig:colorcuts}. We point out that the magnitudes presented are not corrected for lensing, but most sources have estimated magnifications of $\mu<2$, which leads to, at most, a F444W magnitude larger by 0.75\,mag. All the selected sources are then visually inspected to avoid any contamination by noise, { bad pixels, light from nearby objects,} or possible confusion with lower-redshift interlopers. The candidate cutouts and their photometry are shown in Figs.~\ref{fig:spectrosample},~\ref{fig:BrownDwarf},~\ref{fig:SED1},~\ref{fig:SED2}, and~\ref{fig:SED3}. After these steps, our selection contains 20 double-break candidates, of which one is discarded due to discrepant colours compared to the bulk of the sample (see Fig.~\ref{fig:colorcuts}), a point-like appearance in all filters, and an SED that is consistent with that of a cool brown dwarf (see Fig~\ref{fig:BrownDwarf}). This leaves us with a sample of 19 double-break candidates.\footnote{ The photometry and fit properties of the candidate sources can be found on \href{https://niriss.github.io/doublebreak.html}{https://niriss.github.io/doublebreak.html}} 

The distribution of the candidates over the different fields is reported in Table~\ref{tab:fields}. We note variation in the number of candidates from cluster to cluster, with five sources for Abell~370 and MACS~1423 but only two for MACS~1149. If we make the comparison field to field, the largest concentration of sources is found in the Abell~370 cluster field with four sources, with five fields containing only one double-break candidate each. 

Figure~\ref{fig:colorcuts} shows the position of all selected sources in colour-magnitude and colour-colour plots. It also presents the \citetalias{Labbe2023} candidates, with photometry taken from that paper. Considering the low values of cluster magnification for most of our candidates (see Table~\ref{tab:parameters}), most of our sources are intrinsically bright, which means that we are probing the same population of sources as did \citetalias{Labbe2023}, even if we do not see any source as bright as the brightest \citetalias{Labbe2023} object.

Finally, we note that our sample includes the MACS~0416-Y1 source from \cite{Laporte2015} (i.e., id=3107165). This source is extended and has a reported ALMA redshift of $z=8.3118$ \citep{Tamura2019}. We measure its magnitude in ${\rm F444W}=24.74\pm 0.06$ and its magnification to be $\mu=1.63\pm0.01$, thus meeting the selection criteria even accounting for lensing.

\subsection{Source fitting}

For all sources, we first estimate the redshift either with spectroscopy when available or using { a} photo-$z$ template-fitting code. Then, the photometry of each source is fit to measure its stellar mass, accounting for the magnification of the nearby cluster, as described in detail below. 

\subsubsection{Spectroscopy fitting}

\begin{figure}
    \centering
    \includegraphics[width=\linewidth]{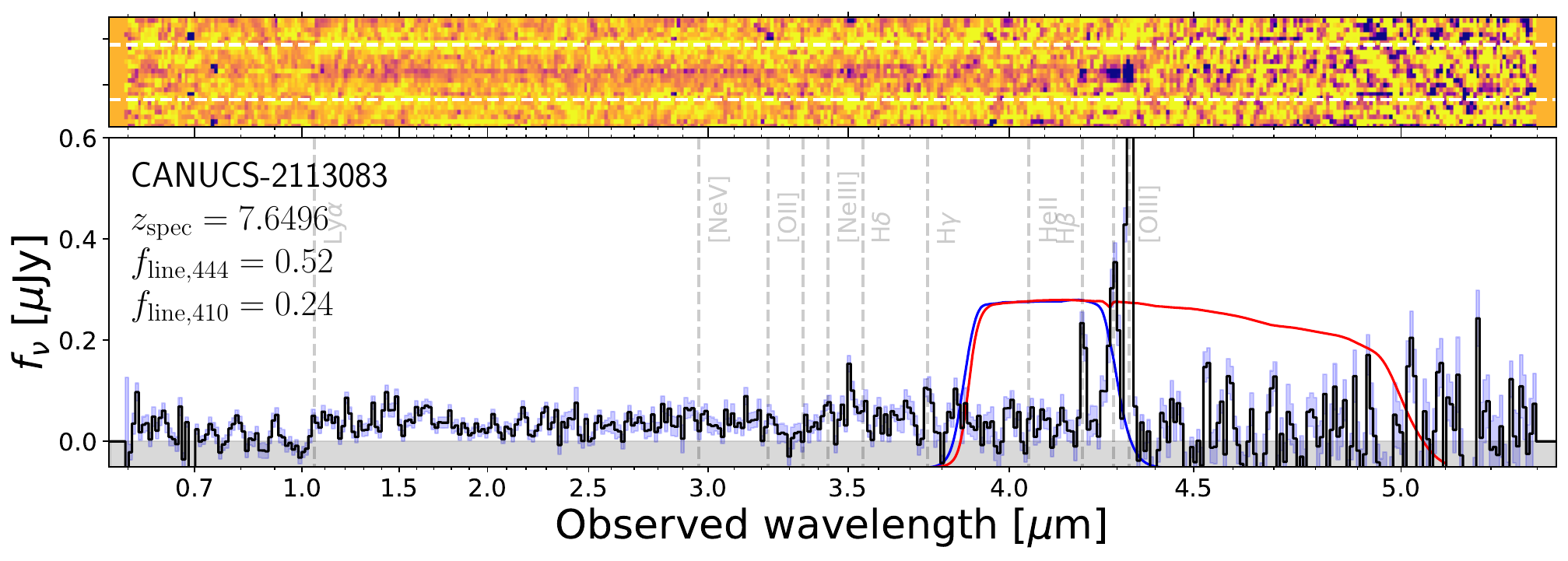}\\
    \includegraphics[width=\linewidth]{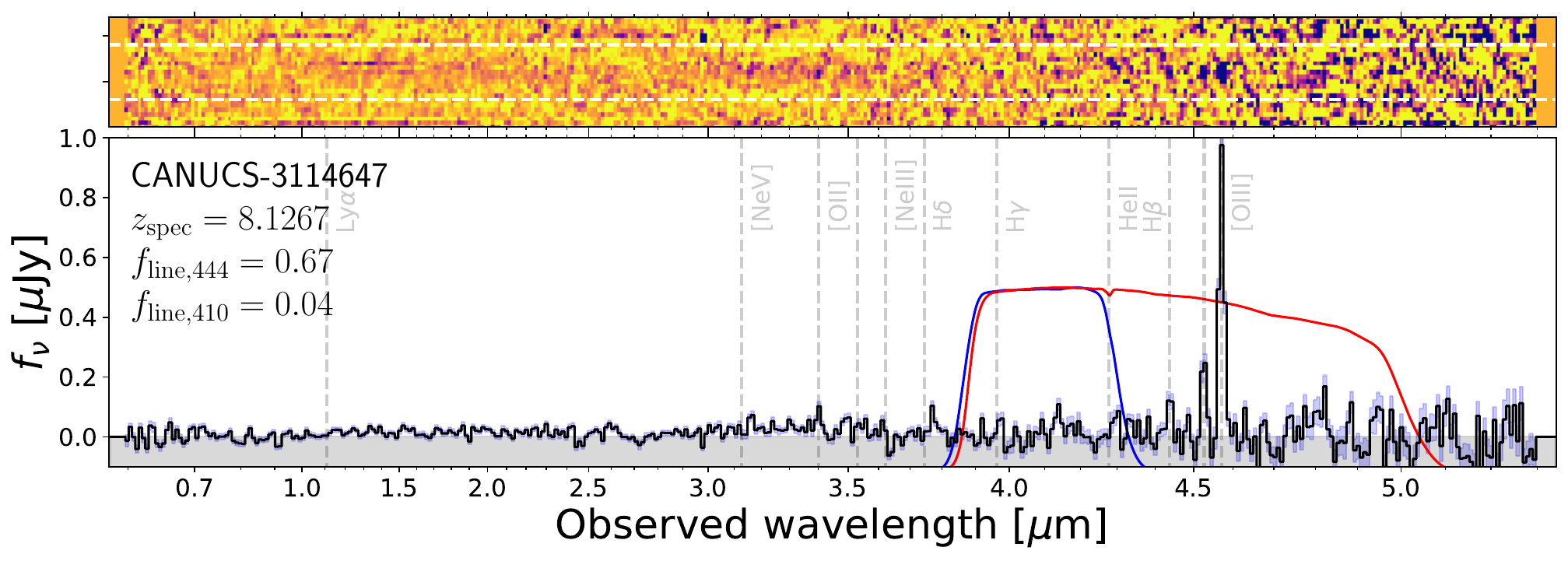}\\
    \includegraphics[width=\linewidth]{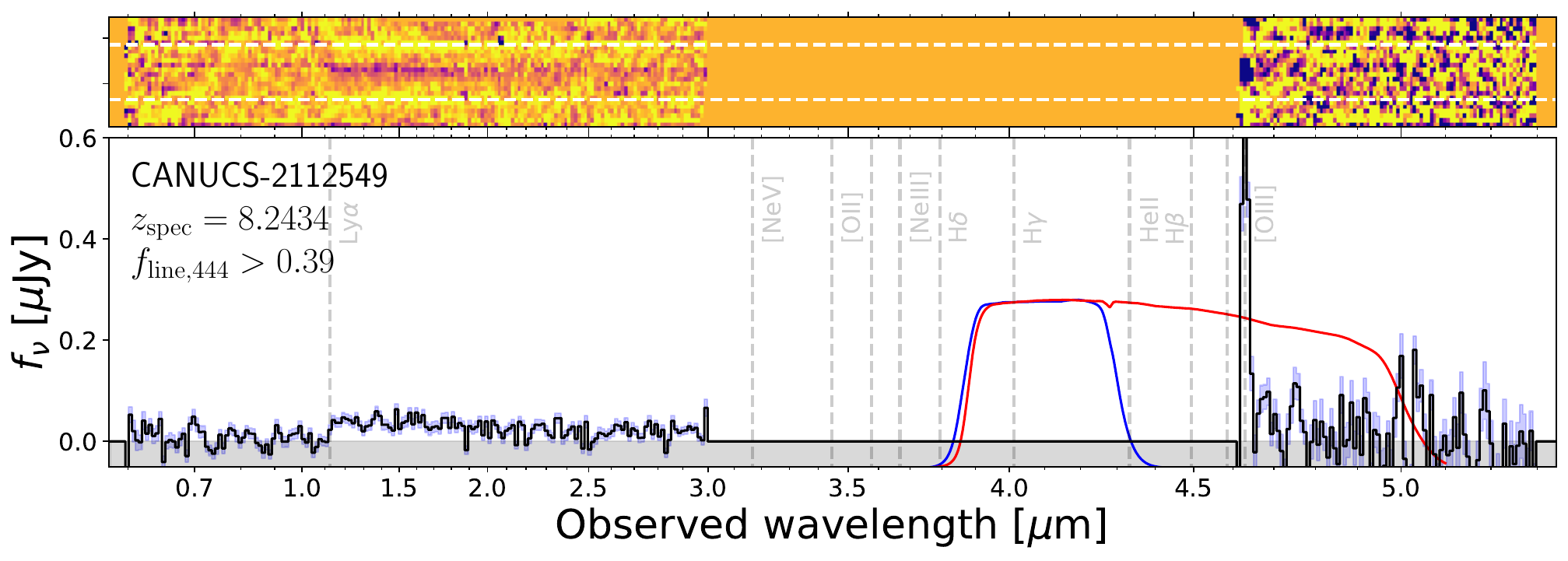}\\
    \includegraphics[width=\linewidth]{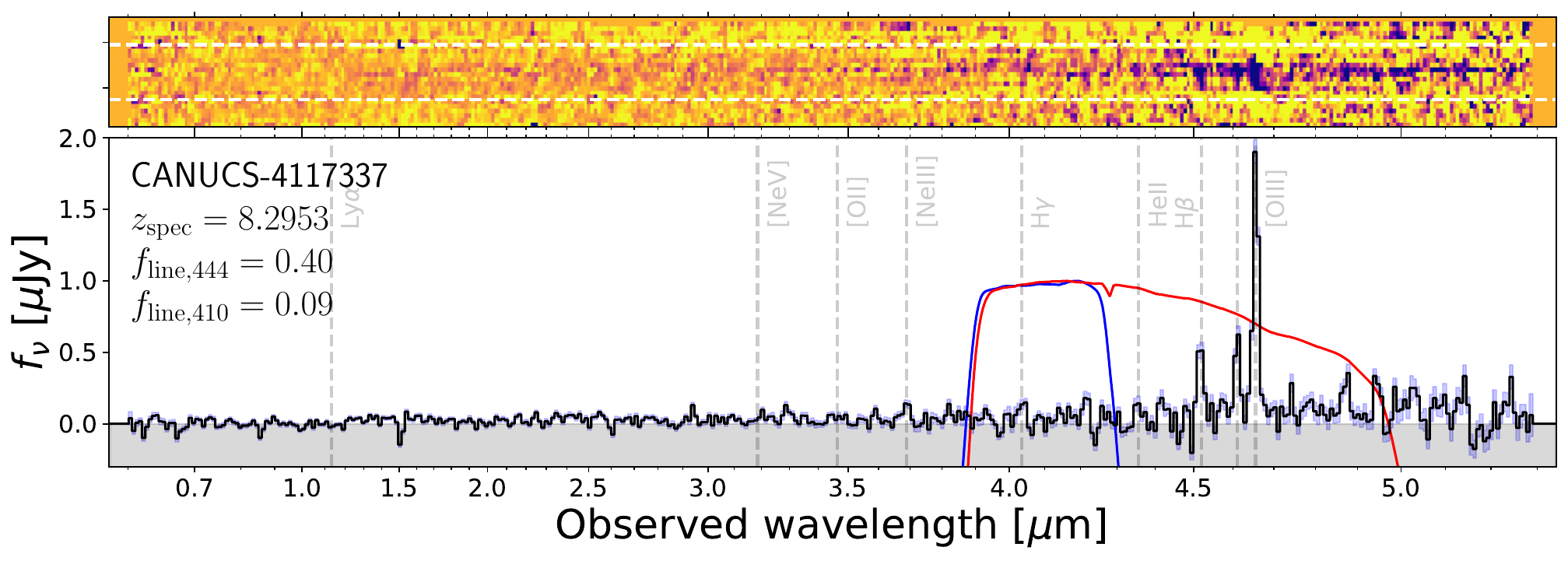}\\
    \includegraphics[width=\linewidth]{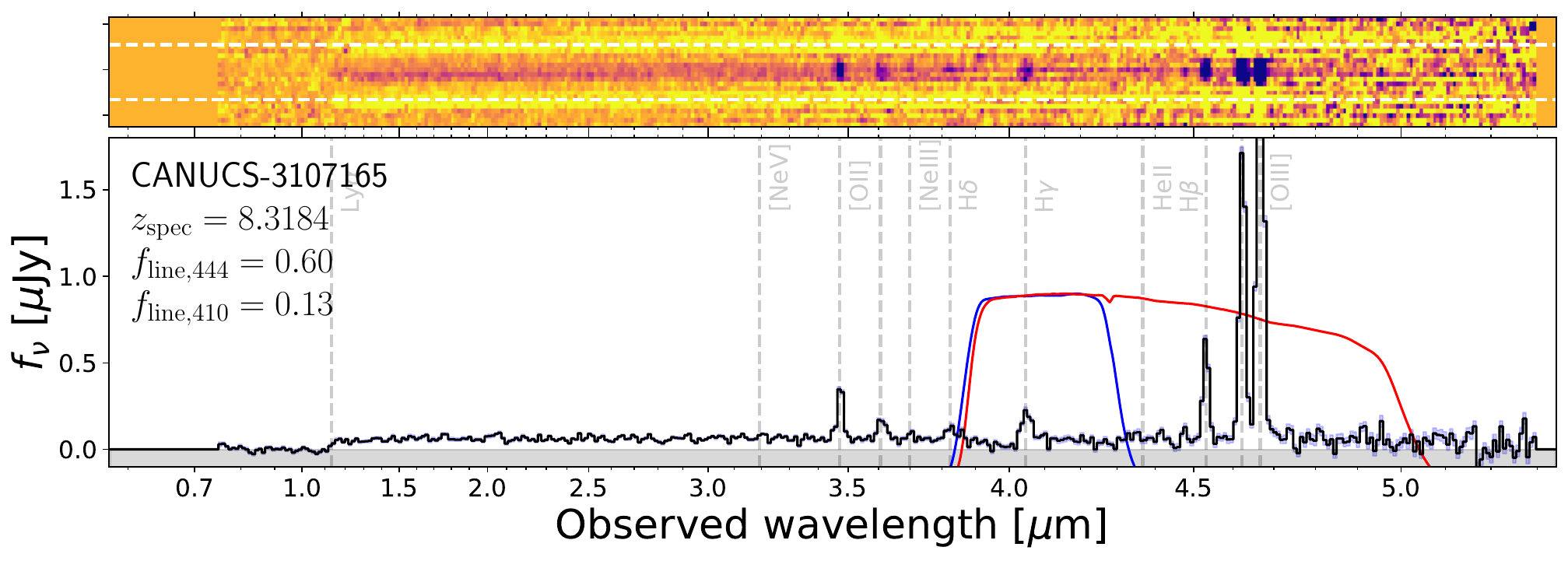}\\
    \caption{ NIRSpec observations of the five targeted double-break sources. The spectra show the fractional contributions to the F410M (blue) and F444W (red) NIRCam filter photometry from strong emission lines. All five galaxies have strong emission lines, especially {\sc [Oiii]}, making up $\sim 50\%$ of the F444W filter flux.}
    \label{fig:spectra}
\end{figure}

The sources selected in the cluster fields have been observed with NIRISS. However,  inspection of the NIRISS data shows that none of the double-break sources have a strong enough detected signal to measure a redshift. Therefore only NIRSpec spectroscopy is considered in the rest of this work.

Among our sample of 19 double-break galaxies, five galaxies were observed with NIRSpec.
Their 1D and 2D spectra are shown in Fig.~\ref{fig:spectra}. Our goals with spectroscopic observations in this paper are to obtain the spectroscopic redshifts (spec-$z$) and to estimate the effect of emission lines on photometry in filters where the Balmer breaks are possibly captured (i.e., F410M and F444W; their transmission curves are also shown in Fig.~\ref{fig:spectra} with blue and red solid lines, respectively). Thus, we focused particularly on emission lines observed at 3.8 $\mu$m to 5.0 $\mu$m, wavelengths that these two filters cover.

Four of the five galaxies have obvious detections of {\sc [Oiii]}$_{\lambda4959,\lambda5007}$ + H$_\beta$ emission lines in this wavelength range, so we first measured the spectroscopic redshifts and the emission line fluxes by performing Gaussian profile fitting on the 1D spectra. { Errors are estimated by perturbing the observed 1D spectra within associated errors and repeating the Gaussian fit a 1000 times, and measuring the 16th and 84th percentiles. We do not include any potential systematic errors here -- these are difficult to estimate but are likely present given the reported offsets of $\delta z=0.004$ between NIRSpec prism and grating spectra \citep{Bunker2023arXivb}.
}
In cases where other, fainter emission lines can be seen in the 2D spectra in the target wavelength range (e.g., H$_\gamma$), we also fit Gaussian profiles to these lines while fixing the redshift to obtain their line fluxes.
We then derived the total fluxes in each of the F410M and F444W filters by integrating the observed spectrum through the filter transmission curves, and calculated ratios of emission-line fluxes to the total broadband fluxes to derive the flux fractions contributed by the emission lines in the F410M and F444W filters. 
It is worth noting that we did not make slit-loss corrections in this analysis, but slit-loss corrections are not expected to significantly affect the spectroscopic redshifts or emission line flux fractions.
The resulting spectroscopic redshifts and emission line flux fractions in the F444W filter ($f_{\rm line,444}$) and in the F410M filter ($f_{\rm line,410}$) are shown in Fig.~\ref{fig:spectra} and Tab.~\ref{tab:parameters}.
One of the five galaxies (id=3107165) 
{ and the ALMA spectroscopic redshift was reported to be $z_{\rm spec}=8.3118$ \citep[][source MACS~0416-Y1]{Tamura2019}, although with some variation over $z=8.3113$--$8.3125$ \citep{Bakx2020,Tamura2023}.
Our NIRSpec redshift of this galaxy is $z_{\rm spec}=8.3184 \pm 0.0001$ 
which is close to but formally discrepant with the ALMA redshift due to our small fit error. Accounting for a possible systematic error in the NIRSpec measurement (\citealt{Bunker2023arXivb} results suggest this may be 
$\delta z\sim0.004$) would give ALMA-JWST redshift agreement at the $\sim 2\sigma$ level.
}

We have one source (id=2112549) whose emission lines mostly fall inside the NIRSpec detector gap. Only one strong emission line at $\lambda_{\rm obs}\sim4.6\ \mu$m is found -- at the edge of the detector but detected at high significance in spite of the proximity to the detector edge. In addition to this line, the spectrum of this source also shows a clear break at $\lambda_{\rm obs}\sim1.1\ \mu$m. This Lyman break would exactly match up with the strong emission line being {\sc [Oiii]}$_{\lambda5007}$ at $z~=~8.2434\pm 0.0011$. We thus fit a single Gaussian profile to the emission line, whose peak and part of blue wing lie on the detector. Since other strong emission lines that fall into the detector gap,  such as H$_\beta$,  are expected to be covered with the F410M/F444W filters, it is impossible to estimate the contributions of all emission lines to the total fluxes in these filters.
For this galaxy, we therefore used the measured {\sc [Oiii]}$_{\lambda5007}$ line flux and estimated its {\sc [Oiii]}$_{\lambda4959}$ flux by assuming {\sc[Oiii]}$_{\lambda5007}$/{\sc [Oiii]}$_{\lambda4959}=3.0$, and took the ratio to the $0\farcs 3$-diameter aperture flux in F444W from photometry to obtain the lower limit of $f_{\rm line,444}$.

\begin{figure*}
    \centering
    \includegraphics[valign=m,width=0.6\linewidth]{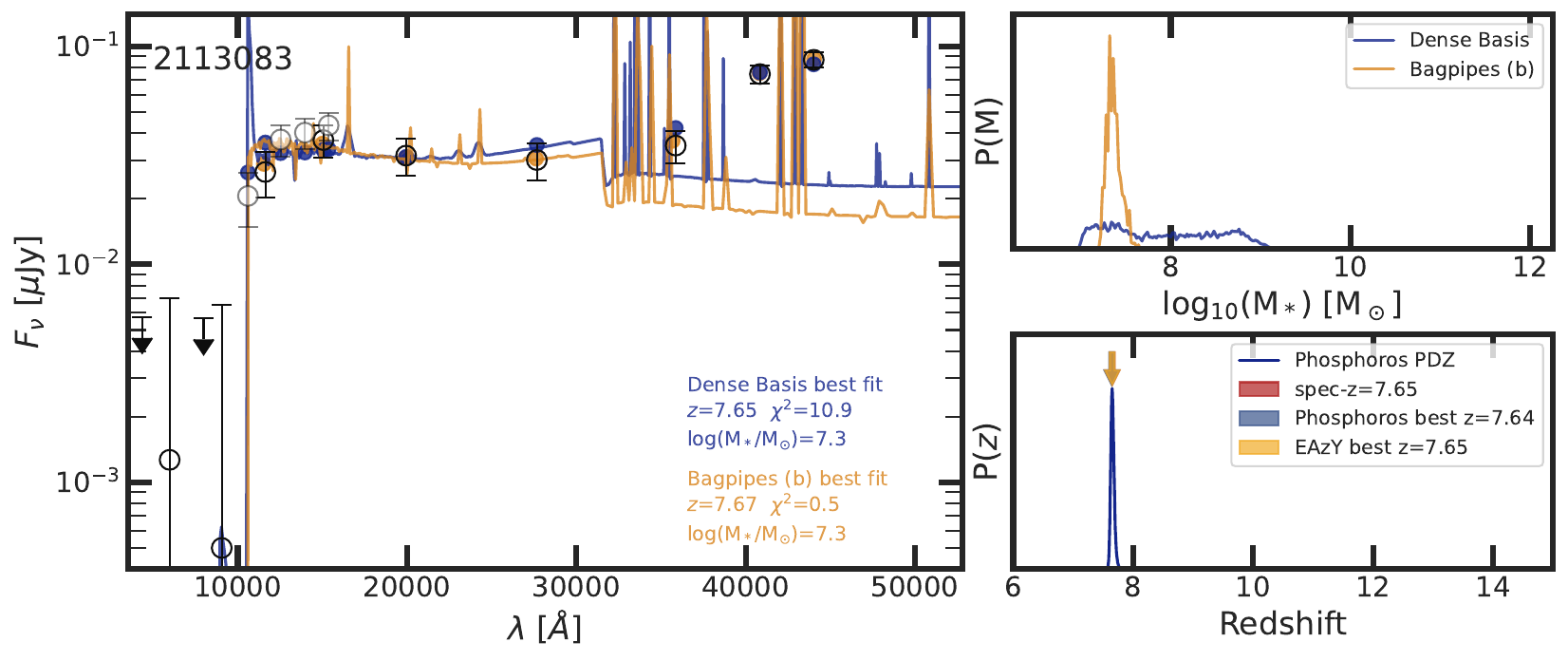}
    \includegraphics[valign=m,width=0.39\linewidth]{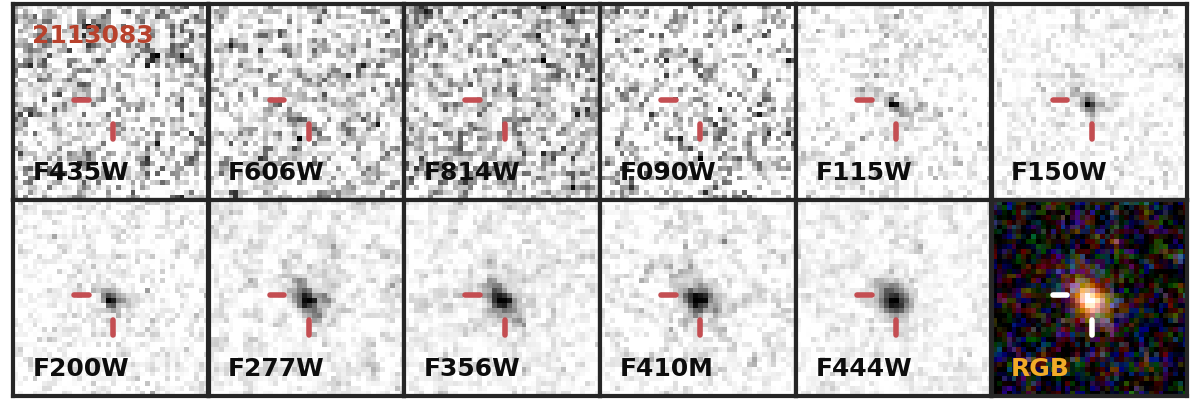}\\
    
    \includegraphics[valign=m,width=0.6\linewidth]{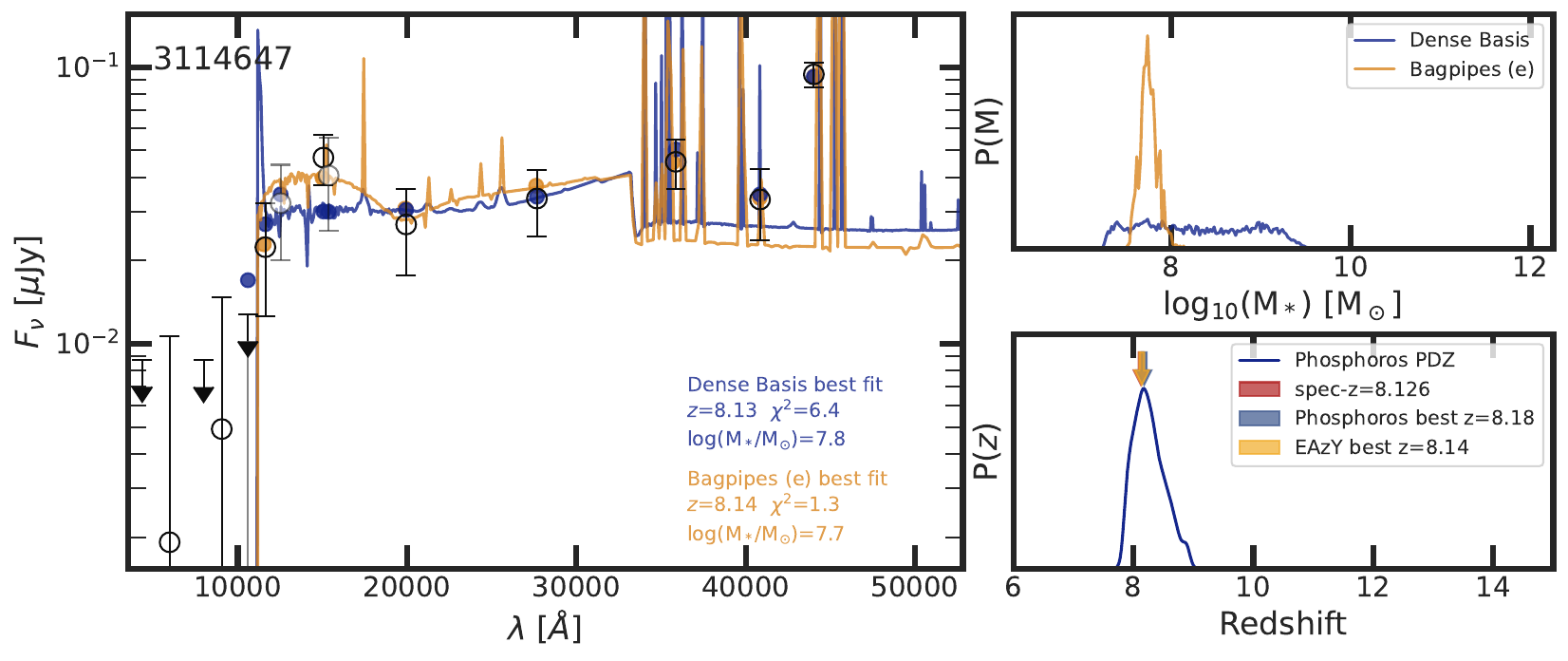}
    \includegraphics[valign=m,width=0.39\linewidth]{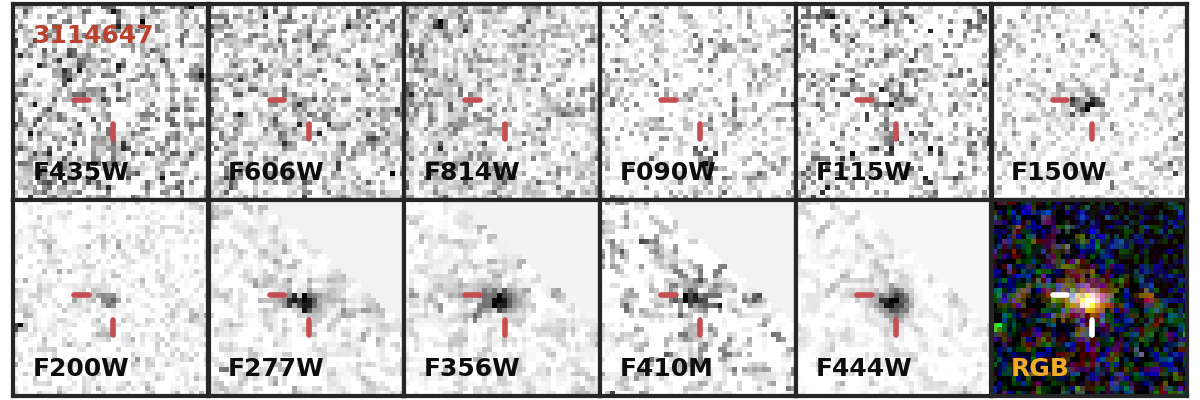} \\
    
    \includegraphics[valign=m,width=0.6\linewidth]{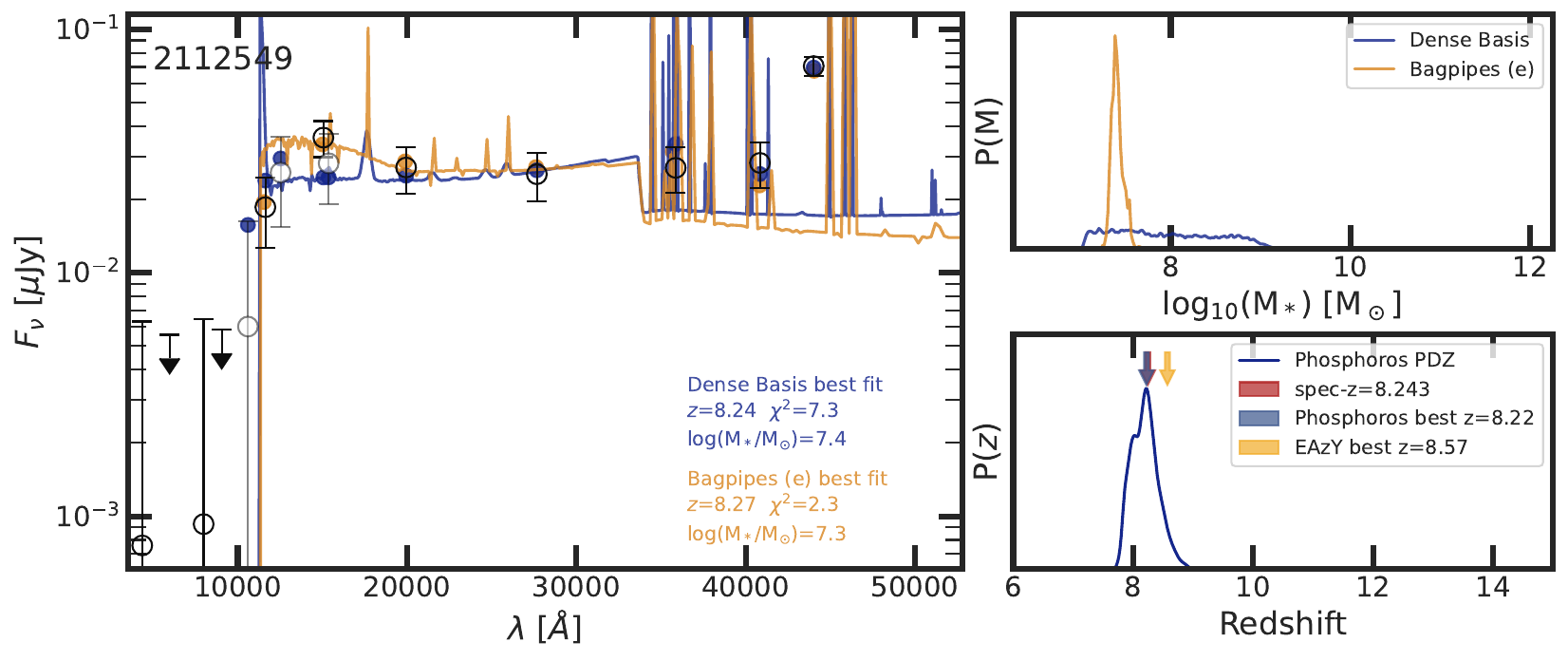}
    \includegraphics[valign=m,width=0.39\linewidth]{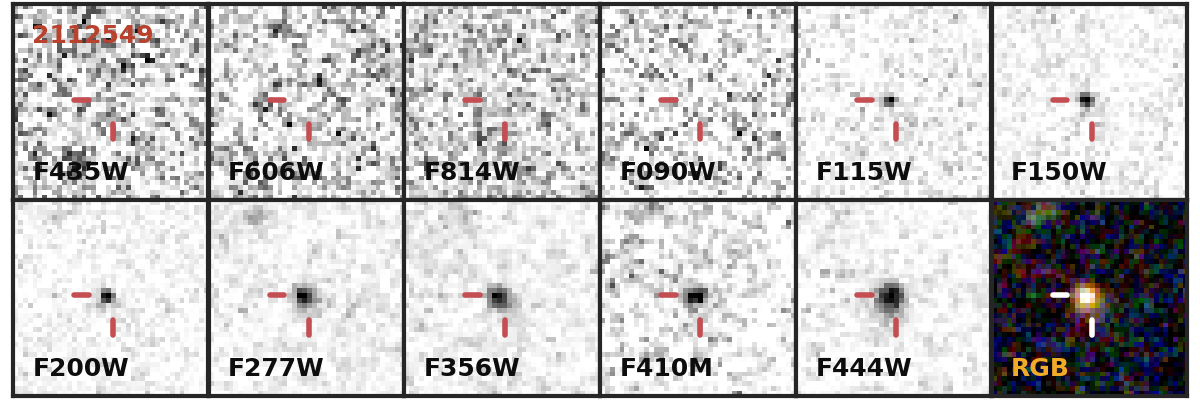}\\
    
    \includegraphics[valign=m,width=0.6\linewidth]{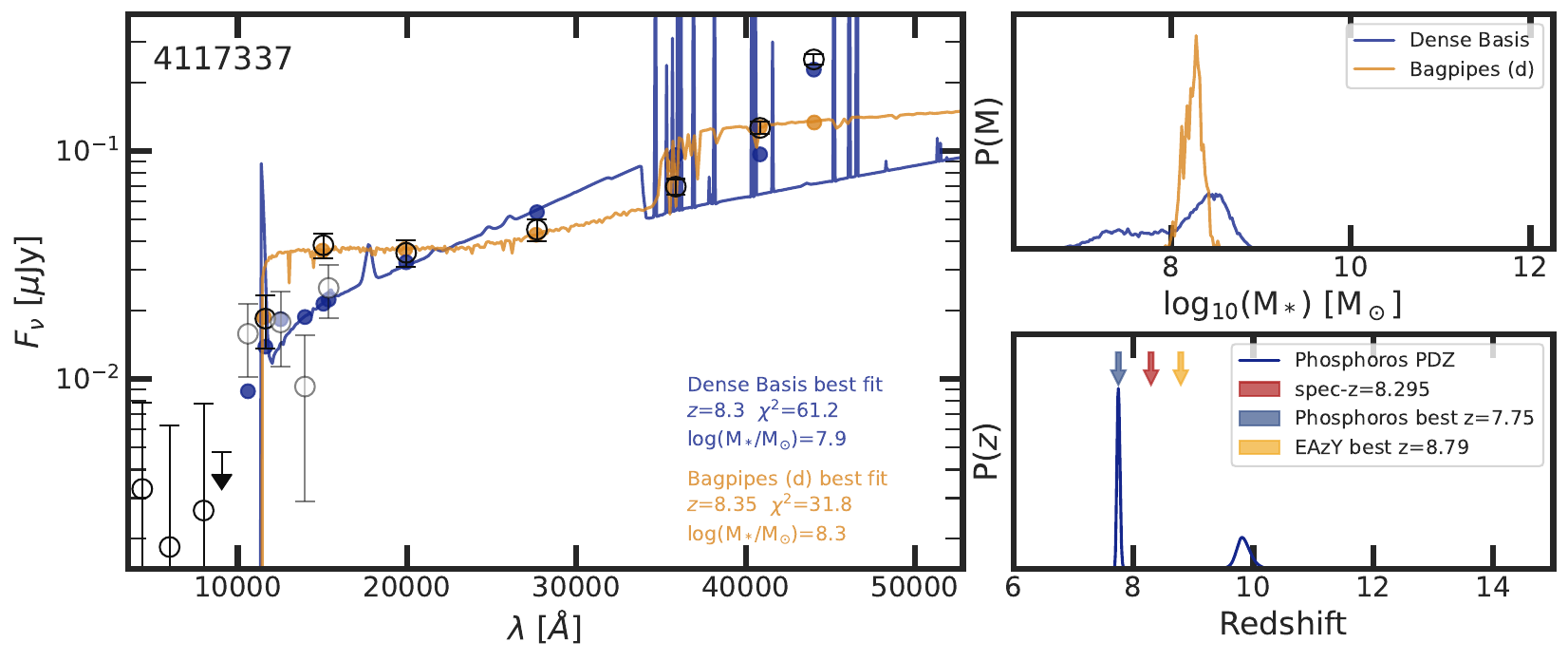}
    \includegraphics[valign=m,width=0.39\linewidth]{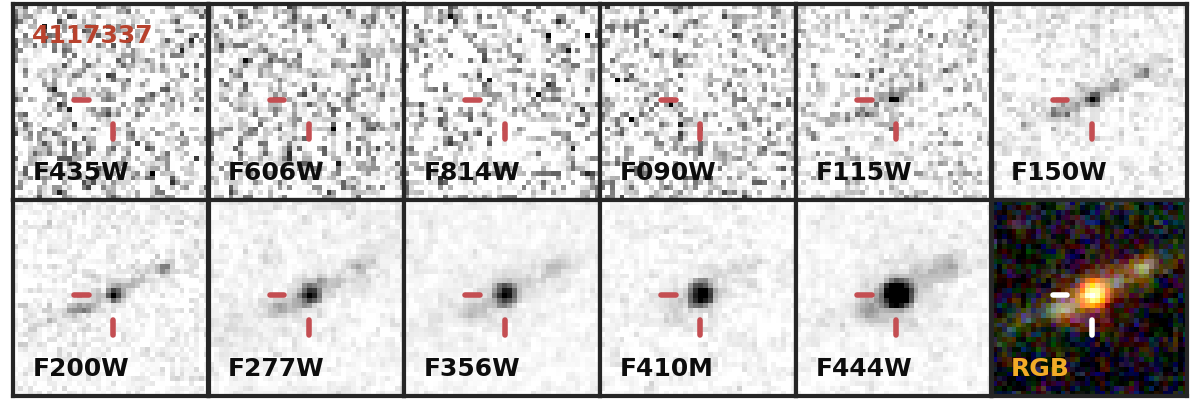}\\
    
    \includegraphics[valign=m,width=0.6\linewidth]{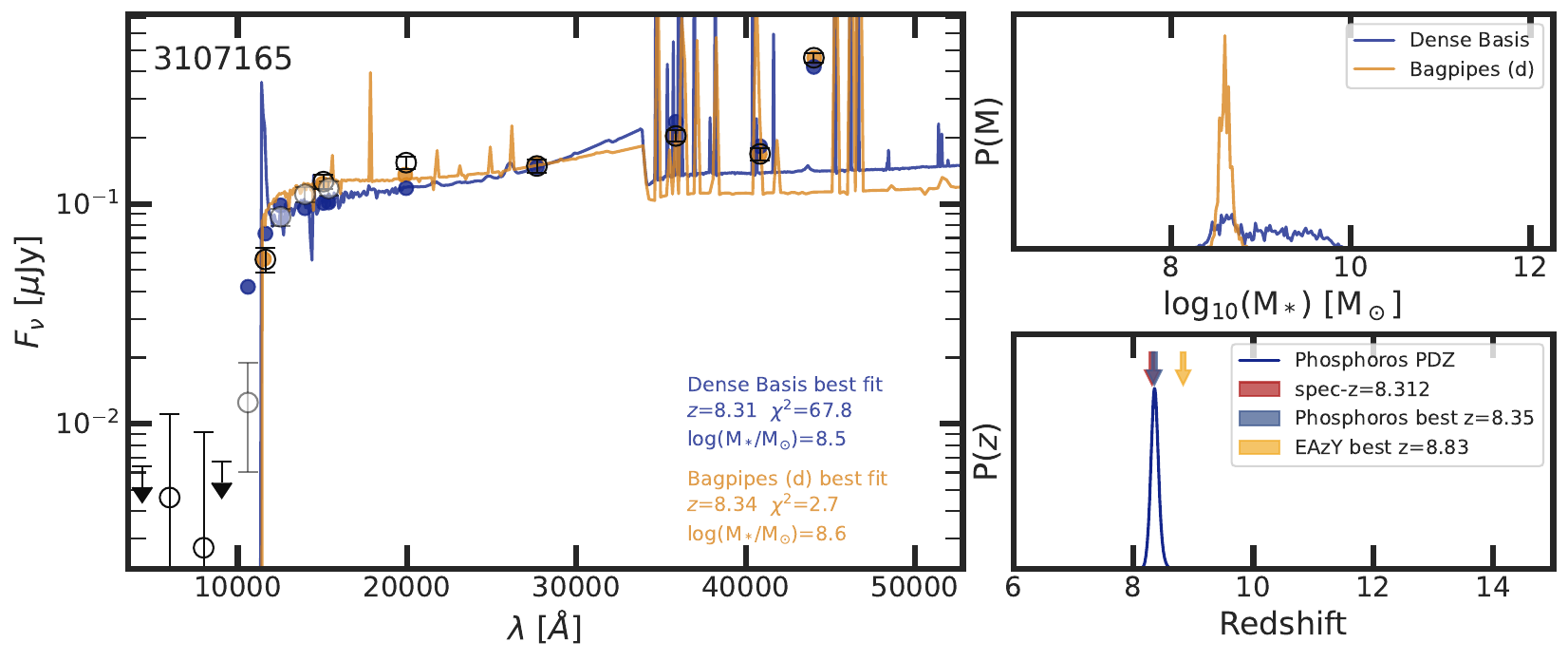}
    \includegraphics[valign=m,width=0.39\linewidth]{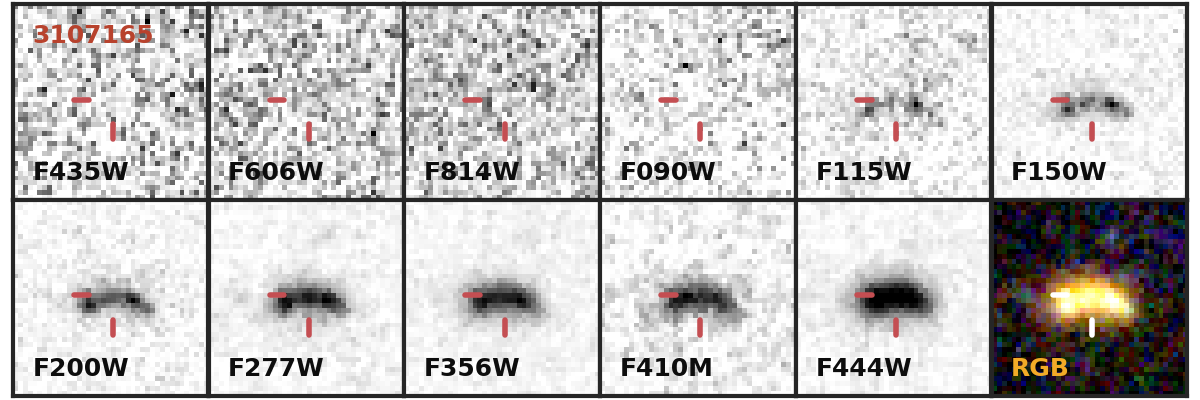}\\

    \caption{ Observed SEDs, fits, and image stamps for the five CANUCS double-break sources observed with NIRSpec. The best model fits of both {\tt Dense Basis} (blue) and {\tt Bagpipes} (orange) are displayed, with the SED-fitting redshift and stellar mass posteriors. In the case of {\tt Bagpipes}, only the best fit configuration (noted a, b, c, d, or e) is shown. The redshift sub-panels show the {\tt Phosphoros} P($z$) (blue curve), and the arrows indicate the {\tt Phosphoros} (blue) and {\tt EAzY} (orange) best fit redshifts and the spec-$z$ (red). The stamps show the sources in the {\it HST} optical bands and {\it JWST} bands.}
    \label{fig:spectrosample}
\end{figure*}

\subsubsection{Photometry fitting}

First, we fit the source SEDs to derive their photometric redshifts. For that task, we use the \texttt{Phosphoros} template-fitting code (\citealt{Desprez2020}; Paltani et al., in prep.). As we are dealing with high-$z$ candidates, we modify the \texttt{Phosphoros} configuration defined in \cite{Desprez2023} to better account for such type of objects. Thus, in addition to the 33 COSMOS templates \citep{Ilbert2013}, we fit the \cite{Larson2023} SED templates (sets 1 and 4). For these templates, intrinsic extinction is allowed to vary in a similar way as for the COSMOS starburst templates, with a reddening excess E(B-V) ranging from $0$--$0.5$ and with two versions of the \cite{Calzetti2000} law with { both with the 2175\AA\ bump, using two different bump intensities}. A prior on all the SEDs is set, weighting their fit results according to their colour-space coverage, thus avoiding over-weighting some solutions due to SEDs that are too similar to each other. No prior is set on source magnitudes to avoid forcing low redshift solutions on what could be highly unusual high-$z$ sources. Redshift probability distribution functions, P($z$), are then estimated over $z=0$--$15$ with $\delta z=0.01$ steps.  The resulting P($z$) are shown in Fig.~\ref{fig:spectrosample},~\ref{fig:SED1},~\ref{fig:SED2}, and~\ref{fig:SED3}.

To ensure the quality of the fit photo-$z$, we compare {\tt Phosphoros} results to those of {\tt EAzY} \citep{Brammer2008}. Here, the CANUCS photometry is fit by the standard {\tt EAzY} templates ({\tt tweak\_fsps\_QSF\_12\_v3}) augmented with the \citet{Larson2023} templates. The results of the best fit {\tt EAzY} redshifts are shown in Fig.~\ref{fig:spectrosample} (and in all figures in Appendix~\ref{sec:extra}) along with the {\tt Phosphoros} results, and show good agreement between the two codes. We also use  {\tt EAzY} to test the brown dwarf hypothesis for source id=4115596. For this,  we fit the Sonora \citet{Marley2021} cool brown dwarf templates to the photometry. The comparison of the best fit $\chi^2$ of these brown dwarf templates ($\chi^2_{\rm BD}=10$) to that for the {\tt EAzY} galaxy templates ($\chi^2_{\rm gal}=62$), along with those of {\tt Dense Basis} and {\tt Bagpipes} (see Fig~\ref{fig:BrownDwarf}) confirms our decision to discarding this source from our sample. All other sources in our sample were also fit with these brown dwarfs models as a test, but they all have $\chi^2_{\rm BD}>60$) and all are much better fit by high-redshift galaxy templates.

\begin{figure*}
    \centering
    \includegraphics[width=\linewidth]{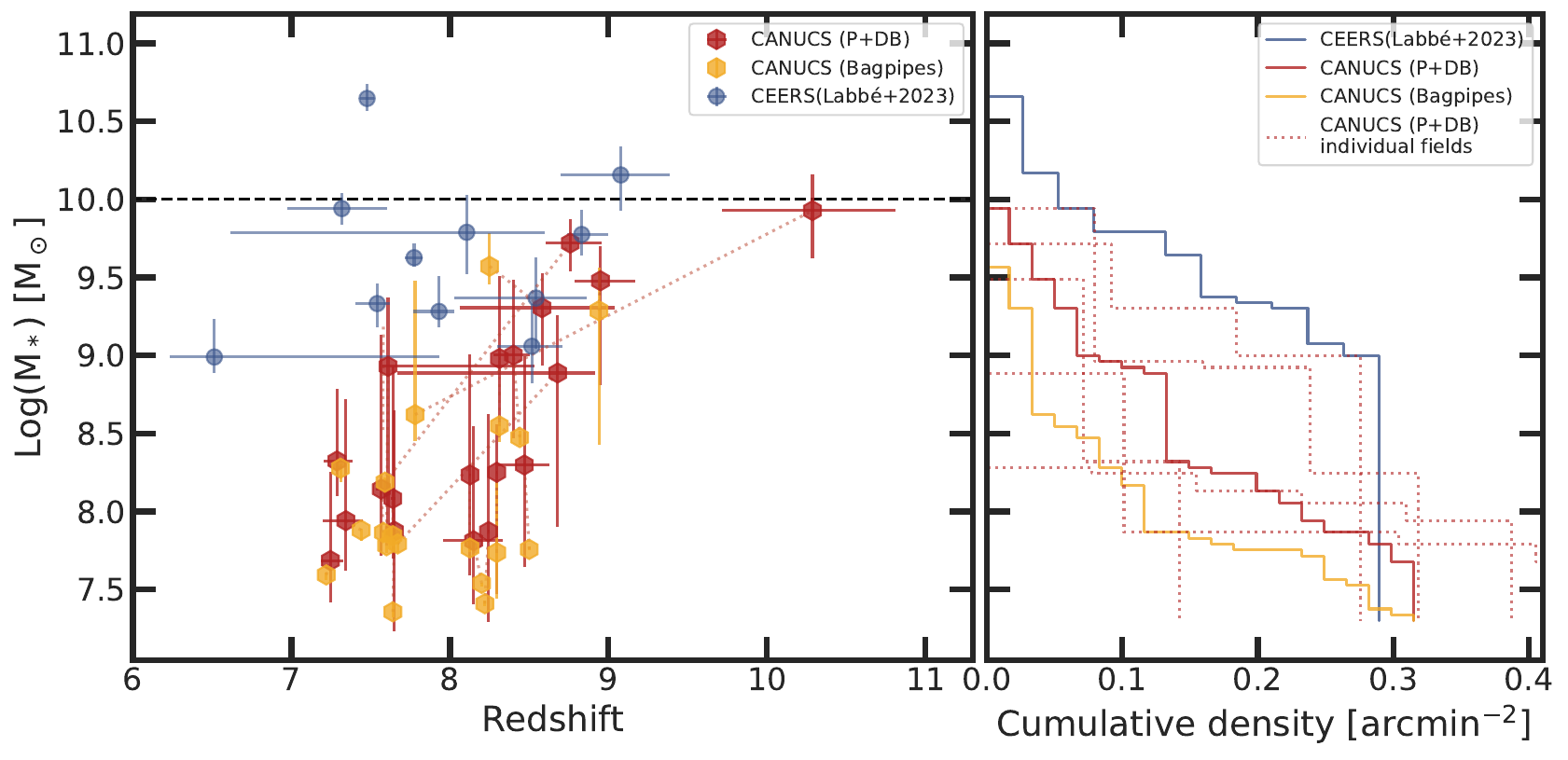}
    \caption{Stellar masses versus redshift (left) and cumulative mass density by decreasing stellar mass (right). In the left panel, the CANUCS sample {\tt Phosphoros}+{\tt DB} redshifts and mass (red hexagons) and the {\tt Bagpipes} ones (orange hexagon) are compared to the CEERS \citetalias{Labbe2023} sample with masses scaled from \citet{Salpeter1955} to \citet{Chabrier2003} IMF (blue circles). As {\tt Bagpipes} and {\tt DB} do not use the same redshift in the fits, the former using only the {\tt Phosphoros} best fit redshift, where the latter uses the full P($z$) (represented in this panel by the median point estimate), their redshift value differs. Thus the dotted lines link the two code results for the same sources. In the right panel, the distribution of cumulative density of sources ranked by stellar masses are shown for both samples and both mass estimates in the CANUCS case.  The red dotted lines in the right panel show the cumulative densities for each of our five clusters (\emph{CLU}+\emph{NCF} fields) separately using the {\tt DB} results, illustrating the large field-to-field variance in the number densities.}
    \label{fig:mass-v-redshift}
\end{figure*}

Next, all sources are fit using \texttt{Dense Basis} \citep[hereafter \texttt{DB},][]{Iyer17,Iyer19} to determine their stellar masses. A collection of \texttt{DB} atlases is created for all redshifts on the same grid as for the \texttt{Phosphoros} P($z$). These atlases are designed to contain $10\,000$ template SEDs, running mostly with the default \texttt{DB} configuration, except for two priors: that of the star formation rate prior, which is set to be flat between $10^{-1}$--$10^{3}~{\rm M_{\sun}.yr}^{-1}$, and that of the mass prior, whose lower limit is set to ${\rm M_{*,min}}=10^7~{\rm M_{\sun}}$. 
For each source, a sample of $100$ redshifts is randomly drawn using the \texttt{Phosphoros} P($z$) as a prior, or the spec-$z$ when available and its photometry is fit to the \texttt{DB} atlases. The cluster magnification is corrected for in each fit result by randomly selecting one instance of the lensing model and measuring the expected magnification for the fit redshift. Combining all the fit results allows us to build a probability distribution function for the mass, P(M), that accounts for both redshift and magnification uncertainties.  The mass P(M) are shown in Fig.~\ref{fig:spectrosample},~\ref{fig:SED1},~\ref{fig:SED2}, and~\ref{fig:SED3}.

To ensure that our \texttt{DB} results can be compared properly to the masses in \citetalias{Labbe2023}, we also fit the photometry of the \citetalias{Labbe2023} sources following the same steps as described before. We use \texttt{Phosphoros} to derive their photo-$z$ and P($z$), and we fit their photometry with \texttt{DB} to obtain their stellar masses. Two of the \citetalias{Labbe2023} sources have spectroscopic redshifts: source id=39575 has $z=7.9932$ \citep{Fujimoto2023} and source id=13050 has $z=5.624$ \citep{Kocevski2023}. Due to it lower spec-$z$ and broad line AGN component, source id=13050 has been discarded from the comparisons. For source id=39575, the stellar mass has been measured with \texttt{DB} at its spec-$z$. {The results of the fits are presented in Table~\ref{tab:parameters}.}

Another set of fits for our sample objects is done using the Bayesian analysis of galaxies for physical inference and parameter estimation software \citep[\texttt{Bagpipes},][]{Carnall2018} following a similar procedure as that outlined in \citetalias{Labbe2023}. The sources are fit using five different configurations used by \citetalias{Labbe2023}, but their redshifts are fixed to their spec-$z$, when available, or at the best-fit redshifts (lowest $\chi^2$) from \texttt{Phosphoros}, only allowed to vary according to a Gaussian prior with $\sigma_z=0.05$. The configurations are: {\tt Bagpipes\_csf\_salim} (noted \emph{a}), {\tt Bagpipes\_rising\_salim} (\emph{b}), {\tt Bagpipes\_csf\_salim\_logage} (\emph{c}), {\tt Bagpipes\_csf\_salim\_logage\_snr10} (\emph{d}), and {\tt Bagpipes\_csf\_smc\_logage} (\emph{e}), and we refer to \citetalias{Labbe2023} for the configuration details. The fit results from the five \texttt{Bagpipes} configurations are combined as follows: the stellar mass reported is the median of all results, and the error limits are given by the lowest and highest results.

\subsection{Results}
\label{sec:results}

The resulting stellar masses and redshifts for the CANUCS double-break sample are presented in Fig.~\ref{fig:mass-v-redshift} and Table~\ref{tab:parameters}. Figure~\ref{fig:mass-v-redshift} also shows the results from \citetalias{Labbe2023}, corrected for the difference in IMF. We note that the CANUCS double-break population remains under the mass threshold of $10^{10}\,{\rm M_{\sun}}$, regardless of the code used for the fit.  We find that the majority of our sources have a photometric redshift $z<9$, with the exception of one source (id=3102668) which has a broad {\tt Phosphoros} P($z$) leading to a large discrepancy between the best-fit redshift and the P($z$) median. This outlier also presents the highest computed mass with {\tt DB}, { with ${\rm M_*}=7.9^{+4.6}_{-4.0}\times10^{9}\,{\rm M}_{\sun}$. The median stellar mass of the CANUCS sample is ${\rm M_{*}}=2.0\times10^{8}\,{\rm M}_{\sun}$ with {\tt DB} and ${\rm M_*}=7.6\times10^{7}\,{\rm M}_{\sun}$ with {\tt Bagpipes}. We note that despite having samples of sources with similar luminosities and colors (see Fig.~\ref{fig:colorcuts}), CANUCS galaxies seems to be generally less massive than the CEERS ones.}

The solid curves in the right panel of Fig.~\ref{fig:mass-v-redshift} show the cumulative source density vs. stellar mass for the different samples and mass fitting procedures. We note a discrepancy between the densities of the \citetalias{Labbe2023} (blue line)  and CANUCS {\tt Phosphoros}+{\tt DB} (red line) samples of roughly $\sim 0.8$\,dex in stellar mass. We also see that the CANUCS {\tt Bagpipes} fits (orange line) provide a lower source density for a given stellar mass than the {\tt Phosphoros}+{\tt DB} fits, but the difference is mainly due to the most massive source with the uncertain redshift.
Figure~\ref{fig:mass-v-redshift} right panel also shows the individual cumulative mass densities for the different CANUCS line-of-sights from the {\tt Phosphoros}+{\tt DB} results. The different field densities exhibit a large scatter, with differences up to $\sim 1.5$\,dex at fixed density. 
We note that the density of the \citetalias{Labbe2023} sample is reached by the densest CANUCS fields, but this is mainly due to sources with high photo-$z$ ($z_{\rm phot}>8.5$ values whose redshift and masses are uncertain (see Sect.~\ref{sec:discussion-redshift}).

\begin{figure}
    \centering
    \includegraphics[width=\linewidth]{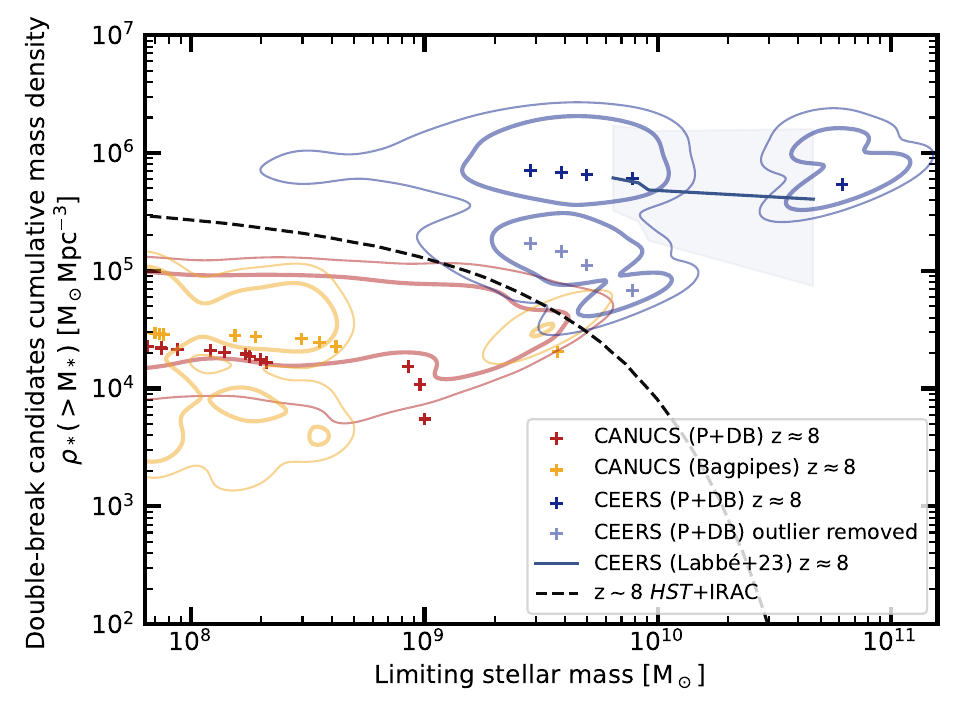}
    \caption{Cumulative stellar mass densities of the double-break source samples in the redshift range $z=7$--$8.5$. The CANUCS sample results are shown by the red crosses for the cumulative mass densities obtained with {\tt DB} and the orange crosses for {\tt Bagpipes}. The CEERS sample \citetalias{Labbe2023} results are displayed as the blue line and the blue crosses show the results obtained with {\tt DB} on that sample. The light shade blue crosses show the CEERS sample {\tt Phosphoros}+{\tt DB} results while removing the contribution of the most massive source. The coloured contours represent the one (thick contours) and two~$\sigma$ (thin contours) errors on the positions of the  different cumulative distribution points computed from resampling. The visible offset of the contours from the points is expected and is due to the population of higher redshift and more massive sources that can fall in the $z\sim8$ range during the random resampling. Inversely, the different modes in the contours are due to the most massive, outlier object in the $z\sim8$ sample leaving during the resampling, thereby leaving only the less massive population. The dashed black line show the expected stellar mass density at $z\sim8$ from {\it HST} and IRAC observations \citep{Stefanon2021}. The masses are all scaled to the \citet{Chabrier2003} IMF.}
    \label{fig:cumulmass}
\end{figure}

From the {\tt DB} and {\tt Bagpipes} fit results, we build the cumulative stellar mass density function of our double-break sample. We focus only on redshifts $z\sim8$, comparable to \citetalias{Labbe2023}'s  $z=7.0$--$8.5$ range, as this is where the majority of our sample is.  For higher redshifts the limited size of our (and \citetalias{Labbe2023}'s) sample would not allow a meaningful comparison. Following \citetalias{Labbe2023}, in Figure~\ref{fig:cumulmass} we show the \citetalias{Labbe2023} cumulative stellar mass density along with CANUCS cumulative stellar mass densities computed with both our template fitting codes, and the results of our {\tt Phosphoros}+{\tt DB} fitting of the \citetalias{Labbe2023} sample photometry. The Figure also displays the \citet{Stefanon2021} estimation of the $z\sim 8$ cumulative mass density function derived with {\it HST} and {\it Spitzer}/IRAC. 

The contours presented in Fig.~\ref{fig:cumulmass} represent the 1- and 2-$\sigma$ errors on the position of the cumulative stellar mass density functions. For the CANUCS samples, they are computed from randomly sampling the joint mass, redshift and magnification posteriors with 10\,000 draws, and building as many cumulative functions considering random sub-samples of the full dataset while accounting for fitting errors and for field-to-field variance in the survey. For the CEERS sample fit with {\tt Phosphoros}+{\tt DB}, we also draw 10\,000 samples from the joint mass and redshift posteriors, although we cannot include field-to-field variance in the same way as we do for CANUCS since the CEERS data cover only one field. Instead, we use bootstrap resampling to account for sample variance. The contours are then built using kernel density estimate with Scott's rule \citep{Scott1992}, encompassing 68 and 90 per cent of the random cumulative mass density point positions. 

Figure~\ref{fig:cumulmass} shows that our {\tt Phosphoros}+{\tt DB} fit results for the CEERS sample are consistent with those in \citetalias{Labbe2023}. We also note that there is good agreement between our two fitting procedures for the CANUCS sample. However, the CEERS and the CANUCS samples provide density distributions that are quite different, the CEERS one lying above the \citet{Stefanon2021} distribution, and the CANUCS one below it. We note that the cumulative distributions are mainly driven by their respective most massive points, which is reflected in the bimodality in mass density  error contours for the lower-mass population of the samples. Both the CEERS sample and the CANUCS {\tt Bagpipes} sample display this multi-modal error distribution due to the presence of one massive outlier in each sample. In the case of the CEERS sample, the mode corresponding to the situation where the massive source is absent from the resampling (shown with ligher-coloured blue crosses) lies close to  the {\it HST}+IRAC cumulative distribution.

In contrast to the CEERS sample, the CANUCS sample lies below the \citet{Stefanon2021} {\it HST}+IRAC curve in all cases. This is not unexpected:  Since there is no reason to think that our sample is complete (it is designed to selected double-break sources only, whilst the \citet{Stefanon2021} curve has no such restriction), it is expected that the CANUCS cumulative double-break source mass distributions are lower limits on the total population.  They are expected to fall below those for the more complete sample from {\it HST}+IRAC. It is the fact that the CEERS double-break sample lies above the {\it HST}+IRAC curve that has been regarded as at odds with previous observations, although -- as we have argued above -- removing the most massive, extreme outlier from that sample goes a long way towards reconciliation with previous, pre-{\it JWST} observations.


\section{Discussion}

The physical properties of the high-$z$ double-break source population has been claimed to be in tension with the $\Lambda$CDM cosmology. This is mainly due to the assumption that this population represents the most massive end of the high redshift population due to their brightness and the putative presence of the Balmer break that indicates an evolved and massive stellar population. Despite finding  in CANUCS sources with similar SED properties in terms of brightness and colours, we do not find stellar masses that appear to be in tension with pre-{\it JWST} literature and cosmology predictions. We discuss here the different reasons that explain why the double-break galaxy population seems not to be problematic from the analysis carried out in this work.

\subsection{Emission lines rather than Balmer Breaks}
\label{sec:discussion-emline}

Double-break photometric selection criteria can select several types of sources \citep{Barro2024,Trussler2023arXiv}: for some the red break can be the Balmer break due to evolved stellar populations; for some, strong optical emission lines can masquerade as a break in the  photometry \citep{Schaerer2009}; and some can be red AGN at high-$z$ \citep{Matthee2023arXiv, Labbe2023arXiv}. The Balmer break galaxies will have greater stellar masses than those dominated by emission lines or an AGN component, so knowing the nature of the photometric break is a strong requirement to ensure the validity of the mass measurements. However, at most redshifts, broad band photometry alone is insufficient to break these degeneracies. To break the degeneracies either spectroscopy or medium band photometry is needed.  

Our sample of double-break sources has been selected through the same photometric criteria prescribed by \citetalias{Labbe2023} (Section~\ref{sec:selection}). { All double-break sources present in the \emph{CLU} fields were identified as high priority candidates for spectroscopic follow-up, but due to limitations on slit positions imposed by the three MSA configurations per \emph{CLU} field centred on the cluster, only five of these were observed.  Given the MSA assignment procedure, there is no reason to think that these five are in any way systematically different than the full sample. } The spectra of all five (Fig~\ref{fig:spectra}) show strong [\textsc{Oiii}] and H$_\beta$ emission lines that fall in the F444W band. Moreover, no Balmer breaks are visible in the continua, although in some cases the continuum S/N is low. The estimated contributions of the emission lines to the observed broadband F444W fluxes range from $\sim~40$ to 70 percent, thus explaining the apparent break observed in the photometry.  

In the CEERS double-break sample, one source has been observed with NIRSpec and has been confirmed to be at redshift $z=7.9932$ \citep{Fujimoto2023}. Similarly to our objects, the reported spectrum also shows strong [\textsc{Oiii}] and H$_\beta$ lines in the F444W wavelength range. 

While we cannot rule out that other galaxies in the double-break sample do not have Balmer breaks, the fact that for all our observed sources spectra are from break-less line emitters seems to indicate that emission lines, rather than Balmer breaks, are a more plausible cause of the observed photometric breaks. The SED fits of some CANUCS galaxies show that both solutions are plausible with {\tt DB} and {\tt Bagpipes} proposing both solutions as best fits (e.g., id=1111752, id=1211968 or id=4117337). In the absence of spectroscopy the emission line/Balmer break degeneracies in these objects cannot be resolved.  Consequently, spectroscopy is required to confirm the nature of all double-break sources { \citep{RobertsBorsani2020}}; alternatively, medium band photometry could be used to properly sample the red parts of the SEDs \citep{Trussler2023arXiv}. Our results are in line with other {\it JWST} photometric and spectroscopic studies that suggest strong rest-optical emission lines are more common than strong Balmer breaks \citep{Endsley2023arXiv,Vikaeus2024MNRAS.tmp.} at the relevant redshifts.

\subsection{Getting the right redshift}
\label{sec:discussion-redshift}

\begin{figure}
    \centering
    \includegraphics[width=\linewidth]{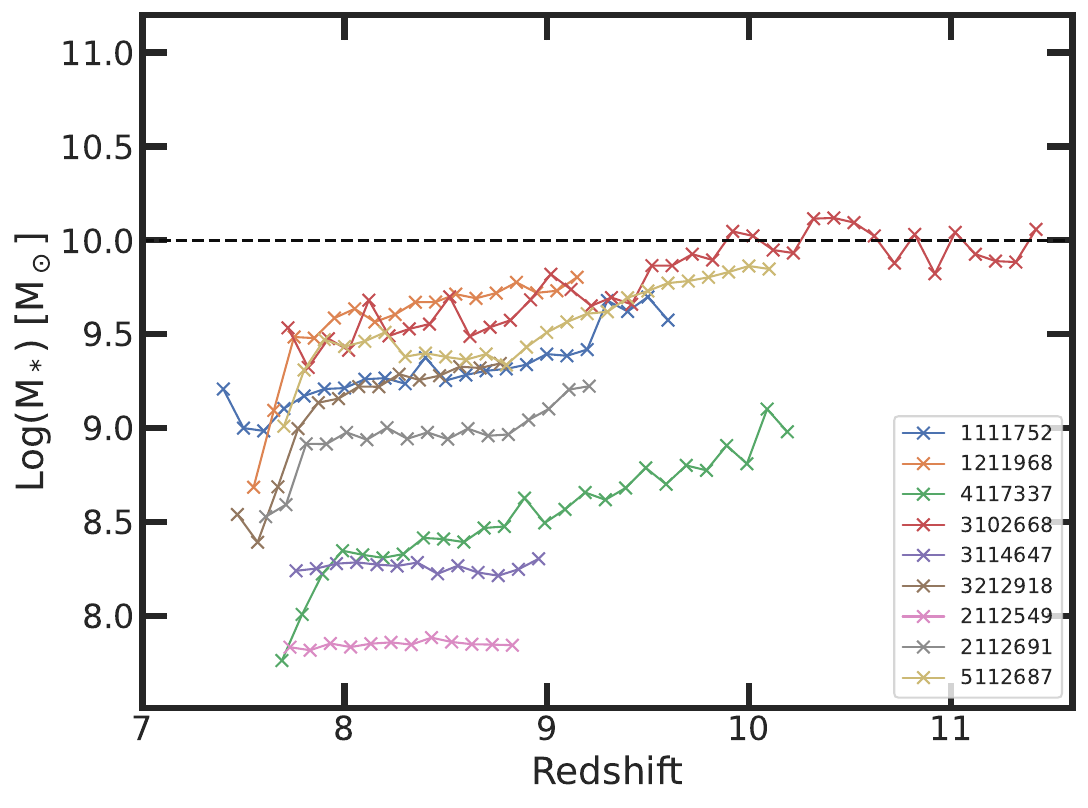}
    \caption{Covariance between stellar mass and redshift in {\tt Dense Basis} fits for sources with a broad P($z$). In most cases, a higher mass is obtained at a higher redshift. Some sources show a steepening at $z>9$ where the strong {\sc [Oiii]} lines move out of the F444W filter.}
    \label{fig:mass-z-correlation}
\end{figure}

To properly assess the stellar masses of galaxies, a reasonable knowledge of the redshift is required, as the mass fits depend on the proper identification of SED features (i.e., Balmer break position and strength) as well as intrinsic brightness of the sources. For instance, Fig.~\ref{fig:mass-z-correlation} shows how the stellar masses vary with the redshift for the CANUCS double-break objects within the broad range of redshift allowed by their  {\tt Phosphoros}+{\tt DB}  P($z$)'s. The Figure illustrates that changing the redshifts moves the positions and strengths of the emission lines and Balmer breaks as the SEDs move through the filters,  resulting in higher inferred masses for higher redshifts. In particular, some sources show a steepening at $z>9$ where the strong {\sc [Oiii]} lines move out of the F444W filter. { For example, source id=4117337 has a spectroscopic redshift at $z=8.295$, and a stellar mass fit of ${\rm log(M_*}/{\rm M_{\sun}} )\sim8.3$  with \texttt{DB} at this redshift. Nevertheless, Fig.~\ref{fig:spectrosample} shows that \texttt{Phosphoros} P($z$) allows for a solution around $z\sim9.7$--$10$, matching a fit of the SED red break with a Balmer break. For these redshift, Fig.~\ref{fig:mass-z-correlation} shows that the stellar mass fit becomes  ${\rm log(M_*}/{\rm M_{\sun}} )\sim8.8$, thus a $0.5\,{\rm dex}$ increase. However, this value is source dependant as it depends on the strength of the emission lines and thus cannot be generalised to the whole sample, but it gives an insight of the correction that can be expected for such effect. This can also explain} some differences between the {\tt DB} and {\tt Bagpipes} results, as they do not use the same redshifts ({\tt Bagpipes} uses the best-fit redshift, which is generally lower than the median redshift, thus leading to lower stellar masses). 

Selection from broadband photometry can lead to the inclusion of outliers in the sample -- outliers for which the photo-$z$ is strongly discrepant with the true redshifts. This is the case for the CEERS source id=13050 in \citetalias{Labbe2023}, with a photo-$z$ estimated at $z_{\rm phot}=8.14^{+0.45}_{-1.71}$ and a stellar mass of ${\rm log(M_*}/{\rm M_{\sun}} )=9.9^{+0.29}_{-0.3}$ but for which spectroscopic observations showed it to be an AGN at $z_{\rm spec}=5.624$ \citep{Kocevski2023}. This underlines the importance of obtaining { spectroscopy} for individual high-$z$ candidates in order to properly characterise the masses (and other properties) of the population. To emphasise:  even though the photometric redshifts may be approximately correct, exact, spectroscopic redshifts are needed for unbiased mass determinations.

\subsection{Field-to-field variance}
\label{sec:discussion-cosmicvariance}

With five widely separated sight-lines, the CANUCS data in Fig.~\ref{fig:mass-v-redshift} and Table~\ref{tab:fields} show different numbers of objects and stellar mass densities in each field. Figure~\ref{fig:mass-v-redshift} (right panel) shows that there is a difference up to $\sim 1.5$\,dex in stellar mass between the most massive double-break sources found in the five different fields. The CEERS field presents a higher mass density than any of the CANUCS fields, and has a mass difference of $\sim 0.5$ to 1.0\,dex with the whole CANUCS sample at fixed source density and considering only the most massive estimates for our sample. Although there may be minor systematic differences between the CANUCS and CEERS samples due to data quality (CANUCS is deeper and has more {\it JWST} filters) and SED fitting procedures (Sections \ref{sec:results}, \ref{sec:discussion-emline}, \ref{sec:discussion-redshift}), it does indeed appear that the four CEERS NIRCam pointings analysed by \citetalias{Labbe2023} are more overdense in massive galaxies at this redshift than the typical random sight-lines probed by CANUCS. 

The variance in mass density observed in these fields is a combination of sample variance and cosmic variance. For the five CANUCS sight-lines, the number of sources observed ranges between 2 and 5. This level of variation could be completely explained by Poisson variance. The higher mass density from CEERS is likely at least partially a consequence of cosmic variance. \citetalias{Labbe2023} estimate the effect of cosmic variance in their sample to be 30\%, but along a single sight-line there will always be an element of uncertainty in such a calculation. For example, the GLASS NIRCam parallel field shows a 3$\times$ to 10$\times$ overdensity in galaxies at redshift $z\sim 10$ \citep{Castellano2023}. High-redshift overdensities have also been observed in {\it JWST} spectroscopic samples \citep{Kashino2023, Morishita2023, Helton2024}, as expected based on the growth of large-scale structure. The CANUCS results presented here show that typical sight-lines do not show a high mass density at these redshifts and serve as a caution that multiple fields should be considered before drawing strong conclusions about what is typical.

\subsection{Effect of outlier(s)}
\label{sec:discussion-outlier}

The cumulative stellar mass density distributions presented in Fig.~\ref{fig:cumulmass} depend strongly on the most massive object in the sample. This is especially the case when a single source is an outlier. This is illustrated by the error contours in the figure showing that in the presence of such massive outliers, a multi-modal distributions results, as is the case for the CEERS sample {\tt DB} results and for the CANUCS {\tt Bagpipes} ones. This is due to the potential absence of outlier sources in the resampling used to create the contours, thus leading to cumulative distributions only considering the lower mass sample. This is further illustrated for the CEERS sample with lighter shade crosses in the Fig~\ref{fig:cumulmass}, showing what the cumulative mass density would be without the most massive source in the sample. Disregarding the most massive source, the results of the CEERS sample are consistent with the \citet{Stefanon2021} stellar mass density distribution, { as suggested by \citet{Akins2023} or \citet{WangYi2023}}.  We note that we can see a similar effect for the {\tt Bagpipes} results on the CANUCS sample that is due to source id=1111752, which is more massive than the rest of the $z\sim8$ sample by $~0.7$\,dex. Disregarding this outlier would draw the density distribution down, but it would still remain consistent with the results of {\tt Phosphoros}+{\tt DB}. This is not a problem, as we expect that the CANUCS sample of double-break galaxies is only a subset of all $z\approx8$ objects.

The effect of these outliers is problematic in small samples { and small probed volumes} regardless of their true nature. It could be that these massive sources are indeed truly massive Balmer break galaxies, but the size of the samples and the volume which is probed are too limited to allow generalising the results obtained with proper statistical significance. As discussed in the previous section (Sect.~\ref{sec:discussion-cosmicvariance}), cosmic variance is an explanation of the potential presence of such outliers in the sample. Larger areas in different, widely-separated sight-lines need to be probed to determine whether the high stellar mass densities they result in are real or  just an artefact of the small samples used to measure the average mass density.

Outliers can also be due to a miss-classification of sources (AGN, brown dwarf, etc.), or even improper fits of their SEDs, as discussed in the previous sections (see Sect.~\ref{sec:discussion-emline} and~\ref{sec:discussion-redshift}). One example of this is CANUCS source id=4115596, whose colours match the double-break criteria but which seems to be a brown dwarf. Including it in the analysis would have added a source { with  ${\rm M_*}=0.3^{+3.7}_{-0.2}\times10^9\,{\rm M_{\sun}}$ according to {\tt DB} and ${\rm M_*}=2.5^{+3.8}_{-2.2}\times10^9\,{\rm M_{\sun}}$ from {\tt Bagpipes} (see Fig~\ref{fig:BrownDwarf}),} which would have resulted in a massive outlier in the sample. Another example in the literature, the \citet{Endsley2023} fit of the most massive galaxy of \citetalias{Labbe2023} suggests that after accounting for emission lines and/or AGN emission, the stellar mass of the source drops by one dex and the source ceases to be an extreme massive galaxy. Sources where the double-break signature has been revealed to be AGN or brown dwarfs { \citep{Greene2023arXiv,kokorev2023,Labbe2023arXiv,Langeroodi2023,Burgasser2024}} should be considered to be such miss-classified outliers.

\section{Conclusions}

This work presents an analysis of double-break sources in the CANUCS survey. Over the effective area $\sim60$\,arcmin$^2$ for $z\sim8$ covering five cluster and parallel fields, we detect 20 sources that correspond to the double-break criteria as defined by \citetalias{Labbe2023}. Removing one brown dwarf from this sample, we are left with 19 double-break galaxies. The redshifts and stellar masses of these 19 galaxies are derived with a combination of {\tt Phosphoros} and either {\tt Dense Basis} or {\tt Bagpipes}. The stellar masses obtained are compared to the ones of the CEERS sample and are found to be lower, and not in contradiction with the stellar mass densities derived with {\it HST} \citep{Stefanon2021} or with galaxy formation models in standard $\Lambda$CDM \citep{Menci2022,Lovell2023}. In other words, CANUCS does not find a large population of massive Balmer-break sources at high $z$.

The difference between the conclusion from the CANUCS sample and that of CEERS can be explained by a combination of four different effects. The first one is the confusion between the contribution of optical emission lines and the Balmer break produced by an evolved galaxy population in the source SEDs. Out of the 19 candidates in our sample, five are observed with NIRSpec and all five of these spectra present strong emission lines that contribute strongly to the red band fluxes. Whilst this does not discard the possibility that other sources in our sample may have pronounced Balmer breaks, emission lines seem more likely to explain the photometry break { than an old and massive stellar population at high-$z$}.

Secondly, a { precise} knowledge on the source redshift is necessary, as the mass can strongly vary with the redshift assumed, especially when strong emission lines (or a pronounced Balmer break) are moving in and out of photometric bands. Thus, unconstrained $P(z)$ can lead to poorly constrained stellar mass estimates. 

Thirdly, field-to-field variance plays a role as we observe significant variation in the properties of the double-break population from one CANUCS field to another, and see an even higher mass density in the CEERS field used by \citetalias{Labbe2023}. A single narrow sight-line is not enough to generalise the properties of the population found in it to the entire universe.

Finally, when handling small samples, the cumulative stellar mass density distribution is highly dependent on the most massive object used to build it, and thus is extremely sensitive to outliers. Whether those outliers are truly rare massive sources, missclassified lower $z$ interlopers, unaccounted-for AGN, brown dwarfs, or even bad fit results, the cumulative mass distribution is driven by their upward reported masses. 

In order to claim that the observed population of high-$z$ galaxies is in tension with our expectations from $\Lambda$CDM, all these points must be addressed. Deep spectroscopy over large areas and different sight-lines in addition to deep broad and medium band photometry are needed to ensure the robustness of the conclusion drawn about the high-end of stellar mass density at $z>7$.

\section*{Acknowledgements}

{ The authors thank the referee for their helpful comments.}
This research was enabled by grant 18JWST-GTO1 from the Canadian Space Agency, and Discovery Grants to MS, AM, and RA from the Natural Sciences and Engineering Research Council (NSERC) of Canada.
YA is supported by a Research Fellowship for Young Scientists from the Japan Society of the Promotion of Science (JSPS).
MB and GR acknowledge support from the ERC Grant FIRSTLIGHT and from the Slovenian national research agency ARRS through grants N1-0238 and P1-0188. MB acknowledges support from the program HST-GO-16667, provided through a grant from the STScI under NASA contract NAS5-26555. 
This research used the Canadian Advanced Network For Astronomy Research (CANFAR) operated in partnership by the Canadian Astronomy Data Centre and The Digital Research Alliance of Canada with support from the National Research Council of Canada the Canadian Space Agency, CANARIE and the Canadian Foundation for Innovation.

\section*{Data Availability}

The CANUCS data is available at DOI: 10.17909/ph4n-6n76. The data presented in this work will be provided on request.



\bibliographystyle{mnras}
\bibliography{references} 

\begin{thebibliography}{}
\makeatletter
\relax
\def\mn@urlcharsother{\let\do\@makeother \do\$\do\&\do\#\do\^\do\_\do\%\do\~}
\def\mn@doi{\begingroup\mn@urlcharsother \@ifnextchar [ {\mn@doi@}
  {\mn@doi@[]}}
\def\mn@doi@[#1]#2{\def\@tempa{#1}\ifx\@tempa\@empty \href
  {http://dx.doi.org/#2} {doi:#2}\else \href {http://dx.doi.org/#2} {#1}\fi
  \endgroup}
\def\mn@eprint#1#2{\mn@eprint@#1:#2::\@nil}
\def\mn@eprint@arXiv#1{\href {http://arxiv.org/abs/#1} {{\tt arXiv:#1}}}
\def\mn@eprint@dblp#1{\href {http://dblp.uni-trier.de/rec/bibtex/#1.xml}
  {dblp:#1}}
\def\mn@eprint@#1:#2:#3:#4\@nil{\def\@tempa {#1}\def\@tempb {#2}\def\@tempc
  {#3}\ifx \@tempc \@empty \let \@tempc \@tempb \let \@tempb \@tempa \fi \ifx
  \@tempb \@empty \def\@tempb {arXiv}\fi \@ifundefined
  {mn@eprint@\@tempb}{\@tempb:\@tempc}{\expandafter \expandafter \csname
  mn@eprint@\@tempb\endcsname \expandafter{\@tempc}}}

\bibitem[\protect\citeauthoryear{{Akins} et~al.,}{{Akins}
  et~al.}{2023}]{Akins2023}
{Akins} H.~B.,  et~al., 2023, \mn@doi [\apj] {10.3847/1538-4357/acef21}, \href
  {https://ui.adsabs.harvard.edu/abs/2023ApJ...956...61A} {956, 61}

\bibitem[\protect\citeauthoryear{{Arrabal Haro} et~al.,}{{Arrabal Haro}
  et~al.}{2023}]{ArrabalHaro2023}
{Arrabal Haro} P.,  et~al., 2023, \mn@doi [\nat] {10.1038/s41586-023-06521-7},
  \href {https://ui.adsabs.harvard.edu/abs/2023Natur.622..707A} {622, 707}

\bibitem[\protect\citeauthoryear{{Asada} et~al.,}{{Asada}
  et~al.}{2023}]{Asada2023}
{Asada} Y.,  et~al., 2023, \mn@doi [\mnras] {10.1093/mnrasl/slad054}, \href
  {https://ui.adsabs.harvard.edu/abs/2023MNRAS.523L..40A} {523, L40}

\bibitem[\protect\citeauthoryear{{Asada} et~al.,}{{Asada}
  et~al.}{2024}]{Asada2024}
{Asada} Y.,  et~al., 2024, \mn@doi [\mnras] {10.1093/mnras/stad3902}, \href
  {https://ui.adsabs.harvard.edu/abs/2024MNRAS.527.11372} {527, 11372}

\bibitem[\protect\citeauthoryear{{Atek} et~al.,}{{Atek}
  et~al.}{2023}]{Atek2023arXiv}
{Atek} H.,  et~al., 2023, \mn@doi [arXiv e-prints] {10.48550/arXiv.2308.08540},
  \href {https://ui.adsabs.harvard.edu/abs/2023arXiv230808540A} {p.
  arXiv:2308.08540}

\bibitem[\protect\citeauthoryear{{Bagley} et~al.,}{{Bagley}
  et~al.}{2023}]{Bagley2023}
{Bagley} M.~B.,  et~al., 2023, \mn@doi [\apjl] {10.3847/2041-8213/acbb08},
  \href {https://ui.adsabs.harvard.edu/abs/2023ApJ...946L..12B} {946, L12}

\bibitem[\protect\citeauthoryear{{Bakx} et~al.,}{{Bakx}
  et~al.}{2020}]{Bakx2020}
{Bakx} T. J.~L.~C.,  et~al., 2020, \mn@doi [\mnras] {10.1093/mnras/staa509},
  \href {https://ui.adsabs.harvard.edu/abs/2020MNRAS.493.4294B} {493, 4294}

\bibitem[\protect\citeauthoryear{{Barro} et~al.,}{{Barro}
  et~al.}{2024}]{Barro2024}
{Barro} G.,  et~al., 2024, \mn@doi [\apj] {10.3847/1538-4357/ad167e}, \href
  {https://ui.adsabs.harvard.edu/abs/2024ApJ...963..128B} {963, 128}

\bibitem[\protect\citeauthoryear{{Bergamini} et~al.,}{{Bergamini}
  et~al.}{2023}]{Bergamini2023}
{Bergamini} P.,  et~al., 2023, \mn@doi [\aap] {10.1051/0004-6361/202244834},
  \href {https://ui.adsabs.harvard.edu/abs/2023A&A...674A..79B} {674, A79}

\bibitem[\protect\citeauthoryear{{Biagetti}, {Franciolini}  \&
  {Riotto}}{{Biagetti} et~al.}{2023}]{Biagetti2023}
{Biagetti} M.,  {Franciolini} G.,   {Riotto} A.,  2023, \mn@doi [\apj]
  {10.3847/1538-4357/acb5ea}, \href
  {https://ui.adsabs.harvard.edu/abs/2023ApJ...944..113B} {944, 113}

\bibitem[\protect\citeauthoryear{{Boylan-Kolchin}}{{Boylan-Kolchin}}{2023}]{Boylan-Kolchin2023}
{Boylan-Kolchin} M.,  2023, \mn@doi [Nature Astronomy]
  {10.1038/s41550-023-01937-7}, \href
  {https://ui.adsabs.harvard.edu/abs/2023NatAs...7..731B} {7, 731}

\bibitem[\protect\citeauthoryear{Bradley et~al.,}{Bradley
  et~al.}{2023}]{Bradley2023}
Bradley L.,  et~al., 2023, astropy/photutils: 1.7.0,
  \mn@doi{10.5281/zenodo.7804137}, \url
  {https://doi.org/10.5281/zenodo.7804137}

\bibitem[\protect\citeauthoryear{Brammer}{Brammer}{2022}]{Brammer2022msaexp}
Brammer G.,  2022, {gbrammer/msaexp: Full working version with 2d drizzling and
  extraction}, \mn@doi{10.5281/zenodo.7299501}, \url
  {https://doi.org/10.5281/zenodo.7299501}

\bibitem[\protect\citeauthoryear{{Brammer}}{{Brammer}}{2023}]{grizli23b}
{Brammer} G.,  2023, {grizli}, Zenodo, \mn@doi{10.5281/zenodo.8210732}, \url
  {https://doi.org/10.5281/zenodo.8210732}

\bibitem[\protect\citeauthoryear{{Brammer}, {van Dokkum}  \& {Coppi}}{{Brammer}
  et~al.}{2008}]{Brammer2008}
{Brammer} G.~B.,  {van Dokkum} P.~G.,   {Coppi} P.,  2008, \mn@doi [\apj]
  {10.1086/591786}, \href
  {https://ui.adsabs.harvard.edu/abs/2008ApJ...686.1503B} {686, 1503}

\bibitem[\protect\citeauthoryear{{Bunker} et~al.,}{{Bunker}
  et~al.}{2023a}]{Bunker2023arXivb}
{Bunker} A.~J.,  et~al., 2023a, \mn@doi [arXiv e-prints]
  {10.48550/arXiv.2306.02467}, \href
  {https://ui.adsabs.harvard.edu/abs/2023arXiv230602467B} {p. arXiv:2306.02467}

\bibitem[\protect\citeauthoryear{{Bunker} et~al.,}{{Bunker}
  et~al.}{2023b}]{Bunker2023}
{Bunker} A.~J.,  et~al., 2023b, \mn@doi [\aap] {10.1051/0004-6361/202346159},
  \href {https://ui.adsabs.harvard.edu/abs/2023A&A...677A..88B} {677, A88}

\bibitem[\protect\citeauthoryear{{Burgasser} et~al.,}{{Burgasser}
  et~al.}{2024}]{Burgasser2024}
{Burgasser} A.~J.,  et~al., 2024, \mn@doi [\apj] {10.3847/1538-4357/ad206f},
  \href {https://ui.adsabs.harvard.edu/abs/2024ApJ...962..177B} {962, 177}

\bibitem[\protect\citeauthoryear{{Calzetti}, {Armus}, {Bohlin}, {Kinney},
  {Koornneef}  \& {Storchi-Bergmann}}{{Calzetti} et~al.}{2000}]{Calzetti2000}
{Calzetti} D.,  {Armus} L.,  {Bohlin} R.~C.,  {Kinney} A.~L.,  {Koornneef} J.,
   {Storchi-Bergmann} T.,  2000, \mn@doi [\apj] {10.1086/308692}, \href
  {https://ui.adsabs.harvard.edu/abs/2000ApJ...533..682C} {533, 682}

\bibitem[\protect\citeauthoryear{{Carnall}, {McLure}, {Dunlop}  \&
  {Dav{\'e}}}{{Carnall} et~al.}{2018}]{Carnall2018}
{Carnall} A.~C.,  {McLure} R.~J.,  {Dunlop} J.~S.,   {Dav{\'e}} R.,  2018,
  \mn@doi [\mnras] {10.1093/mnras/sty2169}, \href
  {https://ui.adsabs.harvard.edu/abs/2018MNRAS.480.4379C} {480, 4379}

\bibitem[\protect\citeauthoryear{{Casey} et~al.,}{{Casey}
  et~al.}{2023a}]{Casey2023arXiv}
{Casey} C.~M.,  et~al., 2023a, \mn@doi [arXiv e-prints]
  {10.48550/arXiv.2308.10932}, \href
  {https://ui.adsabs.harvard.edu/abs/2023arXiv230810932C} {p. arXiv:2308.10932}

\bibitem[\protect\citeauthoryear{{Casey} et~al.,}{{Casey}
  et~al.}{2023b}]{Casey2023}
{Casey} C.~M.,  et~al., 2023b, \mn@doi [\apj] {10.3847/1538-4357/acc2bc}, \href
  {https://ui.adsabs.harvard.edu/abs/2023ApJ...954...31C} {954, 31}

\bibitem[\protect\citeauthoryear{{Castellano} et~al.,}{{Castellano}
  et~al.}{2023}]{Castellano2023}
{Castellano} M.,  et~al., 2023, \mn@doi [\apjl] {10.3847/2041-8213/accea5},
  \href {https://ui.adsabs.harvard.edu/abs/2023ApJ...948L..14C} {948, L14}

\bibitem[\protect\citeauthoryear{{Chabrier}}{{Chabrier}}{2003}]{Chabrier2003}
{Chabrier} G.,  2003, \mn@doi [\pasp] {10.1086/376392}, \href
  {https://ui.adsabs.harvard.edu/abs/2003PASP..115..763C} {115, 763}

\bibitem[\protect\citeauthoryear{{Curtis-Lake} et~al.,}{{Curtis-Lake}
  et~al.}{2023}]{Curtis-Lake2023}
{Curtis-Lake} E.,  et~al., 2023, \mn@doi [Nature Astronomy]
  {10.1038/s41550-023-01918-w}, \href
  {https://ui.adsabs.harvard.edu/abs/2023NatAs...7..622C} {7, 622}

\bibitem[\protect\citeauthoryear{{Dayal} \& {Giri}}{{Dayal} \&
  {Giri}}{2024}]{Dayal2024}
{Dayal} P.,  {Giri} S.~K.,  2024, \mn@doi [\mnras] {10.1093/mnras/stae176},
  \href {https://ui.adsabs.harvard.edu/abs/2024MNRAS.528.2784D} {528, 2784}

\bibitem[\protect\citeauthoryear{{Desprez}, {Richard}, {Jauzac}, {Martinez},
  {Siana}  \& {Cl{\'e}ment}}{{Desprez} et~al.}{2018}]{Desprez2018}
{Desprez} G.,  {Richard} J.,  {Jauzac} M.,  {Martinez} J.,  {Siana} B.,
  {Cl{\'e}ment} B.,  2018, \mn@doi [\mnras] {10.1093/mnras/sty1666}, \href
  {https://ui.adsabs.harvard.edu/abs/2018MNRAS.479.2630D} {479, 2630}

\bibitem[\protect\citeauthoryear{{Desprez} et~al.,}{{Desprez}
  et~al.}{2023}]{Desprez2023}
{Desprez} G.,  et~al., 2023, \mn@doi [\aap] {10.1051/0004-6361/202243363},
  \href {https://ui.adsabs.harvard.edu/abs/2023A&A...670A..82D} {670, A82}

\bibitem[\protect\citeauthoryear{{Doyon} et~al.,}{{Doyon}
  et~al.}{2023}]{Doyon2023}
{Doyon} R.,  et~al., 2023, \mn@doi [\pasp] {10.1088/1538-3873/acd41b}, \href
  {https://ui.adsabs.harvard.edu/abs/2023PASP..135i8001R} {135, 098001}

\bibitem[\protect\citeauthoryear{{Drlica-Wagner} et~al.,}{{Drlica-Wagner}
  et~al.}{2018}]{Drlica-Wagner2018}
{Drlica-Wagner} A.,  et~al., 2018, \mn@doi [\apjs] {10.3847/1538-4365/aab4f5},
  \href {https://ui.adsabs.harvard.edu/abs/2018ApJS..235...33D} {235, 33}

\bibitem[\protect\citeauthoryear{{Ebeling}, {Edge}  \& {Henry}}{{Ebeling}
  et~al.}{2001}]{Ebeling2001}
{Ebeling} H.,  {Edge} A.~C.,   {Henry} J.~P.,  2001, \mn@doi [\apj]
  {10.1086/320958}, \href
  {https://ui.adsabs.harvard.edu/abs/2001ApJ...553..668E} {553, 668}

\bibitem[\protect\citeauthoryear{{El{\'\i}asd{\'o}ttir}
  et~al.,}{{El{\'\i}asd{\'o}ttir} et~al.}{2007}]{Eliasdottir2007arXiv}
{El{\'\i}asd{\'o}ttir} {\'A}.,  et~al., 2007, \mn@doi [arXiv e-prints]
  {10.48550/arXiv.0710.5636}, \href
  {https://ui.adsabs.harvard.edu/abs/2007arXiv0710.5636E} {p. arXiv:0710.5636}

\bibitem[\protect\citeauthoryear{{Endsley} et~al.,}{{Endsley}
  et~al.}{2023a}]{Endsley2023arXiv}
{Endsley} R.,  et~al., 2023a, \mn@doi [arXiv e-prints]
  {10.48550/arXiv.2306.05295}, \href
  {https://ui.adsabs.harvard.edu/abs/2023arXiv230605295E} {p. arXiv:2306.05295}

\bibitem[\protect\citeauthoryear{{Endsley}, {Stark}, {Whitler}, {Topping},
  {Chen}, {Plat}, {Chisholm}  \& {Charlot}}{{Endsley}
  et~al.}{2023b}]{Endsley2023}
{Endsley} R.,  {Stark} D.~P.,  {Whitler} L.,  {Topping} M.~W.,  {Chen} Z.,
  {Plat} A.,  {Chisholm} J.,   {Charlot} S.,  2023b, \mn@doi [\mnras]
  {10.1093/mnras/stad1919}, \href
  {https://ui.adsabs.harvard.edu/abs/2023MNRAS.524.2312E} {524, 2312}

\bibitem[\protect\citeauthoryear{{Euclid Collaboration: Desprez}
  et~al.,}{{Euclid Collaboration: Desprez} et~al.}{2020}]{Desprez2020}
{Euclid Collaboration: Desprez} G.,  et~al., 2020, \mn@doi [\aap]
  {10.1051/0004-6361/202039403}, \href
  {https://ui.adsabs.harvard.edu/abs/2020A&A...644A..31E} {644, A31}

\bibitem[\protect\citeauthoryear{{Forconi}, {Ruchika}, {Melchiorri}, {Mena}  \&
  {Menci}}{{Forconi} et~al.}{2023}]{Forconi2023}
{Forconi} M.,  {Ruchika} {Melchiorri} A.,  {Mena} O.,   {Menci} N.,  2023,
  \mn@doi [\jcap] {10.1088/1475-7516/2023/10/012}, \href
  {https://ui.adsabs.harvard.edu/abs/2023JCAP...10..012F} {2023, 012}

\bibitem[\protect\citeauthoryear{{Franco} et~al.,}{{Franco}
  et~al.}{2023}]{Franco2023arXiv}
{Franco} M.,  et~al., 2023, \mn@doi [arXiv e-prints]
  {10.48550/arXiv.2308.00751}, \href
  {https://ui.adsabs.harvard.edu/abs/2023arXiv230800751F} {p. arXiv:2308.00751}

\bibitem[\protect\citeauthoryear{{Fujimoto} et~al.,}{{Fujimoto}
  et~al.}{2023}]{Fujimoto2023}
{Fujimoto} S.,  et~al., 2023, \mn@doi [\apjl] {10.3847/2041-8213/acd2d9}, \href
  {https://ui.adsabs.harvard.edu/abs/2023ApJ...949L..25F} {949, L25}

\bibitem[\protect\citeauthoryear{{Gaia Collaboration} et~al.,}{{Gaia
  Collaboration} et~al.}{2016}]{Gaia2016a}
{Gaia Collaboration} et~al., 2016, \mn@doi [\aap]
  {10.1051/0004-6361/201629512}, \href
  {https://ui.adsabs.harvard.edu/abs/2016A&A...595A...2G} {595, A2}

\bibitem[\protect\citeauthoryear{{Gaia Collaboration} et~al.,}{{Gaia
  Collaboration} et~al.}{2023}]{Gaia2023}
{Gaia Collaboration} et~al., 2023, \mn@doi [\aap]
  {10.1051/0004-6361/202243940}, \href
  {https://ui.adsabs.harvard.edu/abs/2023A&A...674A...1G} {674, A1}

\bibitem[\protect\citeauthoryear{{Gardner} et~al.,}{{Gardner}
  et~al.}{2023}]{Gardner2023}
{Gardner} J.~P.,  et~al., 2023, \mn@doi [\pasp] {10.1088/1538-3873/acd1b5},
  \href {https://ui.adsabs.harvard.edu/abs/2023PASP..135f8001G} {135, 068001}

\bibitem[\protect\citeauthoryear{{Greene} et~al.,}{{Greene}
  et~al.}{2023}]{Greene2023arXiv}
{Greene} J.~E.,  et~al., 2023, \mn@doi [arXiv e-prints]
  {10.48550/arXiv.2309.05714}, \href
  {https://ui.adsabs.harvard.edu/abs/2023arXiv230905714G} {p. arXiv:2309.05714}

\bibitem[\protect\citeauthoryear{{Helton} et~al.,}{{Helton}
  et~al.}{2024}]{Helton2024}
{Helton} J.~M.,  et~al., 2024, \mn@doi [\apj] {10.3847/1538-4357/ad0da7}, \href
  {https://ui.adsabs.harvard.edu/abs/2024ApJ...962..124H} {962, 124}

\bibitem[\protect\citeauthoryear{{Hoag} et~al.,}{{Hoag}
  et~al.}{2017}]{Hoag2017}
{Hoag} A.,  et~al., 2017, \mn@doi [Nature Astronomy] {10.1038/s41550-017-0091},
  \href {https://ui.adsabs.harvard.edu/abs/2017NatAs...1E..91H} {1, 0091}

\bibitem[\protect\citeauthoryear{Horne}{Horne}{1986}]{Horne_1986}
Horne K.,  1986, \mn@doi [Publications of the Astronomical Society of the
  Pacific] {10.1086/131801}, 98, 609

\bibitem[\protect\citeauthoryear{{H{\"u}tsi}, {Raidal}, {Urrutia}, {Vaskonen}
  \& {Veerm{\"a}e}}{{H{\"u}tsi} et~al.}{2023}]{Hutsi2023}
{H{\"u}tsi} G.,  {Raidal} M.,  {Urrutia} J.,  {Vaskonen} V.,   {Veerm{\"a}e}
  H.,  2023, \mn@doi [\prd] {10.1103/PhysRevD.107.043502}, \href
  {https://ui.adsabs.harvard.edu/abs/2023PhRvD.107d3502H} {107, 043502}

\bibitem[\protect\citeauthoryear{{Ilbert} et~al.,}{{Ilbert}
  et~al.}{2013}]{Ilbert2013}
{Ilbert} O.,  et~al., 2013, \mn@doi [\aap] {10.1051/0004-6361/201321100}, \href
  {http://adsabs.harvard.edu/abs/2013A%26A...556A..55I} {556, A55}

\bibitem[\protect\citeauthoryear{{Iyer} \& {Gawiser}}{{Iyer} \&
  {Gawiser}}{2017}]{Iyer17}
{Iyer} K.,  {Gawiser} E.,  2017, \mn@doi [\apj] {10.3847/1538-4357/aa63f0},
  \href {https://ui.adsabs.harvard.edu/abs/2017ApJ...838..127I} {838, 127}

\bibitem[\protect\citeauthoryear{{Iyer}, {Gawiser}, {Faber}, {Ferguson},
  {Kartaltepe}, {Koekemoer}, {Pacifici}  \& {Somerville}}{{Iyer}
  et~al.}{2019}]{Iyer19}
{Iyer} K.~G.,  {Gawiser} E.,  {Faber} S.~M.,  {Ferguson} H.~C.,  {Kartaltepe}
  J.,  {Koekemoer} A.~M.,  {Pacifici} C.,   {Somerville} R.~S.,  2019, \mn@doi
  [\apj] {10.3847/1538-4357/ab2052}, \href
  {https://ui.adsabs.harvard.edu/abs/2019ApJ...879..116I} {879, 116}

\bibitem[\protect\citeauthoryear{{Jakobsen} et~al.,}{{Jakobsen}
  et~al.}{2022}]{Jakobsen2022}
{Jakobsen} P.,  et~al., 2022, \mn@doi [\aap] {10.1051/0004-6361/202142663},
  \href {https://ui.adsabs.harvard.edu/abs/2022A&A...661A..80J} {661, A80}

\bibitem[\protect\citeauthoryear{{Jauzac} et~al.,}{{Jauzac}
  et~al.}{2019}]{Jauzac2019}
{Jauzac} M.,  et~al., 2019, \mn@doi [\mnras] {10.1093/mnras/sty3312}, \href
  {https://ui.adsabs.harvard.edu/abs/2019MNRAS.483.3082J} {483, 3082}

\bibitem[\protect\citeauthoryear{{Jullo}, {Kneib}, {Limousin},
  {El{\'{\i}}asd{\'o}ttir}, {Marshall}  \& {Verdugo}}{{Jullo}
  et~al.}{2007}]{Jullo2007}
{Jullo} E.,  {Kneib} J.-P.,  {Limousin} M.,  {El{\'{\i}}asd{\'o}ttir} {\'A}.,
  {Marshall} P.~J.,   {Verdugo} T.,  2007, \mn@doi [New Journal of Physics]
  {10.1088/1367-2630/9/12/447}, \href
  {http://adsabs.harvard.edu/abs/2007NJPh....9..447J} {9, 447}

\bibitem[\protect\citeauthoryear{{Kashino}, {Lilly}, {Matthee}, {Eilers},
  {Mackenzie}, {Bordoloi}  \& {Simcoe}}{{Kashino} et~al.}{2023}]{Kashino2023}
{Kashino} D.,  {Lilly} S.~J.,  {Matthee} J.,  {Eilers} A.-C.,  {Mackenzie} R.,
  {Bordoloi} R.,   {Simcoe} R.~A.,  2023, \mn@doi [\apj]
  {10.3847/1538-4357/acc588}, \href
  {https://ui.adsabs.harvard.edu/abs/2023ApJ...950...66K} {950, 66}

\bibitem[\protect\citeauthoryear{{Kneib}}{{Kneib}}{1993}]{Kneib1993}
{Kneib} J.~P.,  1993, PhD thesis, -

\bibitem[\protect\citeauthoryear{{Kocevski} et~al.,}{{Kocevski}
  et~al.}{2023}]{Kocevski2023}
{Kocevski} D.~D.,  et~al., 2023, \mn@doi [\apjl] {10.3847/2041-8213/ace5a0},
  \href {https://ui.adsabs.harvard.edu/abs/2023ApJ...954L...4K} {954, L4}

\bibitem[\protect\citeauthoryear{{Kokorev} et~al.,}{{Kokorev}
  et~al.}{2023}]{kokorev2023}
{Kokorev} V.,  et~al., 2023, \mn@doi [\apjl] {10.3847/2041-8213/ad037a}, \href
  {https://ui.adsabs.harvard.edu/abs/2023ApJ...957L...7K} {957, L7}

\bibitem[\protect\citeauthoryear{{Kron}}{{Kron}}{1980}]{Kron1980}
{Kron} R.~G.,  1980, \mn@doi [\apjs] {10.1086/190669}, \href
  {https://ui.adsabs.harvard.edu/abs/1980ApJS...43..305K} {43, 305}

\bibitem[\protect\citeauthoryear{{Labbe} et~al.,}{{Labbe}
  et~al.}{2023a}]{Labbe2023arXiv}
{Labbe} I.,  et~al., 2023a, \mn@doi [arXiv e-prints]
  {10.48550/arXiv.2306.07320}, \href
  {https://ui.adsabs.harvard.edu/abs/2023arXiv230607320L} {p. arXiv:2306.07320}

\bibitem[\protect\citeauthoryear{{Labb{\'e}} et~al.,}{{Labb{\'e}}
  et~al.}{2023b}]{Labbe2023}
{Labb{\'e}} I.,  et~al., 2023b, \mn@doi [\nat] {10.1038/s41586-023-05786-2},
  \href {https://ui.adsabs.harvard.edu/abs/2023Natur.616..266L} {616, 266}

\bibitem[\protect\citeauthoryear{{Lagattuta} et~al.,}{{Lagattuta}
  et~al.}{2019}]{Lagattuta2019}
{Lagattuta} D.~J.,  et~al., 2019, \mn@doi [\mnras] {10.1093/mnras/stz620},
  \href {https://ui.adsabs.harvard.edu/abs/2019MNRAS.485.3738L} {485, 3738}

\bibitem[\protect\citeauthoryear{{Lagattuta} et~al.,}{{Lagattuta}
  et~al.}{2022}]{Lagattuta2022}
{Lagattuta} D.~J.,  et~al., 2022, \mn@doi [\mnras] {10.1093/mnras/stac418},
  \href {https://ui.adsabs.harvard.edu/abs/2022MNRAS.514..497L} {514, 497}

\bibitem[\protect\citeauthoryear{{Langeroodi} \& {Hjorth}}{{Langeroodi} \&
  {Hjorth}}{2023}]{Langeroodi2023}
{Langeroodi} D.,  {Hjorth} J.,  2023, \mn@doi [\apjl]
  {10.3847/2041-8213/acfeec}, \href
  {https://ui.adsabs.harvard.edu/abs/2023ApJ...957L..27L} {957, L27}

\bibitem[\protect\citeauthoryear{{Laporte} et~al.,}{{Laporte}
  et~al.}{2015}]{Laporte2015}
{Laporte} N.,  et~al., 2015, \mn@doi [\aap] {10.1051/0004-6361/201425040},
  \href {https://ui.adsabs.harvard.edu/abs/2015A&A...575A..92L} {575, A92}

\bibitem[\protect\citeauthoryear{{Laporte}, {Ellis}, {Witten}  \&
  {Roberts-Borsani}}{{Laporte} et~al.}{2023}]{Laporte2023}
{Laporte} N.,  {Ellis} R.~S.,  {Witten} C.~E.~C.,   {Roberts-Borsani} G.,
  2023, \mn@doi [\mnras] {10.1093/mnras/stad1597}, \href
  {https://ui.adsabs.harvard.edu/abs/2023MNRAS.523.3018L} {523, 3018}

\bibitem[\protect\citeauthoryear{{Larson} et~al.,}{{Larson}
  et~al.}{2023}]{Larson2023}
{Larson} R.~L.,  et~al., 2023, \mn@doi [\apj] {10.3847/1538-4357/acfed4}, \href
  {https://ui.adsabs.harvard.edu/abs/2023ApJ...958..141L} {958, 141}

\bibitem[\protect\citeauthoryear{{Lovell}, {Harrison}, {Harikane}, {Tacchella}
  \& {Wilkins}}{{Lovell} et~al.}{2023}]{Lovell2023}
{Lovell} C.~C.,  {Harrison} I.,  {Harikane} Y.,  {Tacchella} S.,   {Wilkins}
  S.~M.,  2023, \mn@doi [\mnras] {10.1093/mnras/stac3224}, \href
  {https://ui.adsabs.harvard.edu/abs/2023MNRAS.518.2511L} {518, 2511}

\bibitem[\protect\citeauthoryear{{Mahler} et~al.,}{{Mahler}
  et~al.}{2019}]{Mahler2019}
{Mahler} G.,  et~al., 2019, \mn@doi [\apj] {10.3847/1538-4357/ab042b}, \href
  {https://ui.adsabs.harvard.edu/abs/2019ApJ...873...96M} {873, 96}

\bibitem[\protect\citeauthoryear{{Malekjani}, {Mc Conville}, {Colg{\'a}in},
  {Pourojaghi}  \& {Sheikh-Jabbari}}{{Malekjani}
  et~al.}{2023}]{Malekjani2023arXiv}
{Malekjani} M.,  {Mc Conville} R.,  {Colg{\'a}in} E.~{\'O}.,  {Pourojaghi} S.,
   {Sheikh-Jabbari} M.~M.,  2023, \mn@doi [arXiv e-prints]
  {10.48550/arXiv.2301.12725}, \href
  {https://ui.adsabs.harvard.edu/abs/2023arXiv230112725M} {p. arXiv:2301.12725}

\bibitem[\protect\citeauthoryear{{Marley} et~al.,}{{Marley}
  et~al.}{2021}]{Marley2021}
{Marley} M.~S.,  et~al., 2021, \mn@doi [\apj] {10.3847/1538-4357/ac141d}, \href
  {https://ui.adsabs.harvard.edu/abs/2021ApJ...920...85M} {920, 85}

\bibitem[\protect\citeauthoryear{{Martis} et~al.,}{{Martis}
  et~al.}{2024}]{Martis2024arXiv}
{Martis} N.~S.,  et~al., 2024, \mn@doi [arXiv e-prints]
  {10.48550/arXiv.2401.01945}, \href
  {https://ui.adsabs.harvard.edu/abs/2024arXiv240101945M} {p. arXiv:2401.01945}

\bibitem[\protect\citeauthoryear{{Matharu} et~al.,}{{Matharu}
  et~al.}{2023}]{Matharu2023}
{Matharu} J.,  et~al., 2023, \mn@doi [\apjl] {10.3847/2041-8213/acd1db}, \href
  {https://ui.adsabs.harvard.edu/abs/2023ApJ...949L..11M} {949, L11}

\bibitem[\protect\citeauthoryear{{Matthee} et~al.,}{{Matthee}
  et~al.}{2023}]{Matthee2023arXiv}
{Matthee} J.,  et~al., 2023, \mn@doi [arXiv e-prints]
  {10.48550/arXiv.2306.05448}, \href
  {https://ui.adsabs.harvard.edu/abs/2023arXiv230605448M} {p. arXiv:2306.05448}

\bibitem[\protect\citeauthoryear{{Mauerhofer} \& {Dayal}}{{Mauerhofer} \&
  {Dayal}}{2023}]{Mauerhofer2023}
{Mauerhofer} V.,  {Dayal} P.,  2023, \mn@doi [\mnras] {10.1093/mnras/stad2734},
  \href {https://ui.adsabs.harvard.edu/abs/2023MNRAS.526.2196M} {526, 2196}

\bibitem[\protect\citeauthoryear{{Menci}, {Castellano}, {Santini}, {Merlin},
  {Fontana}  \& {Shankar}}{{Menci} et~al.}{2022}]{Menci2022}
{Menci} N.,  {Castellano} M.,  {Santini} P.,  {Merlin} E.,  {Fontana} A.,
  {Shankar} F.,  2022, \mn@doi [\apjl] {10.3847/2041-8213/ac96e9}, \href
  {https://ui.adsabs.harvard.edu/abs/2022ApJ...938L...5M} {938, L5}

\bibitem[\protect\citeauthoryear{{Morishita} et~al.,}{{Morishita}
  et~al.}{2023}]{Morishita2023}
{Morishita} T.,  et~al., 2023, \mn@doi [\apjl] {10.3847/2041-8213/acb99e},
  \href {https://ui.adsabs.harvard.edu/abs/2023ApJ...947L..24M} {947, L24}

\bibitem[\protect\citeauthoryear{{Noirot} et~al.,}{{Noirot}
  et~al.}{2023}]{Noirot2023}
{Noirot} G.,  et~al., 2023, \mn@doi [\mnras] {10.1093/mnras/stad1019}, \href
  {https://ui.adsabs.harvard.edu/abs/2023MNRAS.525.1867N} {525, 1867}

\bibitem[\protect\citeauthoryear{{Oesch} et~al.,}{{Oesch}
  et~al.}{2023}]{Oesch2023}
{Oesch} P.~A.,  et~al., 2023, \mn@doi [\mnras] {10.1093/mnras/stad2411}, \href
  {https://ui.adsabs.harvard.edu/abs/2023MNRAS.525.2864O} {525, 2864}

\bibitem[\protect\citeauthoryear{{Oke} \& {Gunn}}{{Oke} \&
  {Gunn}}{1983}]{Oke1983}
{Oke} J.~B.,  {Gunn} J.~E.,  1983, \mn@doi [\apj] {10.1086/160817}, \href
  {https://ui.adsabs.harvard.edu/abs/1983ApJ...266..713O} {266, 713}

\bibitem[\protect\citeauthoryear{{Rieke} et~al.,}{{Rieke}
  et~al.}{2023}]{Rieke2023}
{Rieke} M.~J.,  et~al., 2023, \mn@doi [\pasp] {10.1088/1538-3873/acac53}, \href
  {https://ui.adsabs.harvard.edu/abs/2023PASP..135b8001R} {135, 028001}

\bibitem[\protect\citeauthoryear{{Roberts-Borsani}, {Ellis}  \&
  {Laporte}}{{Roberts-Borsani} et~al.}{2020}]{RobertsBorsani2020}
{Roberts-Borsani} G.~W.,  {Ellis} R.~S.,   {Laporte} N.,  2020, \mn@doi
  [\mnras] {10.1093/mnras/staa2085}, \href
  {https://ui.adsabs.harvard.edu/abs/2020MNRAS.497.3440R} {497, 3440}

\bibitem[\protect\citeauthoryear{{Robertson}}{{Robertson}}{2022}]{Robertson2022}
{Robertson} B.~E.,  2022, \mn@doi [\araa]
  {10.1146/annurev-astro-120221-044656}, \href
  {https://ui.adsabs.harvard.edu/abs/2022ARA&A..60..121R} {60, 121}

\bibitem[\protect\citeauthoryear{{Robertson} et~al.,}{{Robertson}
  et~al.}{2023}]{Robertson2023}
{Robertson} B.~E.,  et~al., 2023, \mn@doi [Nature Astronomy]
  {10.1038/s41550-023-01921-1}, \href
  {https://ui.adsabs.harvard.edu/abs/2023NatAs...7..611R} {7, 611}

\bibitem[\protect\citeauthoryear{{Salpeter}}{{Salpeter}}{1955}]{Salpeter1955}
{Salpeter} E.~E.,  1955, \mn@doi [\apj] {10.1086/145971}, \href
  {https://ui.adsabs.harvard.edu/abs/1955ApJ...121..161S} {121, 161}

\bibitem[\protect\citeauthoryear{{Schaerer} \& {de Barros}}{{Schaerer} \& {de
  Barros}}{2009}]{Schaerer2009}
{Schaerer} D.,  {de Barros} S.,  2009, \mn@doi [\aap]
  {10.1051/0004-6361/200911781}, \href
  {https://ui.adsabs.harvard.edu/abs/2009A&A...502..423S} {502, 423}

\bibitem[\protect\citeauthoryear{{Scott}}{{Scott}}{1992}]{Scott1992}
{Scott} D.,  1992, {Multivariate Density Estimation : Theory, Practice and
  Visualization}.
John Wiley and Sons Ltd, \mn@doi{10.1002/9780470316849}

\bibitem[\protect\citeauthoryear{{Soucail}}{{Soucail}}{1987}]{Soucail1987}
{Soucail} G.,  1987, The Messenger, \href
  {http://adsabs.harvard.edu/abs/1987Msngr..48...43S} {48, 43}

\bibitem[\protect\citeauthoryear{{Speagle}, {Steinhardt}, {Capak}  \&
  {Silverman}}{{Speagle} et~al.}{2014}]{Speagle2014}
{Speagle} J.~S.,  {Steinhardt} C.~L.,  {Capak} P.~L.,   {Silverman} J.~D.,
  2014, \mn@doi [\apjs] {10.1088/0067-0049/214/2/15}, \href
  {https://ui.adsabs.harvard.edu/abs/2014ApJS..214...15S} {214, 15}

\bibitem[\protect\citeauthoryear{{Stefanon}, {Bouwens}, {Labb{\'e}},
  {Illingworth}, {Gonzalez}  \& {Oesch}}{{Stefanon}
  et~al.}{2021}]{Stefanon2021}
{Stefanon} M.,  {Bouwens} R.~J.,  {Labb{\'e}} I.,  {Illingworth} G.~D.,
  {Gonzalez} V.,   {Oesch} P.~A.,  2021, \mn@doi [\apj]
  {10.3847/1538-4357/ac1bb6}, \href
  {https://ui.adsabs.harvard.edu/abs/2021ApJ...922...29S} {922, 29}

\bibitem[\protect\citeauthoryear{{Strait} et~al.,}{{Strait}
  et~al.}{2018}]{Strait2018}
{Strait} V.,  et~al., 2018, \mn@doi [\apj] {10.3847/1538-4357/aae834}, \href
  {http://adsabs.harvard.edu/abs/2018ApJ...868..129S} {868, 129}

\bibitem[\protect\citeauthoryear{{Strait} et~al.,}{{Strait}
  et~al.}{2023}]{Strait2023}
{Strait} V.,  et~al., 2023, \mn@doi [\apjl] {10.3847/2041-8213/acd457}, \href
  {https://ui.adsabs.harvard.edu/abs/2023ApJ...949L..23S} {949, L23}

\bibitem[\protect\citeauthoryear{{Su}, {Li}  \& {Feng}}{{Su}
  et~al.}{2023}]{Su2023arXiv}
{Su} B.-Y.,  {Li} N.,   {Feng} L.,  2023, \mn@doi [arXiv e-prints]
  {10.48550/arXiv.2306.05364}, \href
  {https://ui.adsabs.harvard.edu/abs/2023arXiv230605364S} {p. arXiv:2306.05364}

\bibitem[\protect\citeauthoryear{{Tamura} et~al.,}{{Tamura}
  et~al.}{2019}]{Tamura2019}
{Tamura} Y.,  et~al., 2019, \mn@doi [\apj] {10.3847/1538-4357/ab0374}, \href
  {https://ui.adsabs.harvard.edu/abs/2019ApJ...874...27T} {874, 27}

\bibitem[\protect\citeauthoryear{{Tamura} et~al.,}{{Tamura}
  et~al.}{2023}]{Tamura2023}
{Tamura} Y.,  et~al., 2023, \mn@doi [\apj] {10.3847/1538-4357/acd637}, \href
  {https://ui.adsabs.harvard.edu/abs/2023ApJ...952....9T} {952, 9}

\bibitem[\protect\citeauthoryear{{Trussler} et~al.,}{{Trussler}
  et~al.}{2023}]{Trussler2023arXiv}
{Trussler} J. A.~A.,  et~al., 2023, \mn@doi [arXiv e-prints]
  {10.48550/arXiv.2308.09665}, \href
  {https://ui.adsabs.harvard.edu/abs/2023arXiv230809665T} {p. arXiv:2308.09665}

\bibitem[\protect\citeauthoryear{{Vikaeus} et~al.,}{{Vikaeus}
  et~al.}{2024}]{Vikaeus2024MNRAS.tmp.}
{Vikaeus} A.,  et~al., 2024, \mn@doi [\mnras] {10.1093/mnras/stae323}, \href
  {https://ui.adsabs.harvard.edu/abs/2024MNRAS.tmp..316V} {}

\bibitem[\protect\citeauthoryear{{Wang} \& {Liu}}{{Wang} \&
  {Liu}}{2022}]{WangD2022arXiv}
{Wang} D.,  {Liu} Y.,  2022, \mn@doi [arXiv e-prints]
  {10.48550/arXiv.2301.00347}, \href
  {https://ui.adsabs.harvard.edu/abs/2023arXiv230100347W} {p. arXiv:2301.00347}

\bibitem[\protect\citeauthoryear{{Wang}, {Su}, {Zu}, {Yang}  \& {Feng}}{{Wang}
  et~al.}{2023a}]{WangSarXiv2023}
{Wang} P.,  {Su} B.-Y.,  {Zu} L.,  {Yang} Y.,   {Feng} L.,  2023a, \mn@doi
  [arXiv e-prints] {10.48550/arXiv.2307.11374}, \href
  {https://ui.adsabs.harvard.edu/abs/2023arXiv230711374W} {p. arXiv:2307.11374}

\bibitem[\protect\citeauthoryear{{Wang}, {Lei}, {Yuan}  \& {Fan}}{{Wang}
  et~al.}{2023b}]{WangYi2023}
{Wang} Y.-Y.,  {Lei} L.,  {Yuan} G.-W.,   {Fan} Y.-Z.,  2023b, \mn@doi [\apjl]
  {10.3847/2041-8213/acf46c}, \href
  {https://ui.adsabs.harvard.edu/abs/2023ApJ...954L..48W} {954, L48}

\bibitem[\protect\citeauthoryear{{Wang} et~al.,}{{Wang}
  et~al.}{2023c}]{WangB2023}
{Wang} B.,  et~al., 2023c, \mn@doi [\apjl] {10.3847/2041-8213/acfe07}, \href
  {https://ui.adsabs.harvard.edu/abs/2023ApJ...957L..34W} {957, L34}

\bibitem[\protect\citeauthoryear{{Wilkins} et~al.,}{{Wilkins}
  et~al.}{2024}]{Wilkins2024}
{Wilkins} S.~M.,  et~al., 2024, \mn@doi [\mnras] {10.1093/mnras/stad3558},
  \href {https://ui.adsabs.harvard.edu/abs/2024MNRAS.527.7965W} {527, 7965}

\bibitem[\protect\citeauthoryear{{Willott} et~al.,}{{Willott}
  et~al.}{2022}]{Willott2022}
{Willott} C.~J.,  et~al., 2022, \mn@doi [\pasp] {10.1088/1538-3873/ac5158},
  \href {https://ui.adsabs.harvard.edu/abs/2022PASP..134b5002W} {134, 025002}

\bibitem[\protect\citeauthoryear{{Windhorst} et~al.,}{{Windhorst}
  et~al.}{2023}]{Windhorst2023}
{Windhorst} R.~A.,  et~al., 2023, \mn@doi [\aj] {10.3847/1538-3881/aca163},
  \href {https://ui.adsabs.harvard.edu/abs/2023AJ....165...13W} {165, 13}

\bibitem[\protect\citeauthoryear{{Withers} et~al.,}{{Withers}
  et~al.}{2023}]{Withers2023}
{Withers} S.,  et~al., 2023, \mn@doi [\apjl] {10.3847/2041-8213/ad01c0}, \href
  {https://ui.adsabs.harvard.edu/abs/2023ApJ...958L..14W} {958, L14}

\bibitem[\protect\citeauthoryear{{Xiao} et~al.,}{{Xiao}
  et~al.}{2023}]{Xiao2023arXiv}
{Xiao} M.,  et~al., 2023, \mn@doi [arXiv e-prints] {10.48550/arXiv.2309.02492},
  \href {https://ui.adsabs.harvard.edu/abs/2023arXiv230902492X} {p.
  arXiv:2309.02492}

\makeatother
\end{thebibliography}




\appendix

\section{Full sample properties}
\label{sec:extra}

\begin{figure*}
    \centering
    \includegraphics[valign=m,width=0.6\linewidth]{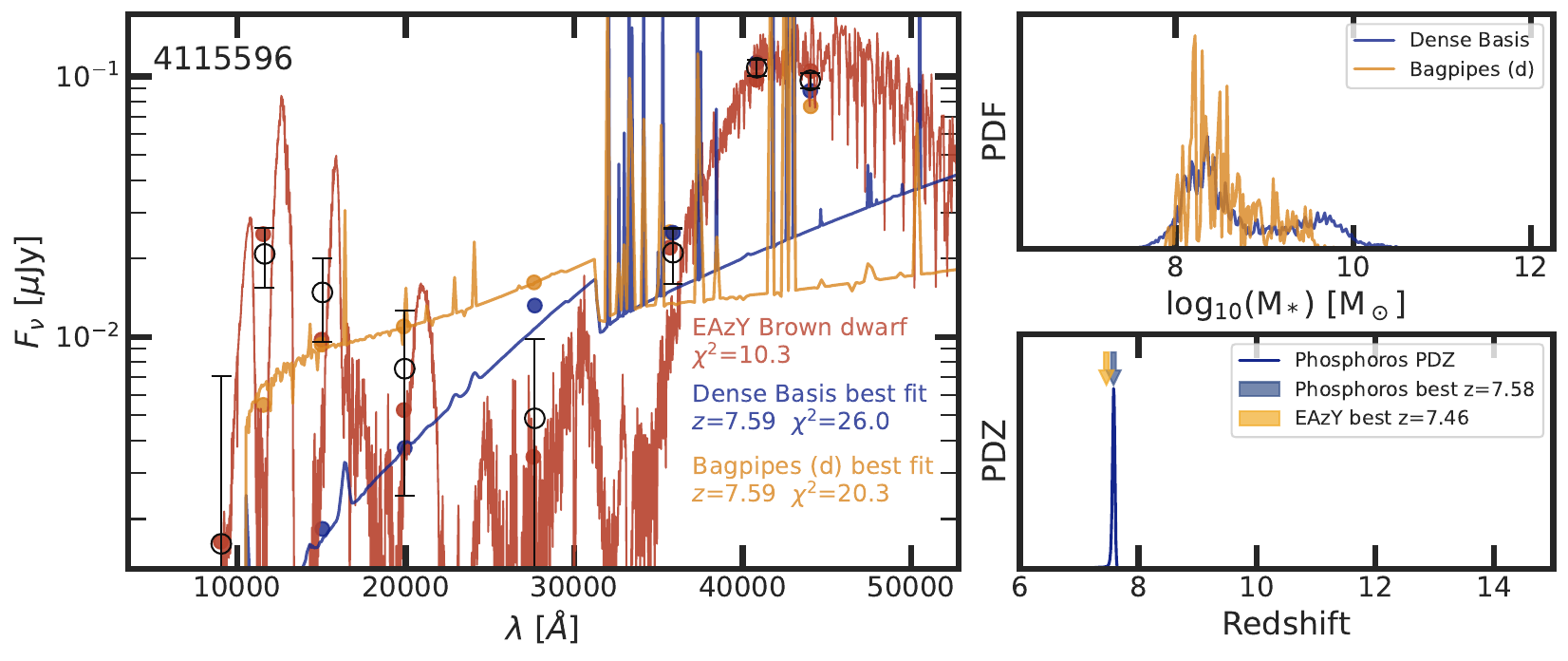}
    \includegraphics[valign=m,width=0.39\linewidth]{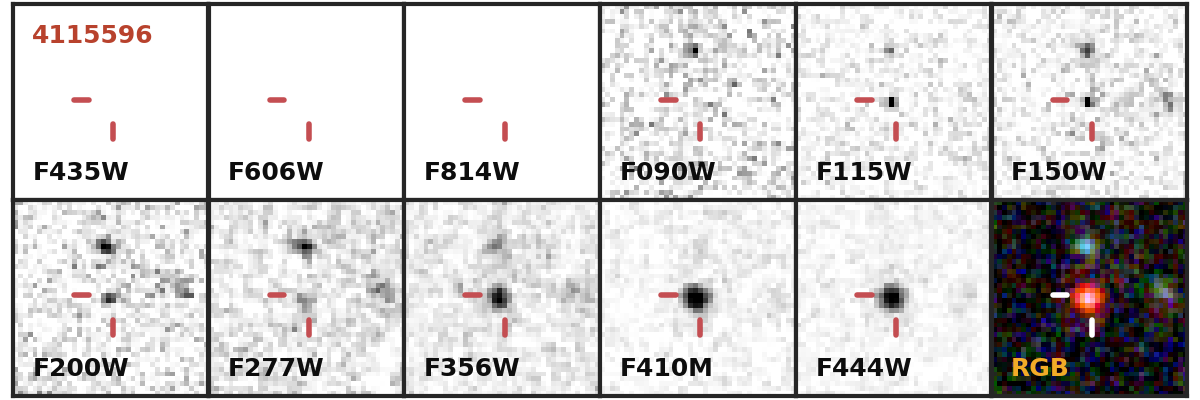}
    \caption{Photometry, SED fit and stamps of the CANUCS source id=4115596 that is much better fit by cool brown dwarf templates than high-redshift galaxy templates.}
    \label{fig:BrownDwarf}
\end{figure*}

\begin{figure*}
    \centering
    \includegraphics[valign=m,width=0.6\linewidth]{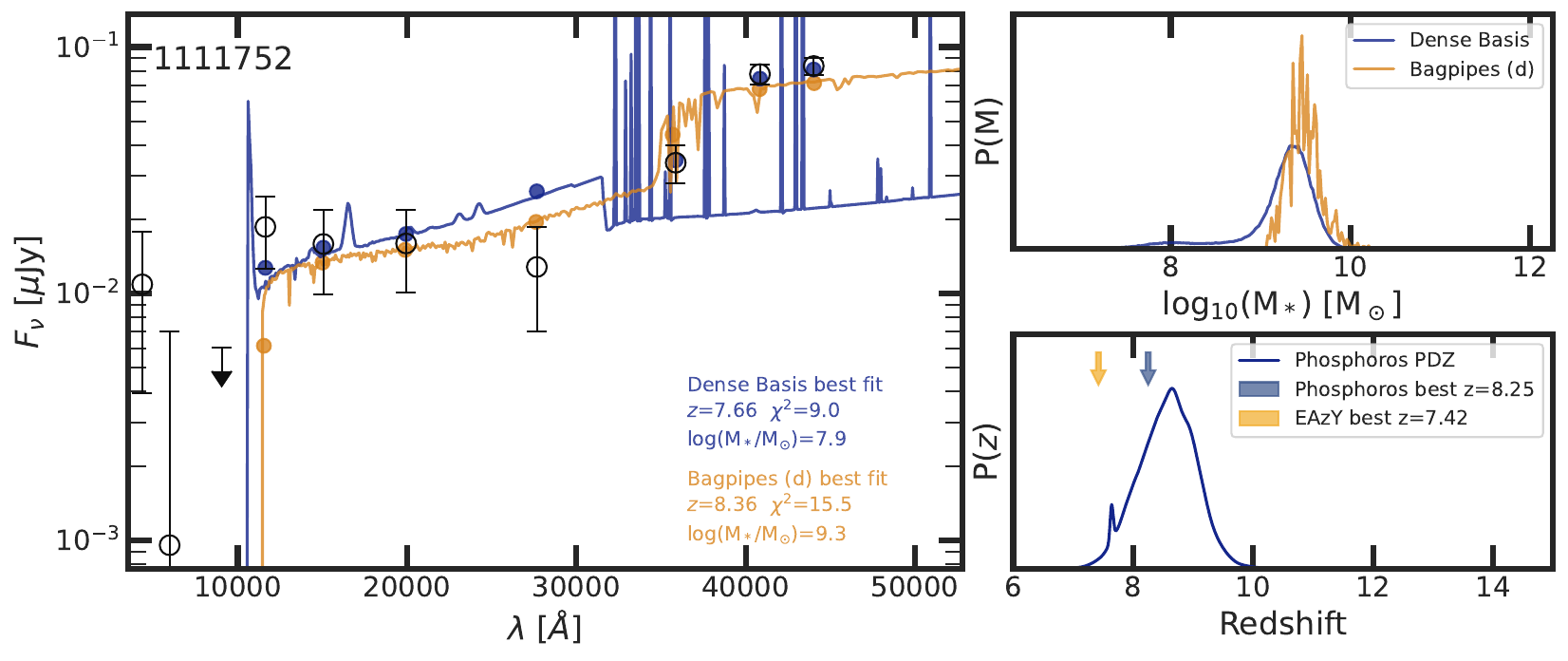}
    \includegraphics[valign=m,width=0.39\linewidth]{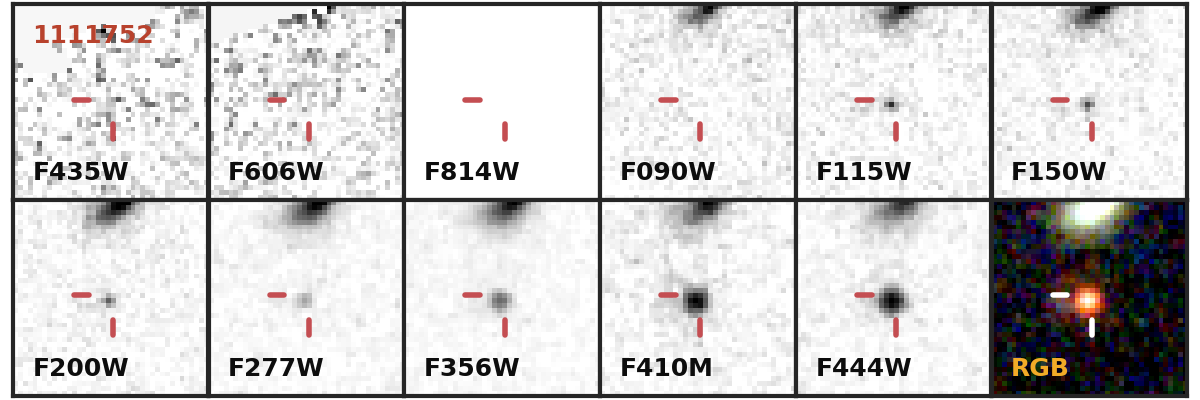}\\
    
    \includegraphics[valign=m,width=0.6\linewidth]{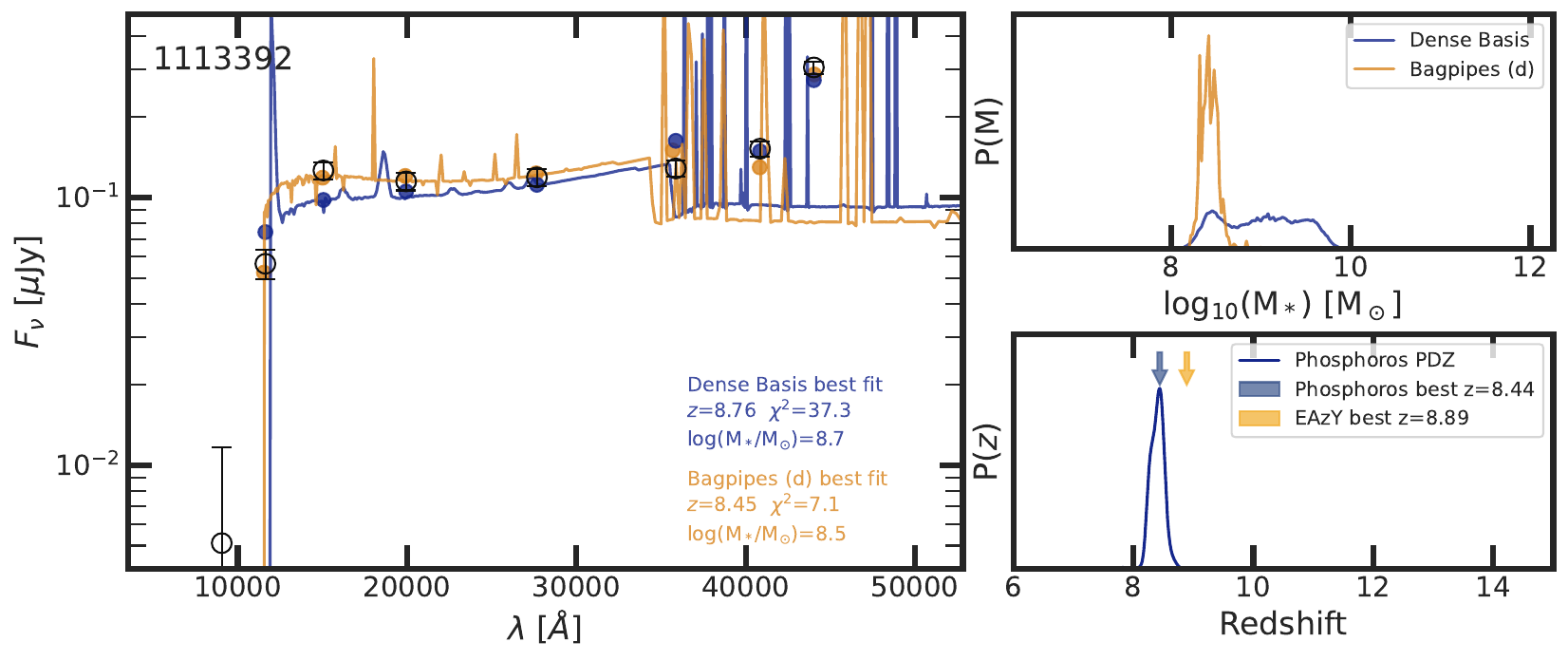}
    \includegraphics[valign=m,width=0.39\linewidth]{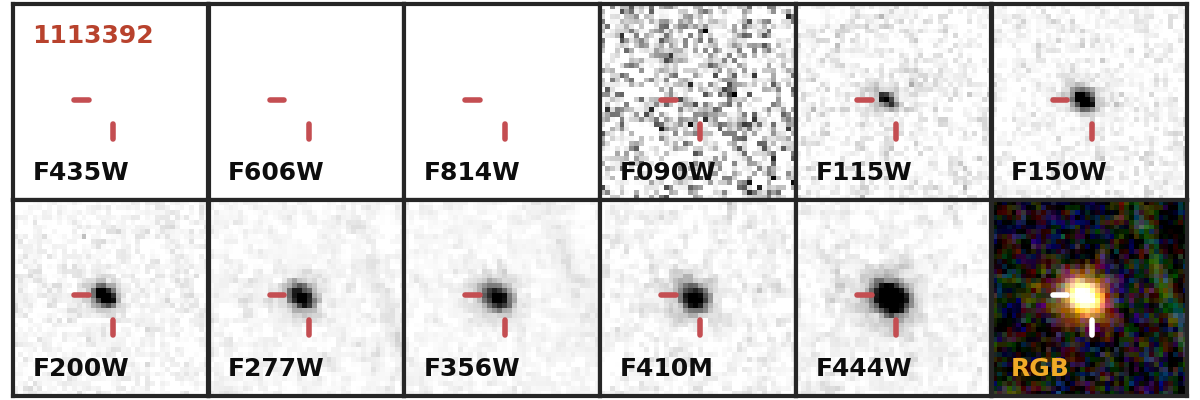}\\
    
    \includegraphics[valign=m,width=0.6\linewidth]{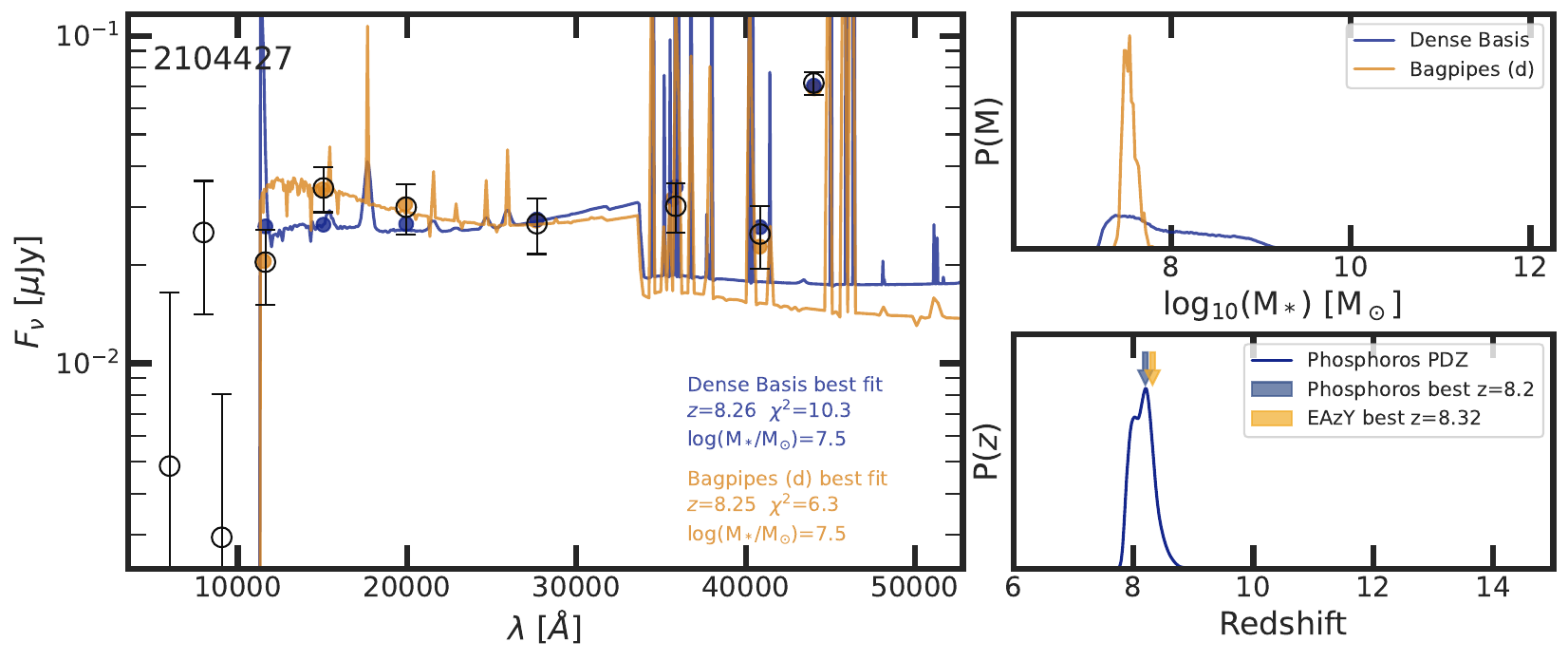}
    \includegraphics[valign=m,width=0.39\linewidth]{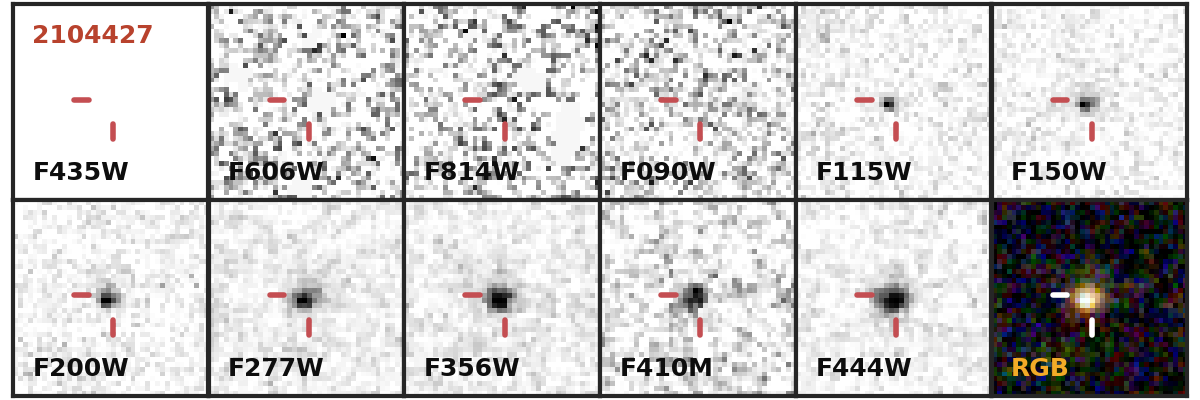}\\

    \includegraphics[valign=m,width=0.6\linewidth]{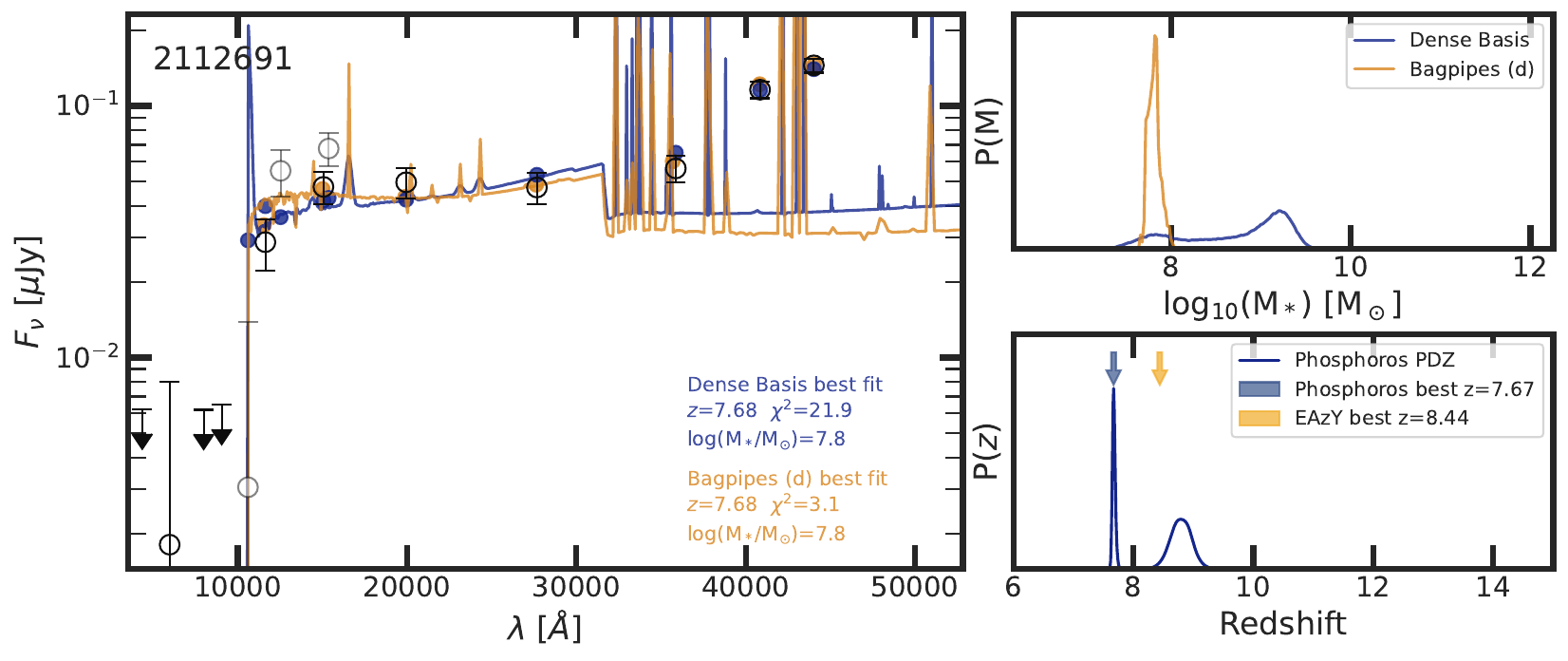}
    \includegraphics[valign=m,width=0.39\linewidth]{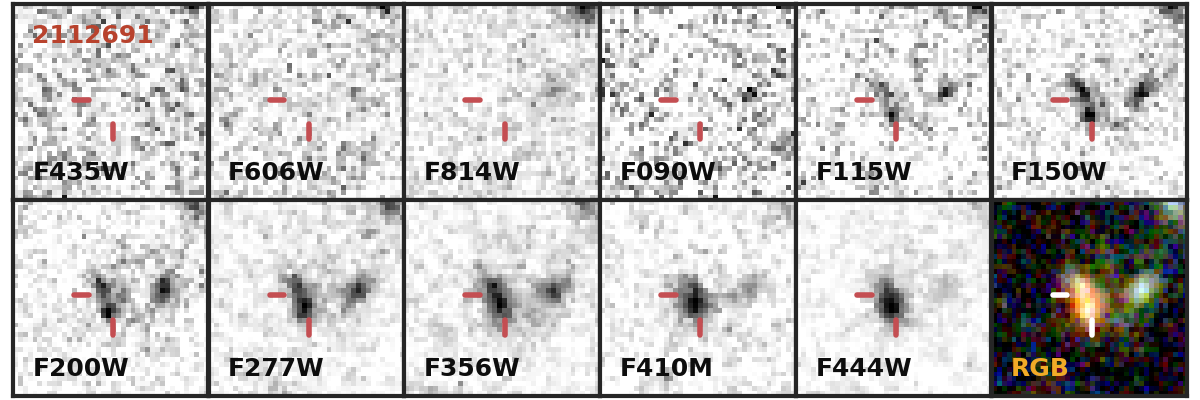}\\

    \caption{Observed photometry, SEDs fits, and image stamps for CANUCS double-break sources. For each source, the best model fits of both {\tt Dense Basis} (blue) and {\tt Bagpipes} (orange) are displayed, as well as the stellar mass posterior. In the case of {\tt Bagpipes}, only the best fit configuration (noted a, b, c, d, or e) is displayed. The redshift sub-panels show for each source the {\tt Phosphoros} P($z$) (blue curve), and the arrows indicate the {\tt Phosphoros} (blue) and {\tt EAzY} (orange) best fit redshifts. The stamps show the sources in the {\it HST} optical bands and in all the observed {\it JWST} bands, and the RGB stamps are using respectively a combination of F444W+F410M+F356W, F356W+F277W+F200W, and F150W+F115W+F090W for sources in \emph{CLU} fields and F444W+F410M+F360M+F335M+F300M, F300M+F277W+F250M+F210M+F182M, and F162M+F150W+F140M+F115W+F090W for sources in the \emph{NCF} fields.}
    \label{fig:SED1}
\end{figure*}

\begin{figure*}
    \centering
    \includegraphics[valign=m,width=0.6\linewidth]{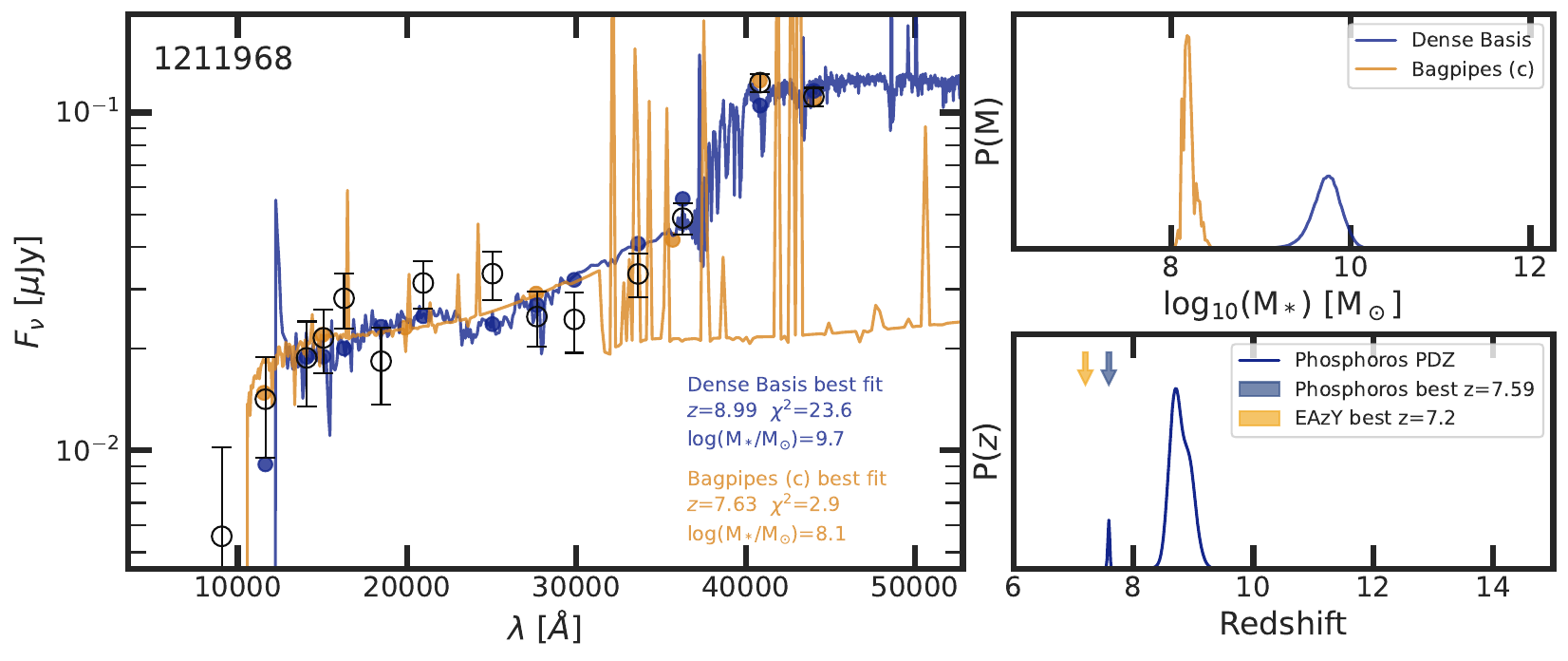}
    \includegraphics[valign=m,width=0.39\linewidth]{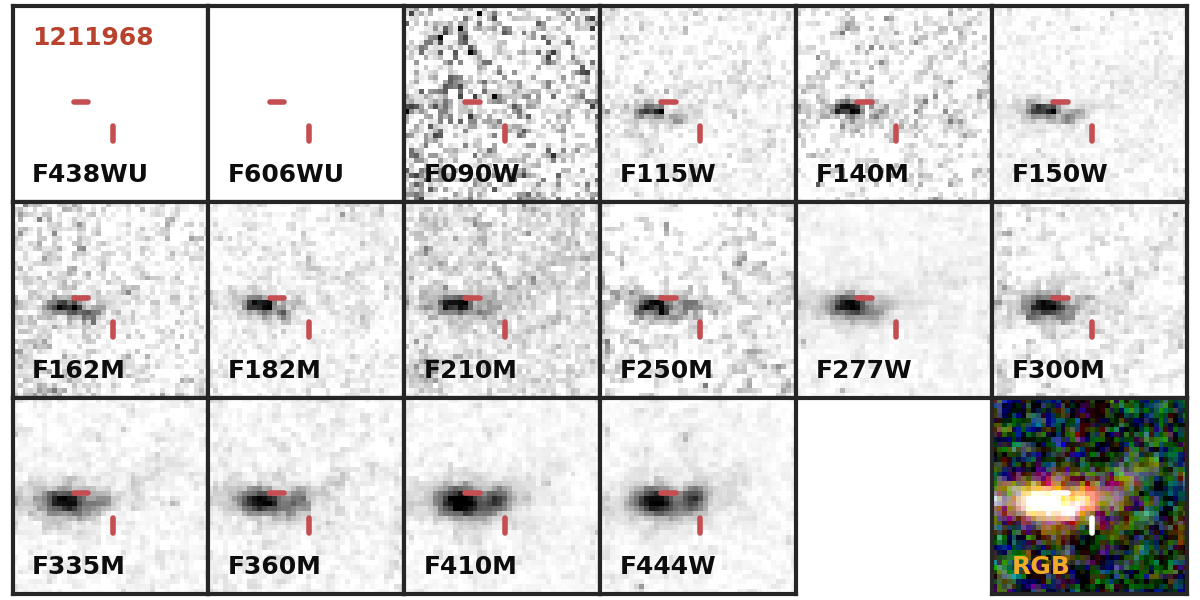}\\
    
    \includegraphics[valign=m,width=0.6\linewidth]{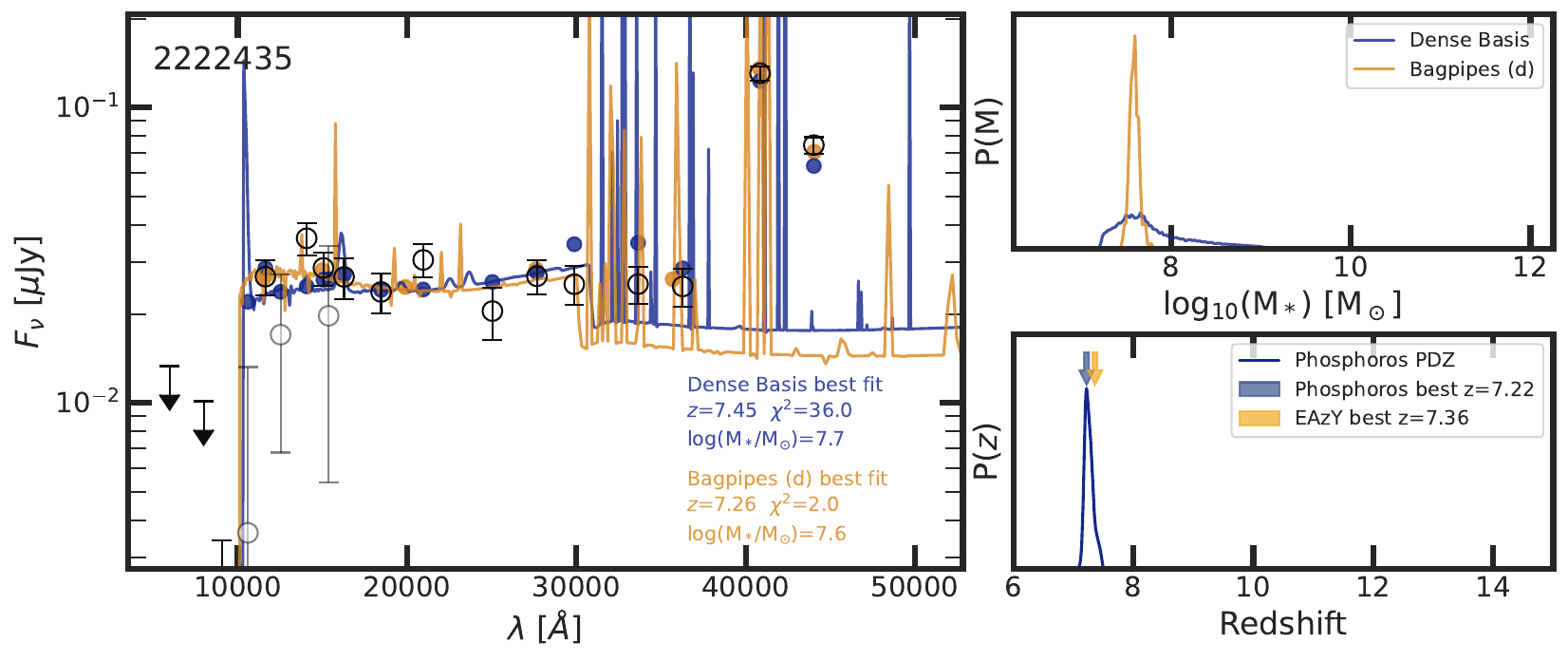}
    \includegraphics[valign=m,width=0.39\linewidth]{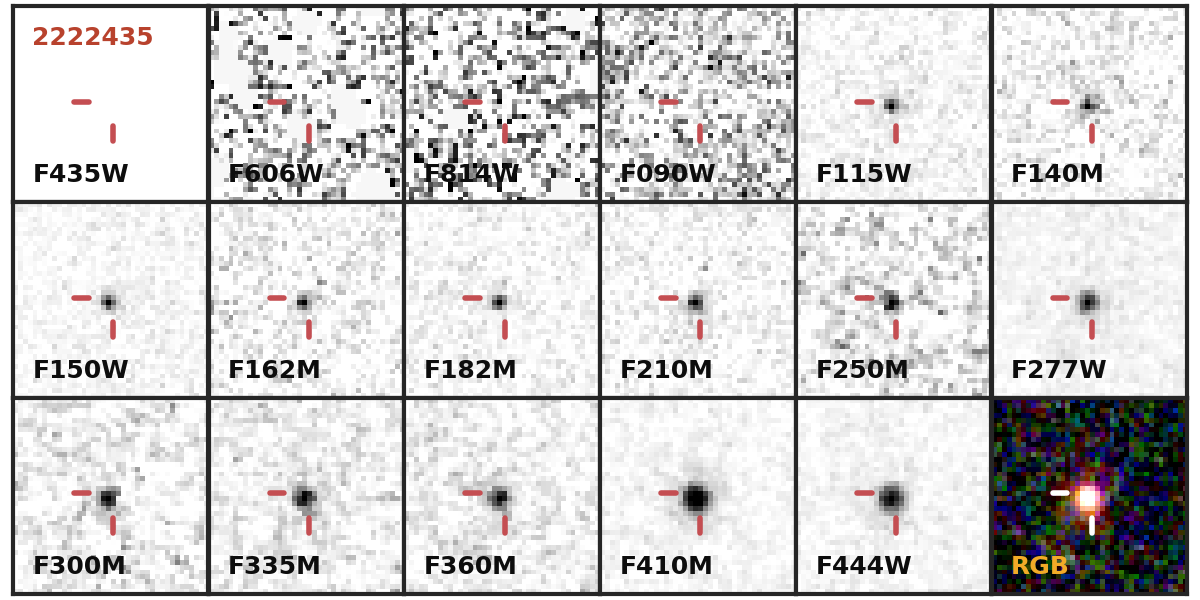}\\
    
    \includegraphics[valign=m,width=0.6\linewidth]{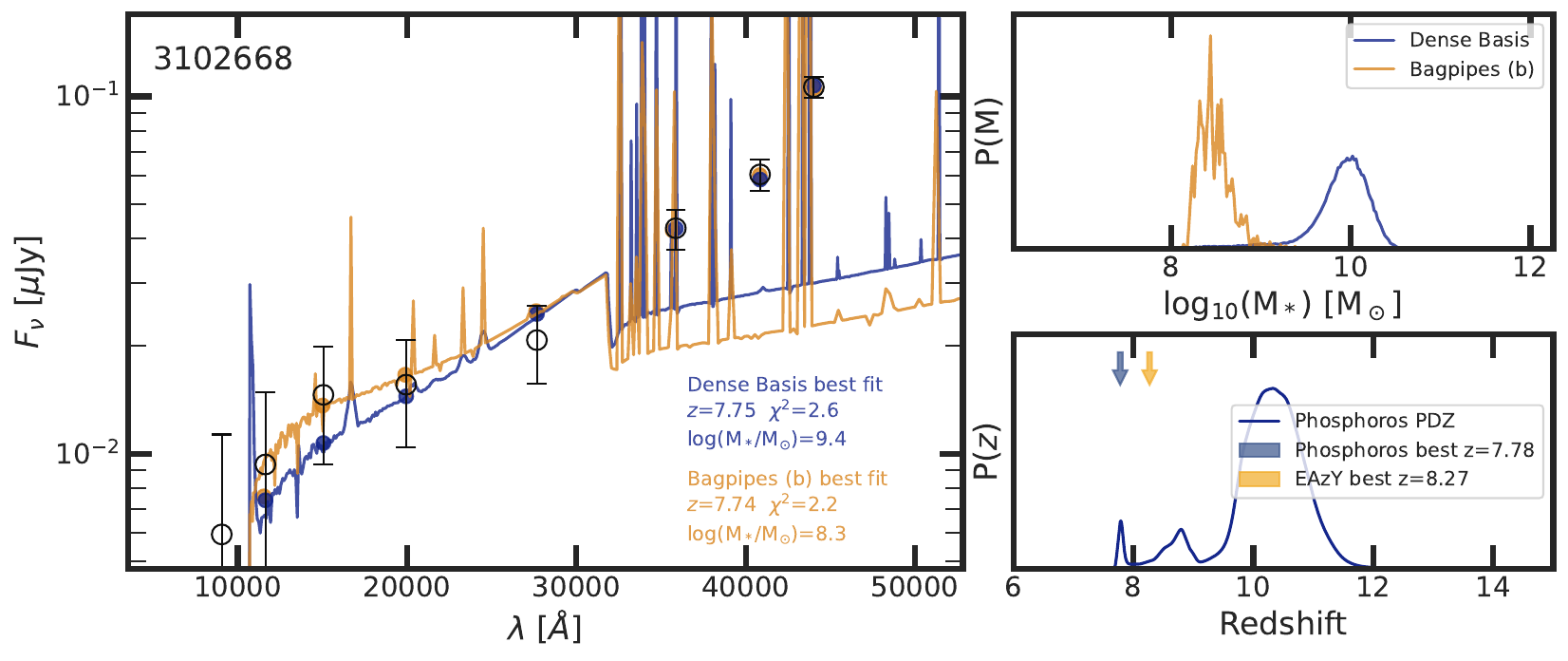}
    \includegraphics[valign=m,width=0.39\linewidth]{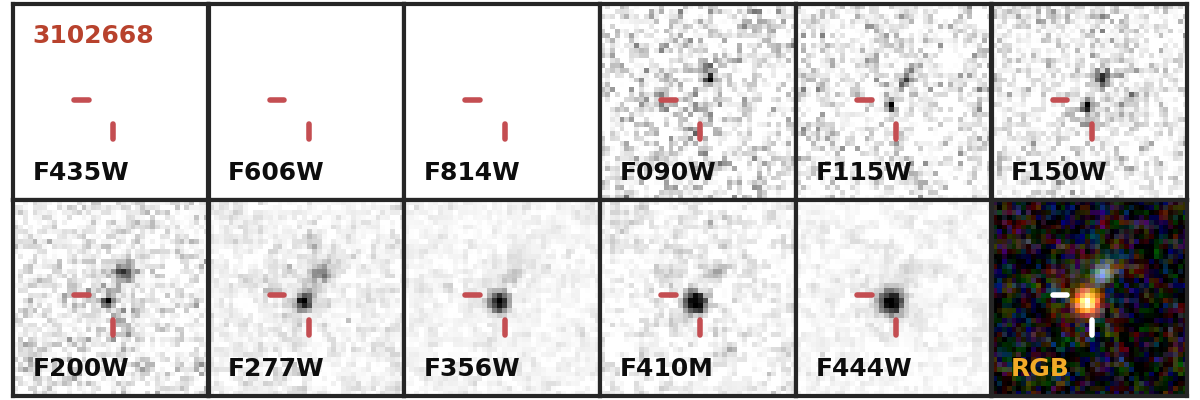}\\

    \includegraphics[valign=m,width=0.6\linewidth]{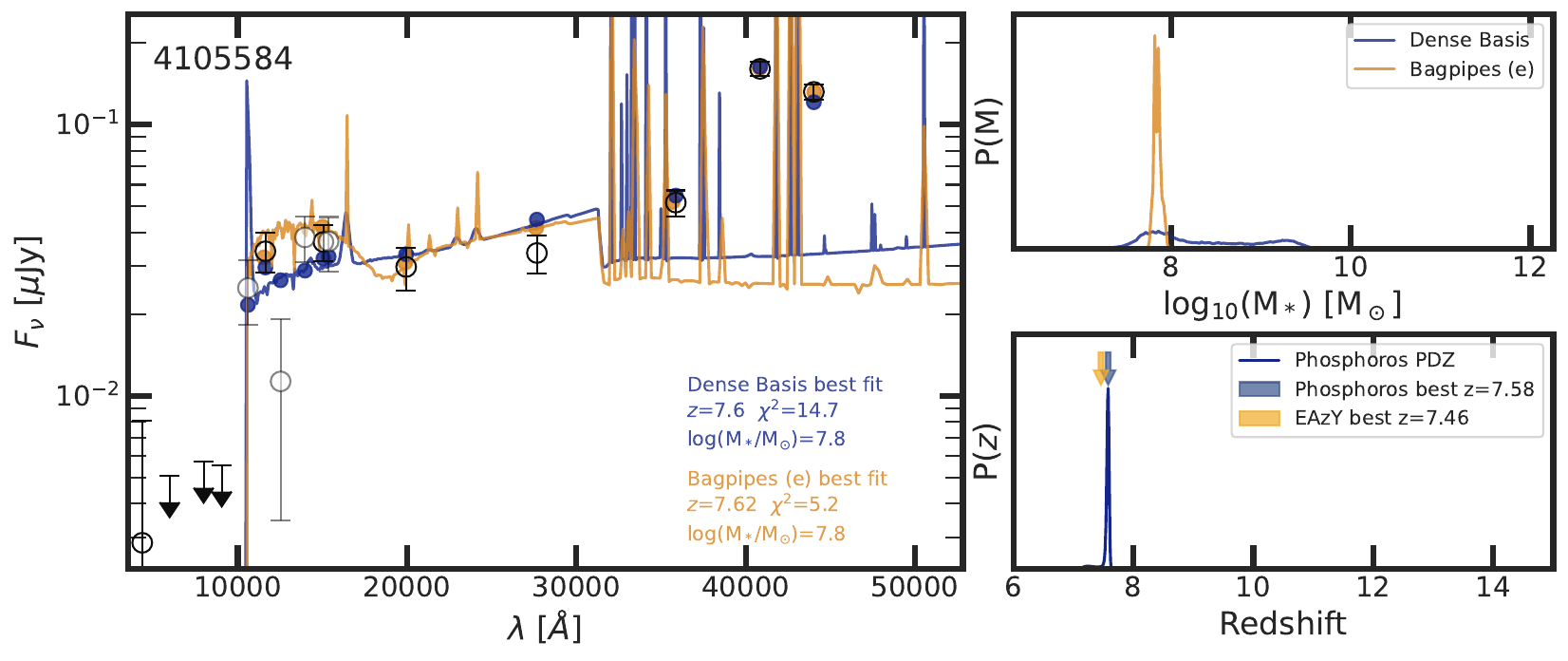}
    \includegraphics[valign=m,width=0.39\linewidth]{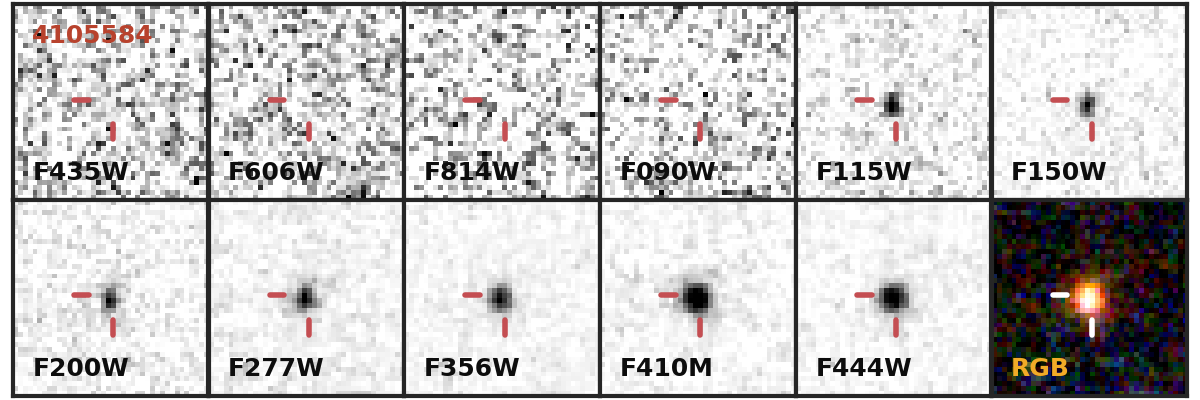}\\

    \includegraphics[valign=m,width=0.6\linewidth]{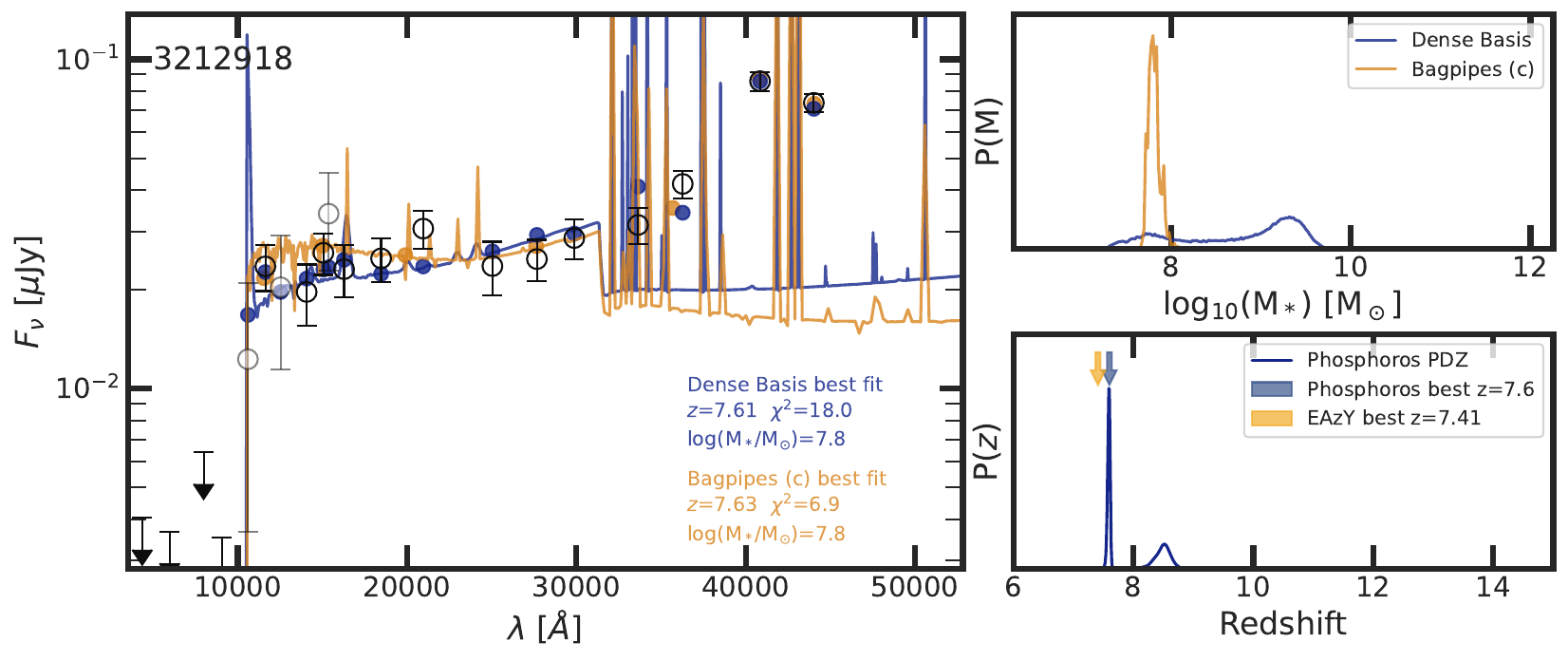}
    \includegraphics[valign=m,width=0.39\linewidth]{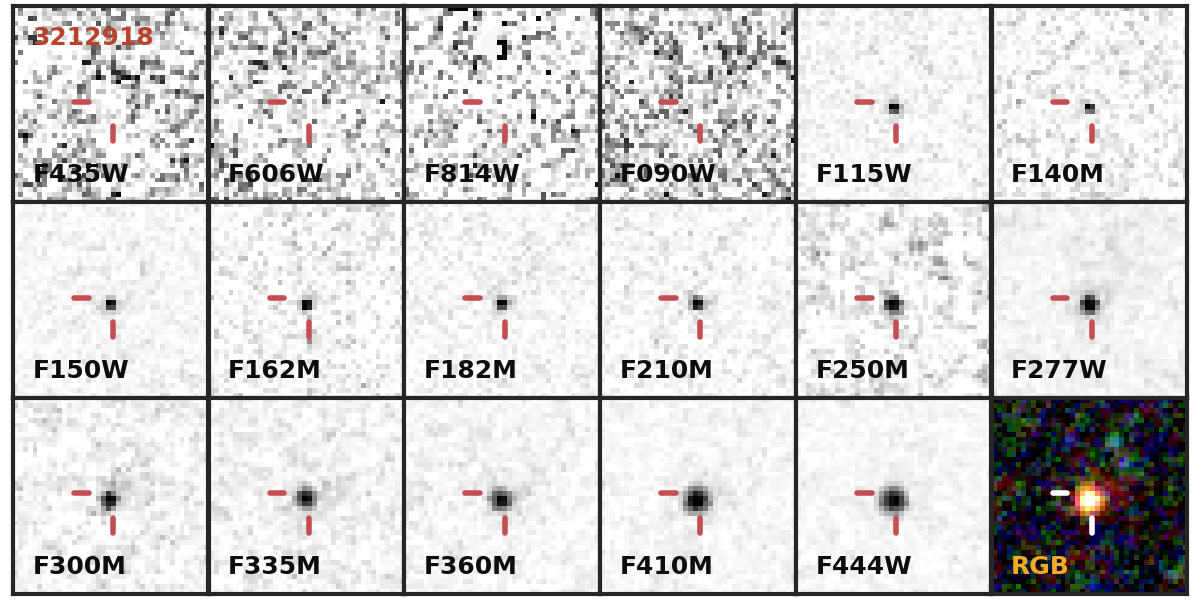}\\

    \caption{Same as Figure~\ref{fig:SED1}.}
    \label{fig:SED2}
\end{figure*}

\begin{figure*}
    \centering
    \includegraphics[valign=m,width=0.6\linewidth]{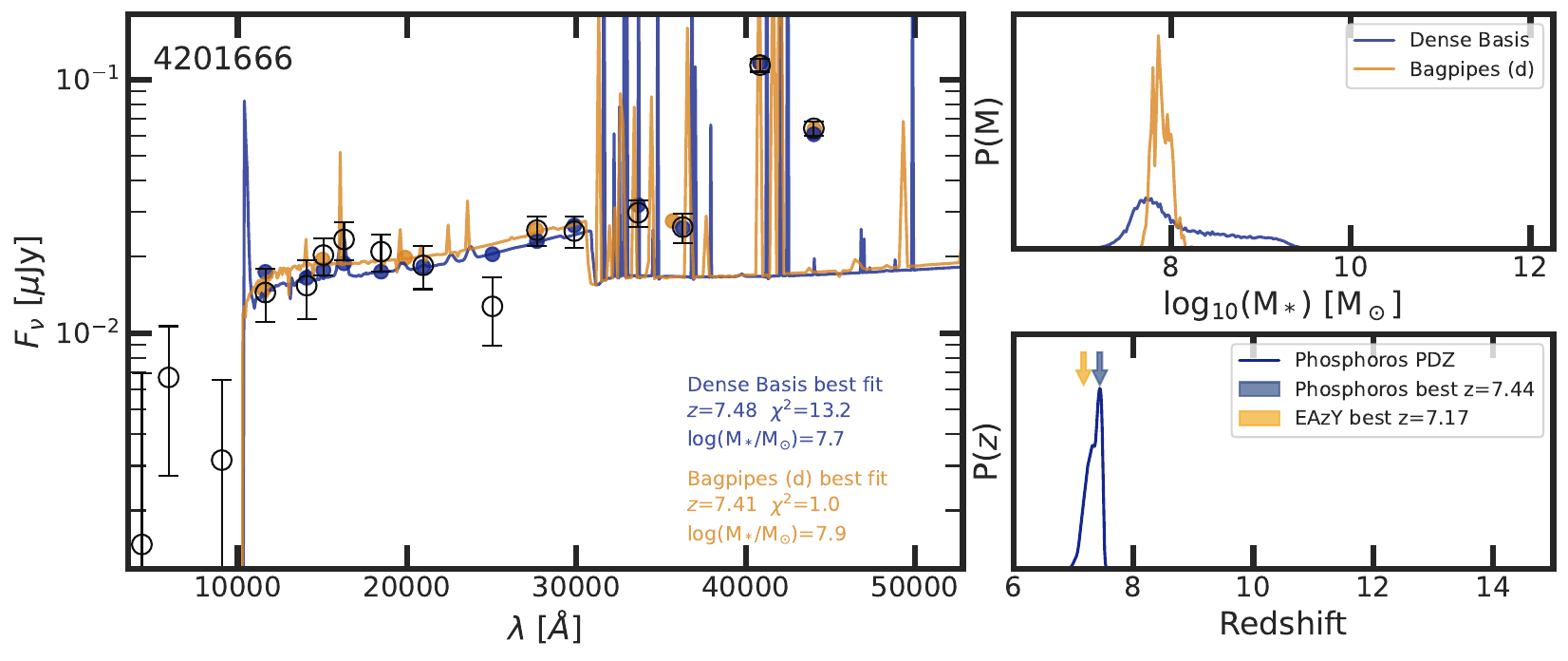}
    \includegraphics[valign=m,width=0.39\linewidth]{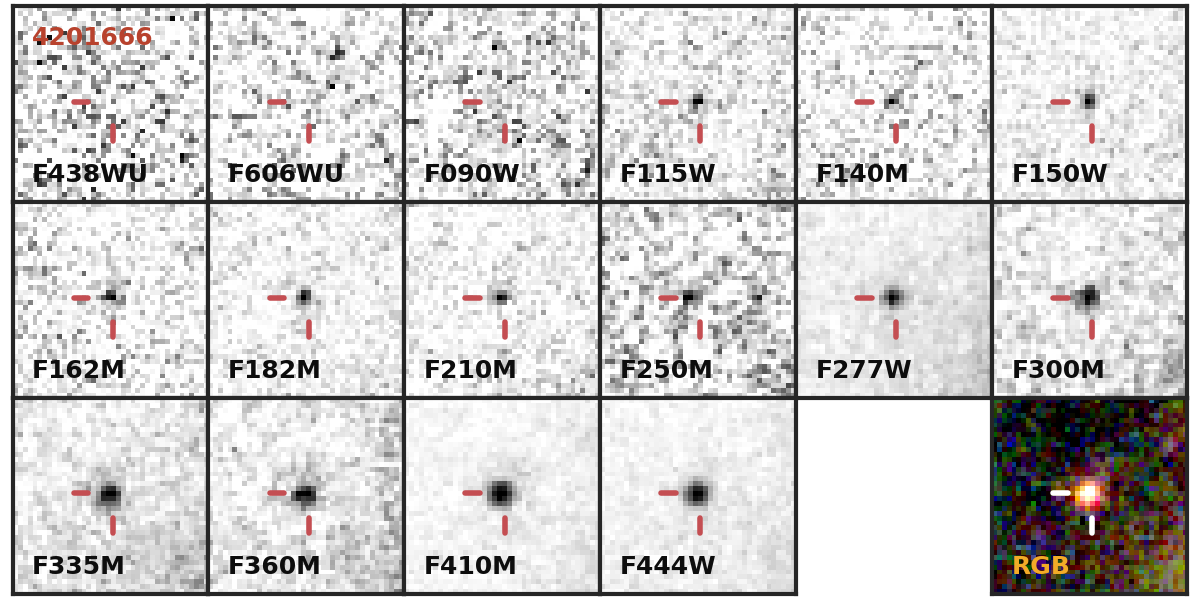}\\
    
    \includegraphics[valign=m,width=0.6\linewidth]{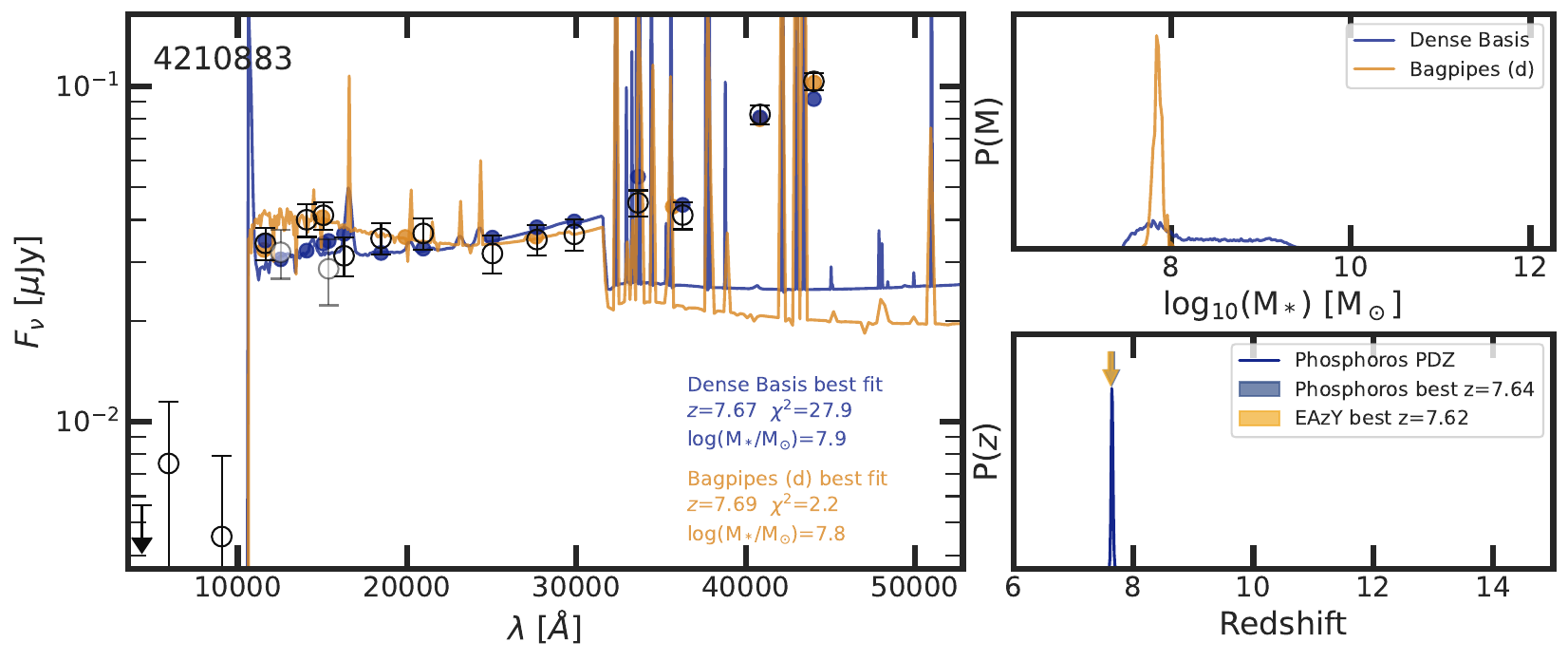}
    \includegraphics[valign=m,width=0.39\linewidth]{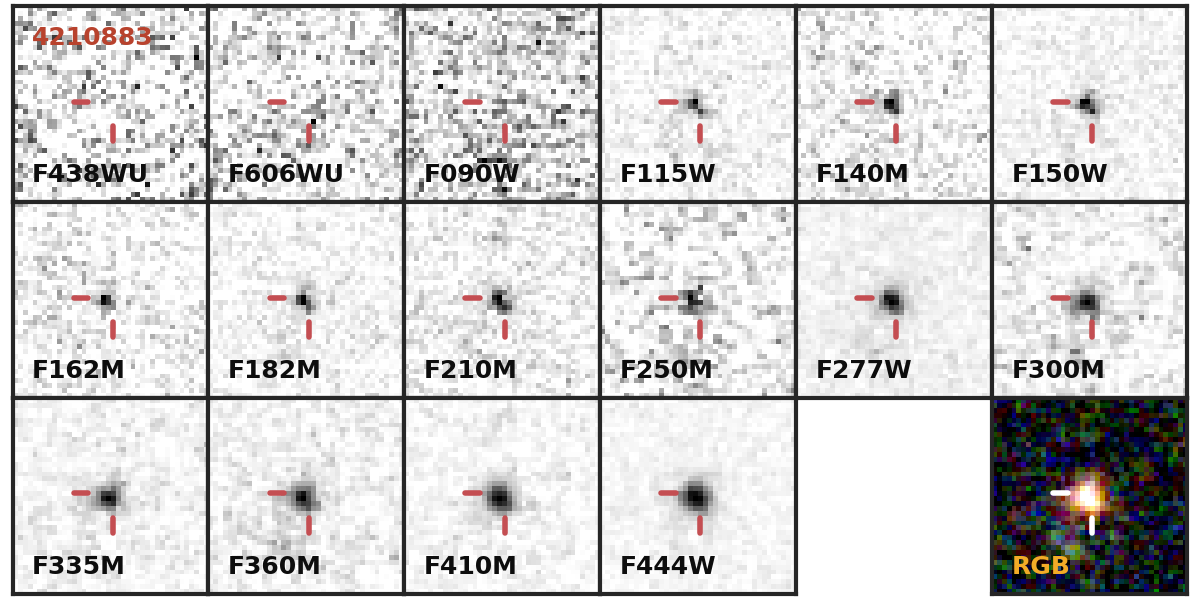}\\

    \includegraphics[valign=m,width=0.6\linewidth]{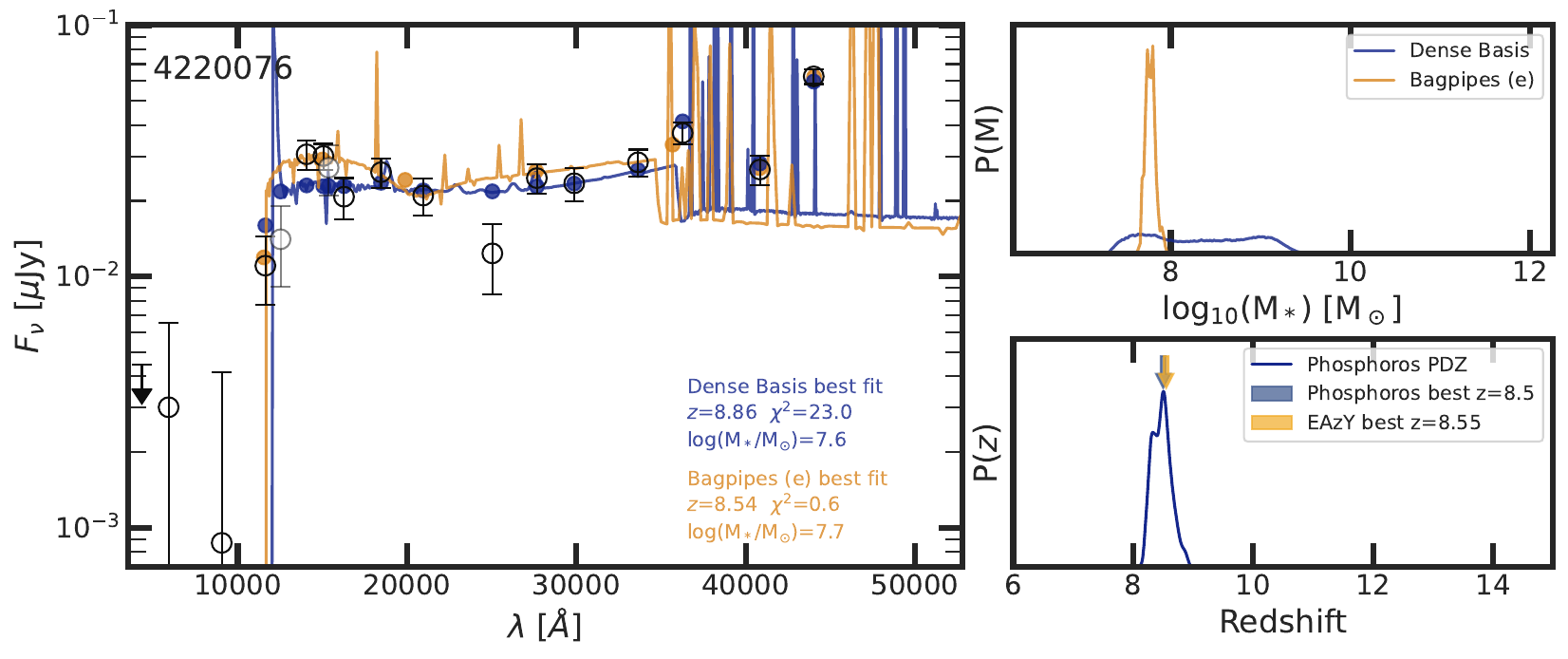}
    \includegraphics[valign=m,width=0.39\linewidth]{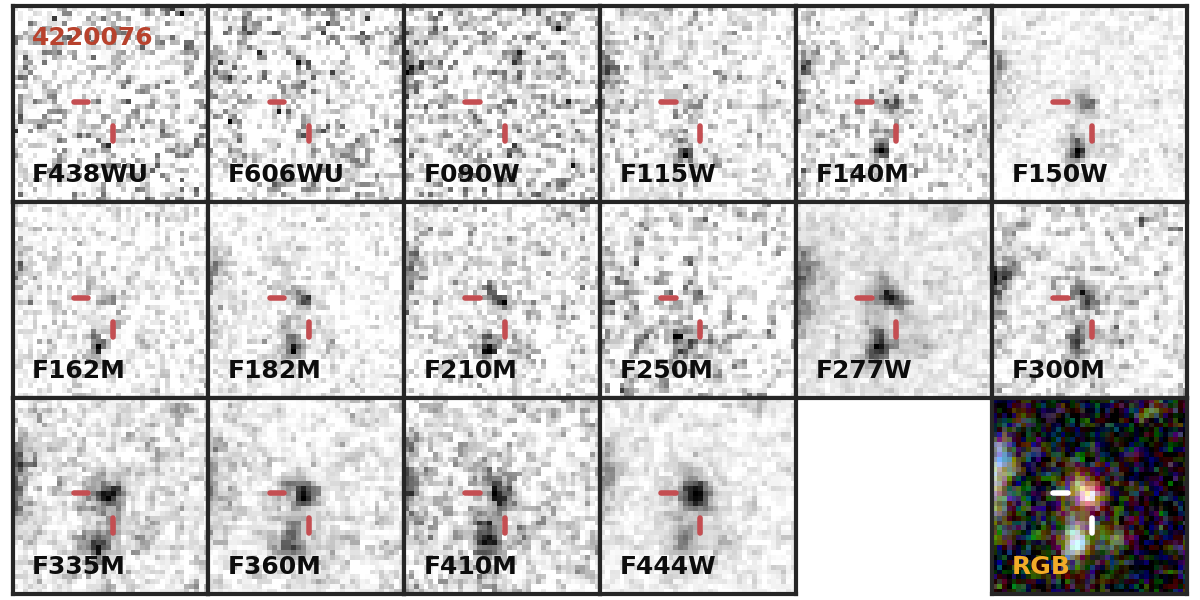}\\
    
    \includegraphics[valign=m,width=0.6\linewidth]{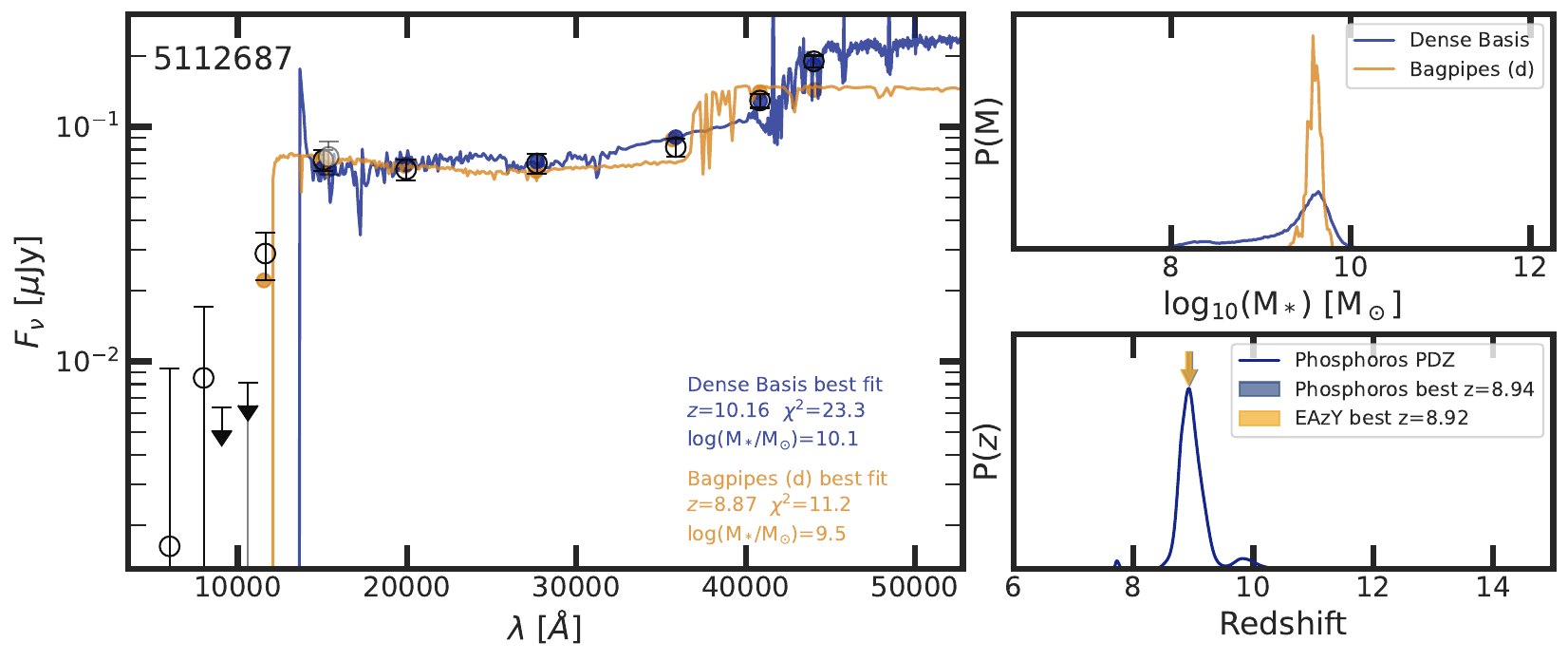}
    \includegraphics[valign=m,width=0.39\linewidth]{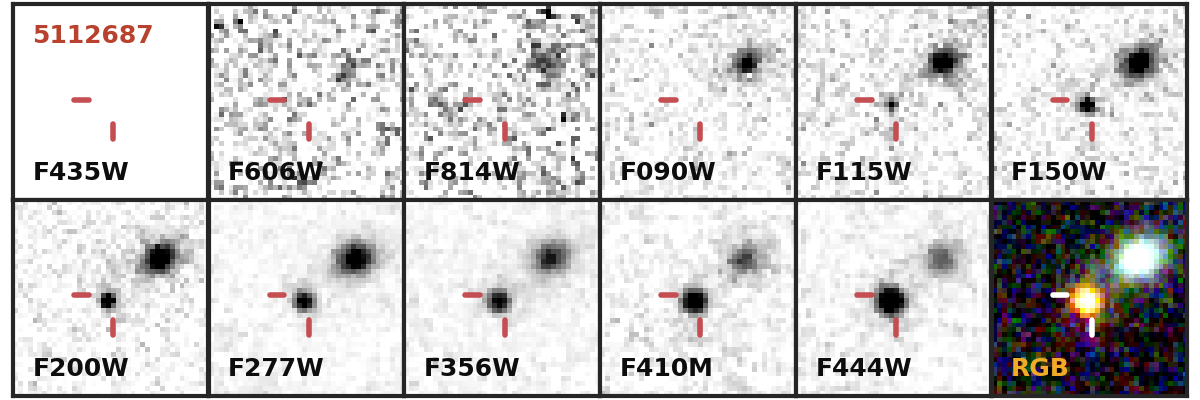}\\

    \includegraphics[valign=m,width=0.6\linewidth]{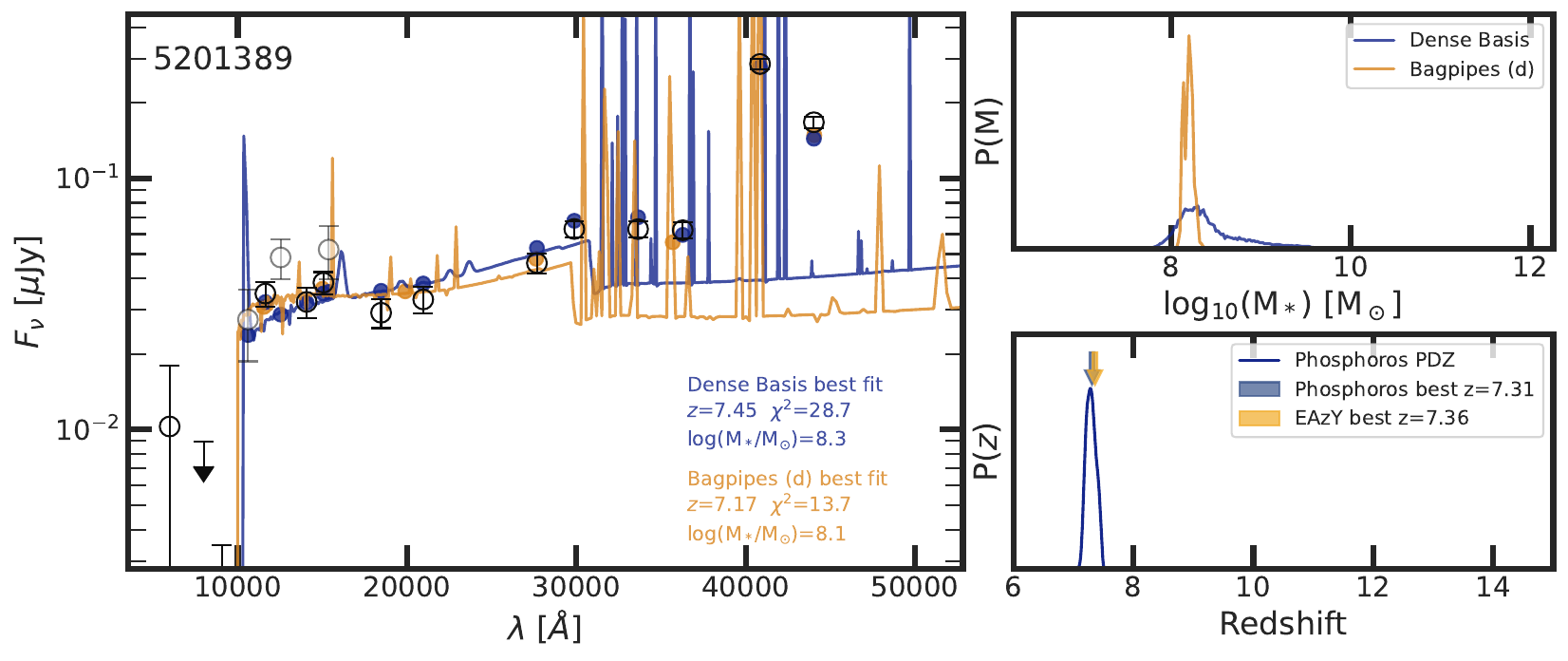}
    \includegraphics[valign=m,width=0.39\linewidth]{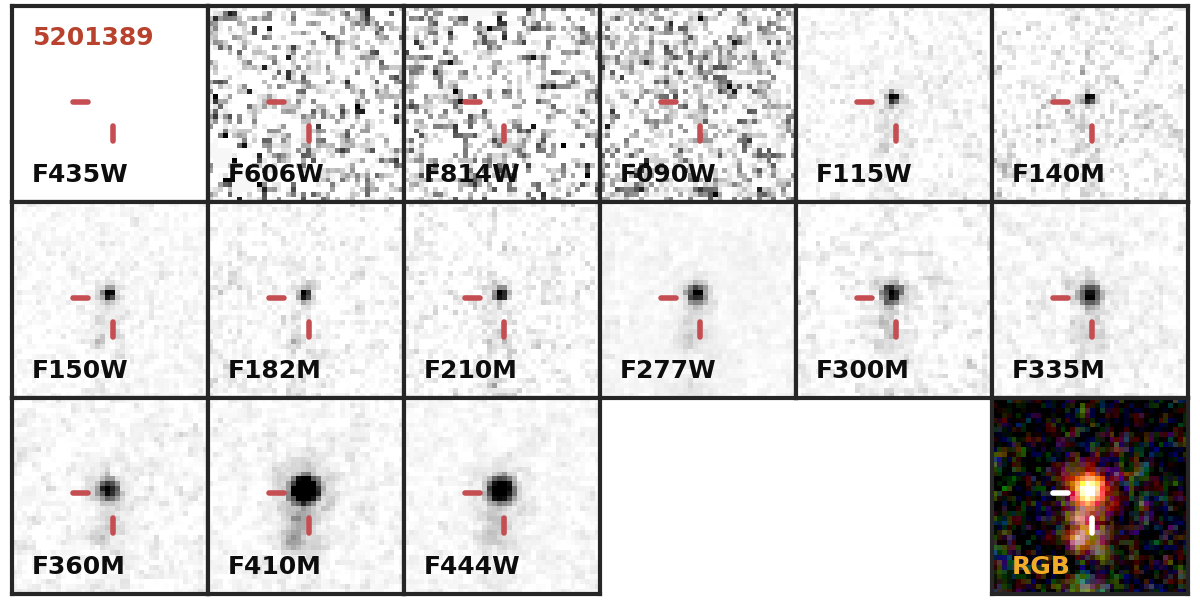}\\
    \caption{Same as Figure~\ref{fig:SED1}.}
    \label{fig:SED3}
\end{figure*}

In this appendix, we report the properties of the double-break galaxies obtained through fitting in Table~\ref{tab:parameters}. We also present the photometry, SED fits, stellar mass posteriors P(M) and photo-$z$ P($z$) and point estimates for all the source that do not have a spec-$z$ in Fig.~\ref{fig:SED1},~\ref{fig:SED2}, and~\ref{fig:SED3}. Figure~\ref{fig:BrownDwarf} show the fits of the CANUCS source id=4115596 which is suspected to be a cool brown dwarf, and thus display the best model fit of {\tt EAzY} for brown dwarfs. { Finally, Table~\ref{tab:depth} quotes the measured $3\sigma$ depths in $0\farcs 3$ apertures in all bands and fields.}

\begin{table*}
    \centering
\caption{ CANUCS and CEERS high-$z$ double-break galaxy fit properties. For CANUCS sources, the first two digits of object ID tag the field+pointing the object belongs to (see Table~\ref{tab:fields} for field+pointing codes). The reported photo-$z$ results are those of {\tt Phosphoros}. {\tt Bagpipes} stellar masses are the median of the 5 configurations, and the given errors are the lowest and highest results of the different configurations.}
    \begin{tabular}{c c c c c c c c c c c}
    \hline
    \hline
    \rule{0pt}{1.2em} id & R.A. & Dec. & $z_{\rm spec}$ & $z_{50}$ & $z_{\rm best}$ & $\mu$ & log(${\rm M}_{*}/{\rm M}_{\sun}$) & log(${\rm M}_{*}/{\rm M}_{\sun}$) & $f_{\rm line,410}$ & $f_{\rm line,444}$ \\
     & [deg] & [deg] & & & & & {\tt DB} & {\tt Bagpipes} &  \\
    \hline
    \rule{0pt}{1.2em} 1111752 & \phantom{0}64.39697\phantom{0} & -11.871979 & --- & \phantom{0}$8.58^{+0.46}_{-0.52}$ & 8.25 & \phantom{0}$1.559^{+0.056}_{-0.195}$ & $9.3^{+0.2}_{-0.4}$ & $9.6_{+0.1}^{+0.1}$ & --- & --- \\
    \rule{0pt}{1.2em} 1113392 & \phantom{0}64.427912 & -11.857791 & --- & \phantom{0}$8.4\phantom{0}^{+0.1}_{-0.12}$ & 8.44 & \phantom{0}$1.118^{+0.011}_{-0.037}$ & $9.0^{+0.5}_{-0.5}$ & $8.5_{+0.1}^{+0.1}$ & --- & --- \\
    \rule{0pt}{1.2em} 1211968 & \phantom{0}64.395294 & -11.799743 & --- & \phantom{0}$8.76^{+0.2}_{-0.15}$ & 7.59 & \phantom{0}$1.046^{+0.004}_{-0.015}$ & $9.7^{+0.2}_{-0.2}$ & $8.2_{+0.1}^{+0.1}$ & --- & --- \\
    \rule{0pt}{1.2em} 2104427 & \phantom{0}40.004549 & \phantom{0}-1.624585 & --- & \phantom{0}$8.15^{+0.18}_{-0.19}$ & 8.2\phantom{0} & \phantom{0}$1.517^{+0.031}_{-0.022}$ & $7.8^{+0.7}_{-0.4}$ & $7.5_{+0.1}^{+0.1}$ & --- & --- \\
    \rule{0pt}{1.2em} 2112549 & \phantom{0}39.994947 & \phantom{0}-1.574828 & $8.2434 \pm 0.0011$ & \phantom{0}$8.19^{+0.23}_{-0.23}$ & 8.22 & \phantom{0}$2.137^{+0.12}_{-0.053}$ & $7.9^{+0.8}_{-0.6}$ & $7.4_{+0.1}^{+0.1}$ & --- & $>0.39$ \\
    \rule{0pt}{1.2em} 2112691 & \phantom{0}39.994837 & \phantom{0}-1.574134 & --- & \phantom{0}$8.68^{+0.24}_{-1.01}$ & 7.67 & \phantom{0}$2.133^{+0.111}_{-0.057}$ & $8.9^{+0.4}_{-1.0}$ & $7.8_{+0.1}^{+0.1}$ & --- & --- \\
    \rule{0pt}{1.2em} 2113083 & \phantom{0}39.955449 & \phantom{0}-1.572618 & $7.6496 \pm 0.0005$ & \phantom{0}$7.65^{+0.03}_{-0.03}$ & 7.64 & \phantom{0}$2.932^{+0.056}_{-0.053}$ & $7.9^{+0.8}_{-0.6}$ & $7.4_{+0.1}^{+0.1}$ & 0.24 & \phantom{$>$}0.52 \\
    \rule{0pt}{1.2em} 2222435 & \phantom{0}40.046757 & \phantom{0}-1.586579 & --- & \phantom{0}$7.25^{+0.08}_{-0.06}$ & 7.22 & \phantom{0}$1.196^{+0.008}_{-0.005}$ & $7.7^{+0.6}_{-0.3}$ & $7.6_{+0.1}^{+0.1}$ & --- & --- \\
    \rule{0pt}{1.2em} 3102668 & \phantom{0}64.079044 & -24.125957 & --- & $10.29^{+0.52}_{-0.57}$ & 7.78 & \phantom{0}$1.099^{+0.002}_{-0.002}$ & $9.9^{+0.2}_{-0.3}$ & $8.9_{+0.6}^{+0.5}$ & --- & --- \\
    \rule{0pt}{1.2em} 3107165 & \phantom{0}64.039231 & -24.093198 & $8.3184 \pm 0.0001$ & \phantom{0}$8.35^{+0.07}_{-0.06}$ & 8.35 & \phantom{0}$1.634^{+0.012}_{-0.009}$ & $9.0^{+0.5}_{-0.4}$ & $8.5_{+0.1}^{+0.1}$ & 0.13 & \phantom{$>$}0.60 \\
    \rule{0pt}{1.2em} 3114647 & \phantom{0}64.038865 & -24.04459\phantom{0} & $8.1267 \pm 0.0006$ & \phantom{0}$8.23^{+0.31}_{-0.24}$ & 8.18 & \phantom{0}$1.54\phantom{0}^{+0.022}_{-0.009}$ & $8.2^{+0.8}_{-0.6}$ & $7.8_{+0.1}^{+0.1}$ & 0.04 & \phantom{$>$}0.67 \\
    \rule{0pt}{1.2em} 3212918 & \phantom{0}64.163246 & -24.090125 & --- & \phantom{0}$7.61^{+0.92}_{-0.04}$ & 7.6\phantom{0} & \phantom{0}$1.049^{+0.001}_{-0.001}$ & $8.9^{+0.4}_{-1.1}$ & $7.8_{+0.1}^{+0.1}$ & --- & --- \\
    \rule{0pt}{1.2em} 4105584 & 215.96759\phantom{0} & \phantom{-}24.071808 & --- & \phantom{0}$7.57^{+0.02}_{-0.03}$ & 7.58 & \phantom{0}$1.665^{+0.056}_{-0.118}$ & $8.1^{+1.0}_{-0.4}$ & $7.9_{+0.1}^{+0.1}$ & --- & --- \\
    \rule{0pt}{1.2em} 4117337 & 215.944453 & \phantom{-}24.068738 & $8.2953 \pm 0.0005$ & \phantom{0}$7.78^{+2.09}_{-0.05}$ & 7.75 & $27.808^{+11.826}_{-4.327}$ & $8.3^{+0.3}_{-0.8}$ & $7.7_{+0.5}^{+0.3}$ & 0.09 & \phantom{$>$}0.40 \\
    \rule{0pt}{1.2em} 4201666 & 215.870262 & \phantom{-}24.11036\phantom{0} & --- & \phantom{0}$7.34^{+0.1}_{-0.15}$ & 7.44 & \phantom{0}$1.023^{+0.002}_{-0.003}$ & $7.9^{+0.8}_{-0.3}$ & $7.9_{+0.1}^{+0.1}$ & --- & --- \\
    \rule{0pt}{1.2em} 4210883 & 215.874228 & \phantom{-}24.162117 & --- & \phantom{0}$7.64^{+0.02}_{-0.02}$ & 7.64 & \phantom{0}$1.011^{+0.001}_{-0.002}$ & $8.1^{+0.8}_{-0.4}$ & $7.8_{+0.1}^{+0.1}$ & --- & --- \\
    \rule{0pt}{1.2em} 4220076 & 215.860227 & \phantom{-}24.15628\phantom{0} & --- & \phantom{0}$8.47^{+0.16}_{-0.17}$ & 8.5\phantom{0} & \phantom{0}$1.009^{+0.001}_{-0.001}$ & $8.3^{+0.7}_{-0.7}$ & $7.8_{+0.1}^{+0.1}$ & --- & --- \\
    \rule{0pt}{1.2em} 5112687 & 177.390929 & \phantom{-}22.349767 & --- & \phantom{0}$8.95^{+0.22}_{-0.16}$ & 8.94 & \phantom{0}$1.057^{+0.001}_{-0.003}$ & $9.5^{+0.2}_{-0.7}$ & $9.5_{+0.1}^{+1.1}$ & --- & --- \\
    \rule{0pt}{1.2em} 5201389 & 177.380669 & \phantom{-}22.268024 & --- & \phantom{0}$7.29^{+0.1}_{-0.08}$ & 7.31 & \phantom{0}$1.004^{+0.001}_{-0.001}$ & $8.3^{+0.5}_{-0.2}$ & $8.3_{+0.1}^{+0.1}$ & --- & --- \\
        \hline
    \rule{0pt}{1.2em} 2859 & 214.840534 & 52.817942 & --- & $10.21^{+0.69}_{-0.31}$ & \phantom{0}9.86 & --- & $10.2^{+0.3}_{-0.3}$ & --- & --- & --- \\
\rule{0pt}{1.2em} 7274 & 214.806671 & 52.837802 & --- & \phantom{0}$7.41^{+0.38}_{-0.17}$ & \phantom{0}7.42 & --- & \phantom{0}$9.7^{+0.1}_{-0.2}$ & --- & --- & --- \\
\rule{0pt}{1.2em} 11184 & 214.892475 & 52.856892 & --- & \phantom{0}$7.12^{+0.29}_{-0.14}$ & \phantom{0}7.11 & --- & \phantom{0}$9.9^{+0.1}_{-0.1}$ & --- & --- & --- \\
\rule{0pt}{1.2em} 14924 & 214.87615\phantom{0} & 52.880833 & --- & \phantom{0}$9.36^{+0.73}_{-0.28}$ & \phantom{0}9.23 & --- & $10.0^{+0.1}_{-0.2}$ & --- & --- & --- \\
\rule{0pt}{1.2em} 16624 & 214.844772 & 52.892108 & --- & \phantom{0}$8.55^{+0.33}_{-0.15}$ & \phantom{0}8.73 & --- & \phantom{0}$9.4^{+0.3}_{-0.8}$ & --- & --- & --- \\
\rule{0pt}{1.2em} 21834 & 214.902227 & 52.93937 & --- & $10.12^{+0.68}_{-0.32}$ & 10.12 & --- & \phantom{0}$9.7^{+0.2}_{-0.2}$ & --- & --- & --- \\
\rule{0pt}{1.2em} 25666 & 214.956837 & 52.973153 & --- & \phantom{0}$5.59^{+0.02}_{-0.01}$ & \phantom{0}5.6\phantom{0} & --- & \phantom{0}$8.3^{+0.6}_{-0.2}$ & --- & --- & --- \\
\rule{0pt}{1.2em} 28984 & 215.002843 & 53.007594 & --- & \phantom{0}$8.33^{+0.51}_{-0.27}$ & \phantom{0}8.23 & --- & \phantom{0}$9.6^{+0.1}_{-0.2}$ & --- & --- & --- \\
\rule{0pt}{1.2em} 35300 & 214.830662 & 52.887777 & --- & \phantom{0}$9.24^{+0.86}_{-0.39}$ & \phantom{0}9.16 & --- & \phantom{0}$9.8^{+0.2}_{-0.2}$ & --- & --- & --- \\
\rule{0pt}{1.2em} 37888 & 214.91251\phantom{0} & 52.949435 & --- & \phantom{0}$6.57^{+1.76}_{-0.99}$ & \phantom{0}5.59 & --- & \phantom{0}$9.1^{+0.3}_{-1.1}$ & --- & --- & --- \\
\rule{0pt}{1.2em} 38094 & 214.983019 & 52.955999 & --- & \phantom{0}$7.72^{+0.3}_{-0.14}$ & \phantom{0}7.67 & --- & $10.8^{+0.1}_{-0.1}$ & --- & --- & --- \\
\rule{0pt}{1.2em} 39575 & 215.0054\phantom{00} & 52.996706 & 7.993 & $10.26^{+1.77}_{-1.25}$ & \phantom{0}8.97 & --- & \phantom{0}$9.5^{+0.2}_{-0.6}$ & --- & --- & --- \\
        \hline
    \end{tabular}
    \label{tab:parameters}
\end{table*}

\begin{table*}
    \centering
    \caption{ List of 3$\sigma$ magnitude depths reached in each field in $0\farcs 3$ circular apertures for all available filters. {\it HST}/ACS values marked with an asterisk are actually {\it HST}/WFC3/UVIS data, where F435W row presents the UVIS/F438W band depth and the F606W row shows the UVIS/F606W one.}
    \begin{tabular}{ccccccccccc}
        \hline
        \hline
         Band & \multicolumn{2}{c}{MACS~0417} & \multicolumn{2}{c}{Abell~370} & \multicolumn{2}{c}{MACS~0416} & \multicolumn{2}{c}{MACS~1423} & \multicolumn{2}{c}{MACS~1149} \\
         & \emph{CLU} & \emph{NCF} & \emph{CLU} & \emph{NCF} & \emph{CLU} & \emph{NCF} & \emph{CLU} & \emph{NCF} & \emph{CLU} & \emph{NCF}\\
         \hline
         \rule{0pt}{1.2em} F090W & 29.1 & 29.7 & 29.5 & 29.9 & 29.3 & 29.8 & 29.3 & 29.8 & 29.4 & 30.0  \\
         F115W & 29.4 & 29.7 & 29.5 & 29.9 & 29.3 & 29.8 & 29.3 & 29.8 & 29.4 & 29.9 \\
         F140M & ---  & 29.0 & ---  & 29.2 & ---  & 29.2 & ---  & 29.2 & ---  &  29.1 \\
         F150W & 29.4  & 30.0 & 29.6 & 30.1 & 29.5 & 30.0 & 29.6 & 30.1 & 29.6 & 30.3 \\
         F162M & ---  & 29.1 & ---  & 29.3 & ---  & 29.2 & ---  & 29.3 & ---  & --- \\
         F182M & ---  & 29.8 & ---  & 29.9 & ---  & 29.8 & ---  & 29.8 & ---  & 29.8 \\
         F200W & 29.5 & --- & 29.7 & --- & 29.7 & --- & 29.8 & --- & 29.7 & --- \\
         F210M & ---  & 29.5 & ---  & 29.7 & ---  & 29.7 & ---  & 29.7 & ---  & 29.7 \\
         F250M & ---  & 29.0 & ---  & 29.2 & ---  & 29.2 & ---  & 29.3 & ---  & --- \\
         F277W & 29.8  & 30.4 & 30.2 & 30.6 & 30.2 & 30.6 & 30.2 & 30.7 & 30.2 & 30.8 \\
         F300M & ---  & 29.4 & ---  & 29.6 & ---  & 29.7 & ---  & 29.7 & ---  & 29.6 \\
         F335M & ---  & 29.8 & ---  & 29.9 & ---  & 29.9 & ---  & 30.0 & ---  & 30.1 \\
         F356W & 30.0  & --- & 30.2 & --- & 30.2 & ---  & 30.3 & ---  & 30.2 & ---  \\
         F360M & ---  & 29.9 & ---  & 29.9 & ---  & 29.9 & ---  & 30.0 & ---  & 29.9 \\
         F410M & 29.1  & 29.8 & 29.5 & 29.9 & 29.5 & 29.9 & 29.6 & 29.9 & 29.5 & 29.9 \\
         F444W & 29.6  & 30.0 & 29.7 & 30.1 & 29.7 & 30.2 & 29.8 & 30.2 & 29.7 & 30.1 \\
         \hline
         \rule{0pt}{1.2em} F435W & 29.3  & 28.6* & 30.0 & 30.1 & 30.2 & 30.2 & 29.4 & 28.5* & 29.8 & 30.2 \\
         F606W & 29.7  & 29.3* & 30.0 & 30.2 & 30.4 & 29.9 & 29.8 & 29.2* & 30.2 & 30.2 \\
         F814W & 28.2  & ---  & 30.3 & 30.2 & 30.5 & 30.4 & 29.0 & --- & 30.3 & 30.3 \\
         \hline
         \rule{0pt}{1.2em} F105W & 28.0  & ---  & 30.0 & 30.2 & 30.1 & 30.3 & 28.7 & --- & 30.2 & 30.2 \\
         F125W & 27.2  & ---  & 29.5 & 29.6 & 29.8 & 29.8 & 28.5 & 28.3 & 30.0 & 29.8 \\
         F140W & 27.2  & ---  & 29.3 & 29.6 & 29.7 & 30.0 & 28.4 & --- & 29.5 & 29.6 \\
         F160W & 27.7  & ---  & 29.5 & 29.7 & 29.7 & 29.9 & 28.4 & 28.1 & 29.8 & 29.6 \\
         \hline
    \end{tabular}
    \label{tab:depth}
\end{table*}


\bsp	
\label{lastpage}
\end{document}